\documentclass[
    aps,
    10pt,
    prab,
    twocolumn,
    amsmath,
    superscriptaddress,
    preprintnumbers,
    floatfix,
    nofootinbib,
    ]{revtex4-2}

\makeatletter
\def\@fnsymbol#1{\ensuremath{\ifcase#1\or *\or \dagger\or \ddagger\or
   \mathsection\or \mathparagraph\or \|\or **\or \dagger\dagger
   \or \ddagger\ddagger \or \mathsection\mathsection \or \mathparagraph\mathparagraph \else\@ctrerr\fi}}
\makeatother

\usepackage[utf8]{inputenc}
\usepackage[detect-all,print-unity-mantissa = false]{siunitx}  
\usepackage{filemod}
\usepackage{graphicx}

\usepackage{wrapfig}
\usepackage{amsmath}
\usepackage{bbold}
\usepackage{nicefrac}
\usepackage{soul}
\usepackage[normalem]{ulem}

\newcommand{\ie}{\emph{i.e.} }
\newcommand{\frf}{f_\mathrm{rf} }
\newcommand{\phirf}{\varphi_\mathrm{rf} }
\newcommand{\fs}{f_\mathrm{s} }
\newcommand{\phis}{\varphi_\mathrm{s} }
\newcommand{\phirel}{\varphi_\mathrm{rel} }
\newcommand{\fc}{f_\mathrm{rev} }
\newcommand{\fv}{f_\mathrm{v} }
\newcommand{\nus}{\nu_\mathrm{s} }
\newcommand{\pv}{p_\mathrm{v} }
\newcommand{\ph}{p_\mathrm{h} }

\allowdisplaybreaks
\usepackage{hyperref}

\begin{document}

\title{Maintaining a Resonance Condition of an RF Spin Rotator Through a Feedback Loop in a Storage Ring}

\author{V.\,Hejny}
\affiliation{Institut f\"ur Kernphysik, Forschungszentrum J\"ulich, 52425 J\"ulich, Germany}

\author{A.\,Andres}
\thanks{Present address: GSI Helmholtzzentrum für Schwerionenforschung GmbH,
Planckstr. 1, 64291 Darmstadt, Germany}
\affiliation{III. Physikalisches Institut B, RWTH Aachen University, 52056 Aachen, Germany}
\affiliation{Institut f\"ur Kernphysik, Forschungszentrum J\"ulich, 52425 J\"ulich, Germany}

\author{J.\,Pretz}
\affiliation{III. Physikalisches Institut B, RWTH Aachen University, 52056 Aachen, Germany}
\affiliation{Institut f\"ur Kernphysik, Forschungszentrum J\"ulich, 52425 J\"ulich, Germany}	
	
\author{F.\,Abusaif}
\thanks{Present  address: Karlsruhe Institute of Technology, Hermann-von-Helmholtz-Platz 1, 76344 Eggenstein-Leopoldshafen,
	Germany}
\affiliation{III. Physikalisches Institut B, RWTH Aachen University, 52056 Aachen, Germany}
\affiliation{Institut f\"ur Kernphysik, Forschungszentrum J\"ulich, 52425 J\"ulich, Germany}

\author{A.\,Aggarwal}
\affiliation{Marian Smoluchowski Institute of Physics, Jagiellonian University, 30348 Cracow, Poland}

\author{A.\,Aksentev}
\affiliation{Institute for Nuclear Research, Russian Academy of Sciences, 117312 Moscow, Russia}

\author{B.\,Alberdi}
\thanks{Present address: Humboldt-Universität zu Berlin, Institut für Physik, Newton-Straße 15, 12489 Berlin, Germany}
\affiliation{III. Physikalisches Institut B, RWTH Aachen University, 52056 Aachen, Germany}
\affiliation{Institut f\"ur Kernphysik, Forschungszentrum J\"ulich, 52425 J\"ulich, Germany}	

\author{L.\,Barion}
\affiliation{University of Ferrara and Istituto Nazionale di Fisica Nucleare, 44100 Ferrara, Italy}

\author{I.\,Bekman}
\thanks{Present address: Institute of Technology and Engineering, Forschungszentrum J\"ulich, 52425 J\"ulich, Germany}
\affiliation{Institut f\"ur Kernphysik, Forschungszentrum J\"ulich, 52425 J\"ulich, Germany}

\author{M.\, Bey\ss}
\affiliation{III. Physikalisches Institut B, RWTH Aachen University, 52056 Aachen, Germany}
\affiliation{Institut f\"ur Kernphysik, Forschungszentrum J\"ulich, 52425 J\"ulich, Germany}

\author{C.\,B\"ohme}
\affiliation{Institut f\"ur Kernphysik, Forschungszentrum J\"ulich, 52425 J\"ulich, Germany}

\author{B.\,Breitkreutz}
\affiliation{Institut f\"ur Kernphysik, Forschungszentrum J\"ulich, 52425 J\"ulich, Germany}

\author{N.\,Canale}
\affiliation{University of Ferrara and Istituto Nazionale di Fisica Nucleare, 44100 Ferrara, Italy}

\author{G.\,Ciullo}
\affiliation{University of Ferrara and Istituto Nazionale di Fisica Nucleare, 44100 Ferrara, Italy}

\author{S.\,Dymov}
\affiliation{University of Ferrara and Istituto Nazionale di Fisica Nucleare, 44100 Ferrara, Italy}

\author{N.-O.\, Fr\"ohlich}
\thanks{Present address: Deutsches Elektronen-Synchrotron, 22607 Hamburg, Germany}
\affiliation{Institut f\"ur Kernphysik, Forschungszentrum J\"ulich, 52425 J\"ulich, Germany}

\author{R.\,Gebel}
\affiliation{Institut f\"ur Kernphysik, Forschungszentrum J\"ulich, 52425 J\"ulich, Germany}

\author{M.\,Gaisser}
\affiliation{III. Physikalisches Institut B, RWTH Aachen University, 52056 Aachen, Germany}

\author{K.\,Grigoryev}
\thanks{Present address: GSI Helmholtzzentrum für Schwerionenforschung GmbH,
Planckstr. 1, 64291 Darmstadt, Germany}
\affiliation{Institut f\"ur Kernphysik, Forschungszentrum J\"ulich, 52425 J\"ulich, Germany}

\author{D.\,Grzonka}
\affiliation{Institut f\"ur Kernphysik, Forschungszentrum J\"ulich, 52425 J\"ulich, Germany}

\author{J.\, Hetzel}
\thanks{Present address: GSI Helmholtzzentrum für Schwerionenforschung GmbH,
Planckstr. 1, 64291 Darmstadt, Germany}
\affiliation{Institut f\"ur Kernphysik, Forschungszentrum J\"ulich, 52425 J\"ulich, Germany}

\author{O.\,Javakhishvili}
\thanks{Present address: Faculty of Nuclear Sciences and Physical Engineering, Czech Technical University in Prague, 160 00 Praha 6, Czech Republic}
\affiliation{Department of Electrical and Computer Engineering, Agricultural University of Georgia, 0159 Tbilisi, Georgia}

\author{A.\,Kacharava}
\affiliation{Institut f\"ur Kernphysik, Forschungszentrum J\"ulich, 52425 J\"ulich, Germany}

\author{V.\,Kamerdzhiev}
\thanks{Present address: GSI Helmholtzzentrum für Schwerionenforschung GmbH,
Planckstr. 1, 64291 Darmstadt, Germany}
\affiliation{Institut f\"ur Kernphysik, Forschungszentrum J\"ulich, 52425 J\"ulich, Germany}

\author{S.\,Karanth}
\affiliation{Marian Smoluchowski Institute of Physics, Jagiellonian University, 30348 Cracow, Poland}

\author{I.\,Keshelashvili}
\thanks{Present address: GSI Helmholtzzentrum für Schwerionenforschung GmbH,
Planckstr. 1, 64291 Darmstadt, Germany}
\affiliation{Institut f\"ur Kernphysik, Forschungszentrum J\"ulich, 52425 J\"ulich, Germany}

\author{A.\,Kononov}
\affiliation{University of Ferrara and Istituto Nazionale di Fisica Nucleare, 44100 Ferrara, Italy}

\author{K.\,Laihem}
\thanks{Present address: GSI Helmholtzzentrum für Schwerionenforschung GmbH,
Planckstr. 1, 64291 Darmstadt, Germany}
\affiliation{III. Physikalisches Institut B, RWTH Aachen University, 52056 Aachen, Germany}

\author{A.\,Lehrach}
\affiliation{III. Physikalisches Institut B, RWTH Aachen University, 52056 Aachen, Germany}
\affiliation{Institut f\"ur Kernphysik, Forschungszentrum J\"ulich, 52425 J\"ulich, Germany}

\author{P.\,Lenisa}
\affiliation{University of Ferrara and Istituto Nazionale di Fisica Nucleare, 44100 Ferrara, Italy}

\author{N.\,Lomidze}
\affiliation{High Energy Physics Institute, Tbilisi State University, 0186 Tbilisi, Georgia}

\author{B.\,Lorentz}
\affiliation{GSI Helmholtzzentrum für Schwerionenforschung GmbH,
Planckstr. 1, 64291 Darmstadt, Germany}

\author{G.\,Macharashvili}
\affiliation{High Energy Physics Institute, Tbilisi State University, 0186 Tbilisi, Georgia}

\author{A.\,Magiera}
\affiliation{Marian Smoluchowski Institute of Physics, Jagiellonian University, 30348 Cracow, Poland}

\author{D.\,Mchedlishvili}
\affiliation{High Energy Physics Institute, Tbilisi State University, 0186 Tbilisi, Georgia}

\author{A.\,Melnikov}
\affiliation{Institute for Nuclear Research, Russian Academy of Sciences, 117312 Moscow, Russia}

\author{F.\,Müller}
\affiliation{III. Physikalisches Institut B, RWTH Aachen University, 52056 Aachen, Germany}
\affiliation{Institut f\"ur Kernphysik, Forschungszentrum J\"ulich, 52425 J\"ulich, Germany}

\author{A.\,Nass}
\affiliation{Institut f\"ur Kernphysik, Forschungszentrum J\"ulich, 52425 J\"ulich, Germany}

\author{N.N.\,Nikolaev}
\affiliation{L.D. Landau Institute for Theoretical Physics, 142432 Chernogolovka, Russia}

\author{D.\,Okropiridze}
\thanks{Present address: Ruhr-Universit\"at Bochum, Institut f\"ur Experimentalphysik I, 44801 Bochum, Germany}
\affiliation{High Energy Physics Institute, Tbilisi State University, 0186 Tbilisi, Georgia}

\author{A.\,Pesce}
\affiliation{Institut f\"ur Kernphysik, Forschungszentrum J\"ulich, 52425 J\"ulich, Germany}

\author{A.\,Piccoli}
\affiliation{University of Ferrara and Istituto Nazionale di Fisica Nucleare, 44100 Ferrara, Italy}

\author{V.\,Poncza}
\affiliation{Institut f\"ur Kernphysik, Forschungszentrum J\"ulich, 52425 J\"ulich, Germany}

\author{D.\,Prasuhn}
\affiliation{Institut f\"ur Kernphysik, Forschungszentrum J\"ulich, 52425 J\"ulich, Germany}

\author{F.\,Rathmann}
\thanks{Present address: Brookhaven National Laboratory, Upton, NY 11973, USA}
\affiliation{Institut f\"ur Kernphysik, Forschungszentrum J\"ulich, 52425 J\"ulich, Germany}

\author{A.\,Saleev}
\thanks{Present address: Institut f\"ur nachhaltige Wasserstoffwirtschaft, Forschungszentrum J\"ulich, 52425 J\"ulich, Germany}
\affiliation{University of Ferrara and Istituto Nazionale di Fisica Nucleare, 44100 Ferrara, Italy}

\author{D.\,Shergelashvili}
\affiliation{High Energy Physics Institute, Tbilisi State University, 0186 Tbilisi, Georgia}	

\author{V.\,Shmakova}
\thanks{Present address: Brookhaven National Laboratory, Upton, NY 11973, USA}
\affiliation{University of Ferrara and Istituto Nazionale di Fisica Nucleare, 44100 Ferrara, Italy}

\author{R.\,Shankar}
\affiliation{University of Ferrara and Istituto Nazionale di Fisica Nucleare, 44100 Ferrara, Italy}

\author{N.\,Shurkhno}
\thanks{Present address: GSI Helmholtzzentrum für Schwerionenforschung GmbH,
Planckstr. 1, 64291 Darmstadt, Germany}
\affiliation{Institut f\"ur Kernphysik, Forschungszentrum J\"ulich, 52425 J\"ulich, Germany}

\author{S.\,Siddique}
\thanks{Present address: GSI Helmholtzzentrum für Schwerionenforschung GmbH,
Planckstr. 1, 64291 Darmstadt, Germany}
\affiliation{III. Physikalisches Institut B, RWTH Aachen University, 52056 Aachen, Germany}
\affiliation{Institut f\"ur Kernphysik, Forschungszentrum J\"ulich, 52425 J\"ulich, Germany}

\author{A.\,Silenko}
\affiliation{Bogoliubov Laboratory of Theoretical Physics, International Intergovernmental Organization Joint Institute for Nuclear Research, 141980 Dubna, Russia}

\author{J. Slim}
\thanks{Present address: Deutsches Elektronen-Synchrotron, 22607 Hamburg, Germany}
\affiliation{III. Physikalisches Institut B, RWTH Aachen University, 52056 Aachen, Germany}

\author{H.\,Soltner}
\affiliation{Institute of Technology and Engineering, Forschungszentrum J\"ulich, 52425 J\"ulich, Germany}

\author{R.\,Stassen}
\affiliation{Institut f\"ur Kernphysik, Forschungszentrum J\"ulich, 52425 J\"ulich, Germany}

\author{E.J.\,Stephenson}		
\affiliation{Indiana University, Department of Physics, Bloomington, Indiana 47405, USA}

\author{H.\,Ströher}
\affiliation{Institut f\"ur Kernphysik, Forschungszentrum J\"ulich, 52425 J\"ulich, Germany}

\author{M.\,Tabidze}
\affiliation{High Energy Physics Institute, Tbilisi State University, 0186 Tbilisi, Georgia}

\author{G.\,Tagliente}
\affiliation{Istituto Nazionale di Fisica Nucleare sez.\ Bari, 70125 Bari, Italy}

\author{Y.\,Valdau}
\thanks{Present address: GSI Helmholtzzentrum für Schwerionenforschung GmbH,
Planckstr. 1, 64291 Darmstadt, Germany}
\affiliation{Institut f\"ur Kernphysik, Forschungszentrum J\"ulich, 52425 J\"ulich, Germany}

\author{M.\,Vitz}
\affiliation{III. Physikalisches Institut B, RWTH Aachen University, 52056 Aachen, Germany}
\affiliation{Institut f\"ur Kernphysik, Forschungszentrum J\"ulich, 52425 J\"ulich, Germany}

\author{T.\,Wagner}
\thanks{Present address: GSI Helmholtzzentrum für Schwerionenforschung GmbH,
Planckstr. 1, 64291 Darmstadt, Germany}
\affiliation{III. Physikalisches Institut B, RWTH Aachen University, 52056 Aachen, Germany}
\affiliation{Institut f\"ur Kernphysik, Forschungszentrum J\"ulich, 52425 J\"ulich, Germany}

\author{A.\,Wirzba}
\affiliation{Institute for Advanced Simulation, Forschungszentrum J\"ulich, 52425 J\"ulich, Germany}

\author{A.\,Wro\'{n}ska}
\affiliation{Marian Smoluchowski Institute of Physics, Jagiellonian University, 30348 Cracow, Poland}

\author{P.\,W\"ustner}
\affiliation{Institute of Technology and Engineering, Forschungszentrum J\"ulich, 52425 J\"ulich, Germany}

\author{M. \.{Z}urek}
\thanks{Present address: Argonne National Laboratory, Lemont, Illinois 60439, USA}
\affiliation{Institut f\"ur Kernphysik, Forschungszentrum J\"ulich, 52425 J\"ulich, Germany}

\collaboration{JEDI collaboration}
\noaffiliation

\date{April 9, 2025}

\begin{abstract}
This paper presents the successful application of a phase-lock feedback system to maintain the resonance condition of a radio frequency (rf) spin rotator (specifically, an rf Wien filter) with respect to a \SI{120}{kHz} spin precession in the Cooler Synchrotron (COSY) storage ring. Real-time monitoring of the spin precession and the rf Wien filter signal allows the relative phase between the two to be stabilized at an arbitrary setpoint. The feedback system compensates for deviations in the relative phase by adjusting the frequency and/or phase of the rf device as needed. With this method, a variation in the phase relative to the demanded value with a standard deviation of $\sigma_{\Delta\varphi}\approx\SI{0.2}{rad}$ could be achieved. The system was implemented in two runs aiming at a first direct measurement of the deuteron electric dipole moment in 2018 and 2021. In addition, the difference between a single-bunch beam affected by the spin rotator and a two-bunch system in which only one bunch is exposed to the spin rotator fields is discussed. Both methods have been used during these beam times. The ability to keep the spin precession and the rf fields synchronized is crucial for future investigations of electric dipole moments of charged particles using storage rings.
\end{abstract}

\maketitle

\section{Motivation}

Radio Frequency (rf) spin rotators are essential devices used in particle accelerators, specifically storage rings, to manipulate the spin orientation of particles such as protons, electrons, or polarized ions. Frequent reversals of the beam polarization direction can significantly reduce the systematic errors in experiments with polarized beams. By generating oscillating magnetic and/or electric fields, these devices interact with the particle's magnetic moment, allowing precise control over the spin direction ideally without affecting the particle's trajectory. This control is crucial for precision experiments, \emph{e.g.} the measurement of electric dipole moments, in nuclear and particle physics, where spin polarization plays a significant role in probing fundamental interactions.

In the 1990s, rf spin rotators were first used at the Indiana University Cyclotron Facility (IUCF) Cooler Ring to induce an rf depolarizing resonance by sweeping the frequency of the rf magnet through the resonance frequency to flip the spin of a \SI{139}{MeV} proton beam \cite{Caussyn:1994aea}. For an isolated resonance, the modified Froissart-Stora equation \cite{Froissart:1960zz, Blinov:1998ya, Anferov:1998wr} describes the crossing of an isolated rf resonance and relates the initial beam polarization to its final polarization after crossing. About a decade later, a spin-flip efficiency of more than 99.9$\%$ was achieved with an rf dipole using a \SI{1.94}{GeV/c} vertically polarized proton beam at the Cooler Synchrotron COSY in Jülich \cite{Morozov:2004rb}. Using an rf dipole magnet, spin flipping of a \SI{669}{MeV} horizontally polarized electron beam stored in the MIT-Bates storage ring was also performed with very high efficiency in the presence of an almost full Siberian snake \cite{Morozov:2001ne}. A more sophisticated spin flipper, consisting of nine-dipole magnets, was used to flip the spin of a 255\,GeV polarized proton beam at the Brookhaven National Laboratory (BNL) Relativistic Heavy Ion Collider \cite{Huang:2018jtf} and to measure the spin tune at high energies (\si{24} and \SI{255}{GeV}) with a driven coherent spin motion \cite{Huang:2019mce}. 
Radio Frequency fields have also been successfully used to overcome strong intrinsic spin resonances at the Alternating Gradient Synchrotron (AGS) at BNL. At each of the spin resonances, a coherent oscillation was excited by an rf dipole. The coherent oscillation could enable an adiabatic spin flip by preserving the beam emittance \cite{Bai:1998gj}.

Except when frequency sweeps are used, rf spin rotators have to run on resonance with the spin precession, \ie the oscillation frequency $\frf$ must be a harmonic of the spin-precession frequency $f_\mathrm{s}$. 
Given the spin tune $\nus = \gamma G$ (where $\gamma$ and $G$ are the Lorentz factor and the gyromagnetic anomaly, respectively) one can write 
\begin{equation}\label{eq:res}
    \frf = | \nus + h |\,\fc \,,\quad h\in\mathbb{Z}\,,
\end{equation}
where $\fc$ is the revolution frequency of the beam. $|\nus| = \fs/\fc$ also describes the number of spin revolutions per turn. As the stability of $\nus$ depends on a multitude of storage-ring parameters, the question arises whether a constant frequency $\frf$ is sufficient for maintaining this resonance condition for hundreds of seconds.
This paper discusses this issue in the context of recent experimental efforts at the Cooler Synchrotron COSY at Forschungszentrum Jülich.

The JEDI collaboration (Jülich Electric Dipole Moment Investigations) has installed a novel wave-guide rf Wien filter\cite{Slim:2016pim, Slim:2020ufk} at COSY with the aim of conducting a first direct measurement of the permanent electric dipole moment (EDM) of the deuteron. The existence of such an EDM would be an essential ingredient in explaining one of the unsettled questions in particle physics and cosmology: the dominance of matter over anti-matter in the universe, quantified by the parameter $\eta = (n_B - n_{\bar{B}})/n_\gamma$\cite{ParticleDataGroup:2024cfk}, which exceeds the Standard Model predictions by several orders of magnitude. 

In the absence of an EDM, the invariant spin axis (see Fig.~\ref{fig:angles}) in an ideal magnetic storage ring is vertical, resulting in a precessing in-plane spin component.
According to the Thomas-BMT equation \cite{Bargmann:1959gza,Fukuyama:2013ioa}, a non-zero EDM causes a tilt of the invariant spin axis in radial direction leading to a small oscillation of the vertical  spin component. The experimental approach employs the rf Wien filter to accumulate these spin rotations over a full accelerator cycle resulting in a sizable vertical spin component \cite{Rathmann:2013rqa, Morse:2013hoa,Rathmann:2019lwi}. For a systematic study of this effect, the invariant spin axis can be tilted longitudinally by means of a superconducting solenoid, and the rf Wien filter (\ie the spin rotation axis) can be rotated around the longitudinal axis. In its nominal position the magnetic field of the rf Wien filter points in vertical and the electric field in radial direction. The magnitude of the signal depends on the relative phase $\phirel$ between the spin precession and the field oscillations in the rf Wien filter. This dependence on $\phirel$ requires the ability to set and control this phase along with the resonance condition in real time. 

This paper presents a control system for phase and frequency implemented at COSY for the experimental runs with the rf Wien filter in 2018 and 2021. A proof of principle of a similar system has already been reported in \cite{JEDI:2017bnp}. However, that approach was different: instead of the frequency signal $\frf$ the beam revolution frequency $\fc$ had been controlled to stabilize and control the spin precession itself. The solution discussed here, \ie modifying $\frf$ and its phase, was chosen to avoid higher-order effects for other beam parameters. 
Two scenarios are presented: one where the feedback system acts directly on a particle bunch influenced by the rf spin rotator, and another employing two bunches and a newly developed system with fast switches to operate the rf spin rotator selectively on a specific bunch \cite{JEDI:2023btw} with the feedback acting on the second, unaffected bunch. An analytic model is used to discuss the differences between the two cases.

\section{Introduction}
\label{sec:introduction}
\begin{figure}
    \centering
    \includegraphics{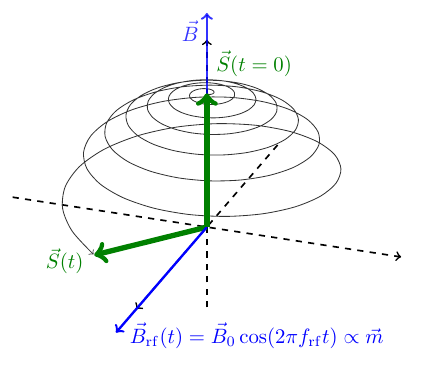}
    \caption{As an example, the initial spin vector $\vec S(t=0)$ points in vertical direction. An rf device providing a magnetic field
    $\vec B_{\mathrm{rf}}(t) = \vec B_0\cos(2\pi f_{\mathrm{rf}} t)$ with $\vec{B_0}=B_0\vec{m}$ then rotates the spin vector by an angle $\chi(t) \propto B_{\mathrm{rf}}(t)$ per pass around the axis $\vec m \parallel \vec B_0$, see Eq.~(\ref{eq:rot_per_turn}). In addition, the holding field $\vec B$ of the accelerator causes the spin vector to precess with the frequency $\fs = \nus \fc$ around the vertical axis. This only works if the resonance condition in Eq.~(\ref{eq:res}) is satisfied.
    } 
    \label{fig:rf_spinrotation}
\end{figure}

For the following discussion on spin rotations, the effects of an electric dipole moment (EDM) are neglected. Furthermore, the term "polarization" is used instead of "spin". Polarization, defined as the statistical average of individual spin vectors, is the measurable representation of the collective spin state of the particles. Strictly speaking, a spin rotator acts on the individual spins. However, the polarization vector follows the same rules. All important variables used in the following sections are listed in Tables~\ref{tab:description_of_variables}
and \ref{tab:description_of_experimental_parameters}.

The polarization vector in a storage ring is comprised of two main components: a stable component parallel to the invariant spin axis $\vec{c}$, and a precessing component in the plane perpendicular to it.  In the case of an ideal magnetic ring without any additional spin rotators, the invariant spin axis is vertical, \ie parallel to the magnetic field of the bending elements. Typical observables include the vertical and in-plane components of the polarization, $\pv$ and $\ph$, respectively, and the spin tune, $\nus$. During the last decade, the JEDI collaboration has developed tools to measure these quantities in real-time with high precision \cite{Bagdasarian:2014ega,JEDI:2015vwa,JEDI:2018txsa,JEDI:2017bnp}.

\begin{figure}
    \centering
    \includegraphics[width=0.48\columnwidth,page=2]{angles.pdf}
    \includegraphics[width=0.48\columnwidth,page=1]{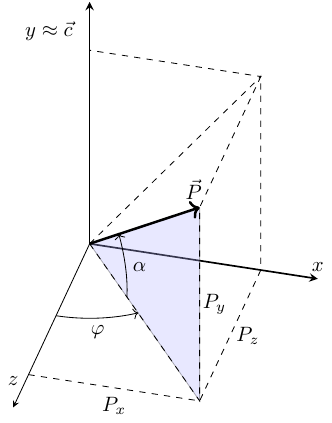}
    \caption{Left: Definition of angle $\xi$,
    $\vec c$ denotes the invariant spin axis, which deviates only by a small amount from the vertical 
    direction. A spin rotator rotates the spin vector around the axis $\vec m$.
    Right: Definition of the out-of-plane angle $\alpha$ and the spin-precession phase $\varphi$.
    }
    \label{fig:angles}
\end{figure}

\begin{table*}[t]
  \centering
  \caption{Overview of frequently used variables and parameters.}
  \begin{tabular*}{\textwidth}{@{\extracolsep{\fill}}lcr}
    Description & Symbol  & Defined in or near\\ \hline \hline
    Time in cycle & $t$ &    \\
    Turn number & $n$ &    \\
    Spin rotation angle per pass through an rf device & $\chi(t)$ & Eq. (\ref{eq:rot_per_turn}) \\
    Initial phase of the rf device  & $\phirf$ & Eq. (\ref{eq:rot_per_turn})  \\
    Invariant spin axis (unit vector)& $\vec{c}$ & Fig. \ref{fig:angles}\\
    Spin rotation axis of an rf device (unit vector) & $\vec{m}$ & Fig. \ref{fig:angles}\\
    Angle between the spin rotation axis $\vec{m}$ and the invariant spin axis $\vec{c}$ & $\xi$ & Eq. (\ref{eq:res_strength}) and Fig. \ref{fig:angles}\\
    Vertical polarization component & $\pv$ & Eq. (\ref{eq:res_strength})\\
    Oscillation frequency of the vertical spin component & $\fv$ & Eq. (\ref{eq:res_strength})\\
    Resonance strength & $\epsilon$ & Eq. (\ref{eq:res_strength})\\
    Out-of-plane angle of the polarization vector & $\alpha$ & Eq. (\ref{eq:alpha}) \\ 
    Spin-precession phase & $\varphi$ & Eq. (\ref{eq:free_spin_precession})\\
    Offset of the spin-precession phase & $\phis$ & Eq. (\ref{eq:free_spin_precession}) and Eq. (\ref{eq:st_phase})\\
    Relative phase between the spin precession and the rf Wien filter oscillation & $\phirel$ & Eq. (\ref{eq:rel_phase})\\
    Measured rf Wien filter phase & $\phirf^{\mathrm{meas}}$ & Eq. (\ref{eq:phi_wf})\\
    Measured relative phase  & $\phirel^{\text{meas}}$ & Eq. (\ref{eq:rel_phase_meas})\\
    Demanded relative phase  & $\phirel^{\text{demand}}$ & Eq. (\ref{eq:phase_correction})\\
    Difference of the relative phase with respect to the demanded value & $\Delta\varphi$ & Eq. (\ref{eq:phase_correction}) \\
    Averaged difference between the measured relative phase and the demanded phase & $\Delta\varphi_{\mathrm{avg}}$ & Eq. (\ref{eq:phase_correction_fit})\\
    Slope of the difference between the measured relative phase and the demanded phase  & $m$ & Eq. (\ref{eq:phase_correction_fit})\\
    Phase offset due to cable delays and latencies during signal processing & $\varphi_{\mathrm{off}}$ &  Eq. (\ref{eq:phioff})\\
    \hline \hline
   \end{tabular*}
  \label{tab:description_of_variables}
\end{table*}

  An rf spin rotator can be characterized by a rotation axis $\vec m$ (defined by the oscillating magnetic and/or electric fields), an oscillation frequency $\frf$, an initial phase $\phirf$ at $t=0$, and a rotation angle per pass
\begin{equation}
    \chi(t) = \chi_0 \sin \left(2\pi\frf t + \phirf\right).
    \label{eq:rot_per_turn}
\end{equation}
If one assumes a bunched beam with bunches passing the rf spin rotator at times $t = n/\fc$ (with $n \in\mathbb{N}$ being the turn number), this becomes 
\begin{equation}
    \chi(n) = \chi_0 \sin \left(2\pi\nus n + \phirf\right)
\end{equation}
if the resonance condition given in Eq.~(\ref{eq:res}) is fulfilled.
The most basic scenario is a rotation out of the stable, non-precessing vertical spin direction as shown exemplarily in Fig.~\ref{fig:rf_spinrotation} resulting in an oscillation of the vertical polarization component $\pv$ with the frequency $\fv$. Other cases depend on the relative phase between the spin precession and $\frf$.
One can define the resonance strength (or resonance spin-flip tune) $\epsilon$ as
\begin{equation} \label{eq:res_strength}
\epsilon = \frac{\fv}{\fc} = \frac{1}{4\pi}\chi_0\,|\vec c\times\vec m| = \frac{1}{4\pi}\chi_0\,\sin\xi 
\end{equation}
with $\xi$ being the tilt angle of the rotation axis with respect to the invariant spin axis, as shown in Fig.~\ref{fig:angles} \cite{JEDI:2017wlr}. The resonance strength is a direct measure of the tilt angle $\xi$. 

In order to decide on the necessity of an active frequency and phase control, one needs to consider the impact of a detuned frequency as well as the phase stability. 
Running off-resonance by a frequency change $\Delta f$ has a direct impact on the induced spin motion. For example, as shown in Sect.~\ref{sec:bw}, Eq.~(\ref{eq:bw}), the amplitude of the oscillating polarization can be described by a Lorentz curve
with the width $\Gamma = 2\epsilon\fc$.  If the variation of the spin-precession frequency $\fs$ over the running period is of similar size or larger, one needs to adjust $\frf$ with time. For the operation of the rf Wien filter at COSY typical run-time parameters are (see \cite{Rathmann:2019lwi})
\begin{eqnarray}
    \chi_0 &\approx& \qty{3.8e-6}{rad} \\
    \xi &\approx& \qtyrange{-30}{30}{mrad} \label{eq:val_xi}\\
    \fc &\approx& \qty{750000}{Hz},
\end{eqnarray}
resulting in $\Gamma < \qty{13.6}{mHz}$.  
The typical variation of the spin tune within one cycle is $\Delta\nus\approx\num{1e-8}$ \cite{JEDI:2015vwa} corresponding to $\Delta\fs \approx \qty{7.5}{mHz}$. 

A more sensitive parameter to monitor is the relative phase $\phirel$ between the rf oscillation and the spin precession. 
Assuming an initial phase $\phis$, the free spin precession can be described by
\begin{equation} \label{eq:free_spin_precession}
    \varphi = (2\pi\nus n + \phis)\,\mathrm{mod}\,2\pi\,,
\end{equation}
and, with the rf spin rotator operated at resonance,
\begin{equation} \label{eq:rel_phase}
    \phirel = \phis - \phirf \equiv \mathrm{const.} \, .
\end{equation}
A frequency mismatch caused by variations of $\fs$ and/or $\frf$ results in a varying relative phase
\begin{equation} \label{eq:dphireldn}
    \frac{\mathrm{d}\phirel}{\mathrm{d}t} = 2\pi (\Delta \fs  - \Delta \frf) \, .
\end{equation}
If one considers a possible frequency change corresponding to a spin tune change of \mbox{$\Delta\nus = \num{1e-8}$}, this would correspond to a phase change of \mbox{$\Delta\phirel$} close to {$2\pi$} over a typical cycle length of $T=\qty{100}{s}$ 
(cf. Fig.~\ref{fig:spin_tune_phase}), which calls for permanent monitoring and control of frequency and relative phase.

\section{Experimental Setup}
\label{sec:setup}

\begin{table}[t]
  \centering
  \caption{Overview of frequently used beam parameters.}
  \resizebox{\columnwidth}{!}{%
  \begin{tabular}{lcrr}
    Description & Symbol & Value & Defined in or near\\ \hline \hline
    Beam revolution frequency  & $\fc$ & \SI{750}{kHz}  & Eq. (\ref{eq:res})  \\
    Spin precession frequency & $\fs$ & \SI{120}{kHz} & Eq. (\ref{eq:res}) \\
    Lorentz factor & $\gamma$ & \num{1.126} & Eq. (\ref{eq:res}) \\
    Gyromagnetic anomaly & $G$ & \num{-0.143} & Eq. (\ref{eq:res}) \\
    Spin tune  & $\nus$ & \num{-0.16} & Eq. (\ref{eq:res}) \\
    Frequency of the rf device  & $\frf$ & \SI{871}{kHz} & Eq. (\ref{eq:res}) and Eq. (\ref{eq:rot_per_turn})   \\
    Deuteron beam momentum & $p$ & $\SI{970}{MeV/c}$ & Section \ref{sec:setup} \\
    Beam polarization & & $\approx\num{0.5}$ & Section \ref{sec:setup} \\
    \hline \hline
  \end{tabular}
  }
  \label{tab:description_of_experimental_parameters}
\end{table}

All measurements were conducted at the Cooler Synchrotron COSY at the Forschungszentrum Jülich \cite{Maier:1997zj, Weidemann:2014qca}. The relevant beam parameters are listed in Table \ref{tab:description_of_experimental_parameters}.  
A typical accelerator cycle proceeds as follows:
About \num{1e9} vector-polarized deuterons with the initial polarization perpendicular to the ring plane are injected and accelerated to a momentum of $p=\qty{970}{MeV/c}$. Throughout the cycle, a radio frequency cavity is used to bunch the beam. After the acceleration, the beam is electron-cooled in order to reduce its emittance. A final orbit correction is then applied to have the beam running through the center of all elements. In the next step, the initial vertical polarization is rotated into the ring plane using an rf solenoid. During beam preparation this setup is used to optimize the in-plane polarization lifetime (or spin-coherence time) by means of sextupoles as discussed in \cite{JEDI:2016swi}. During the regular measurement the remaining part of the cycle is used to operate the rf Wien filter at resonance and a given relative phase $\phirel$ to determine the resonance strength $\epsilon$. The spin rotation axis of the rf Wien filter is set vertically with a small, variable angle $\xi$ relative to the invariant spin axis, see Eqs.~(\ref{eq:res_strength}) and (\ref{eq:val_xi}). At the end of the cycle measurements of the betatron tunes and chromaticities are performed using standard COSY diagnostics.

\begin{figure}
    \centering
    \includegraphics[width=\columnwidth]{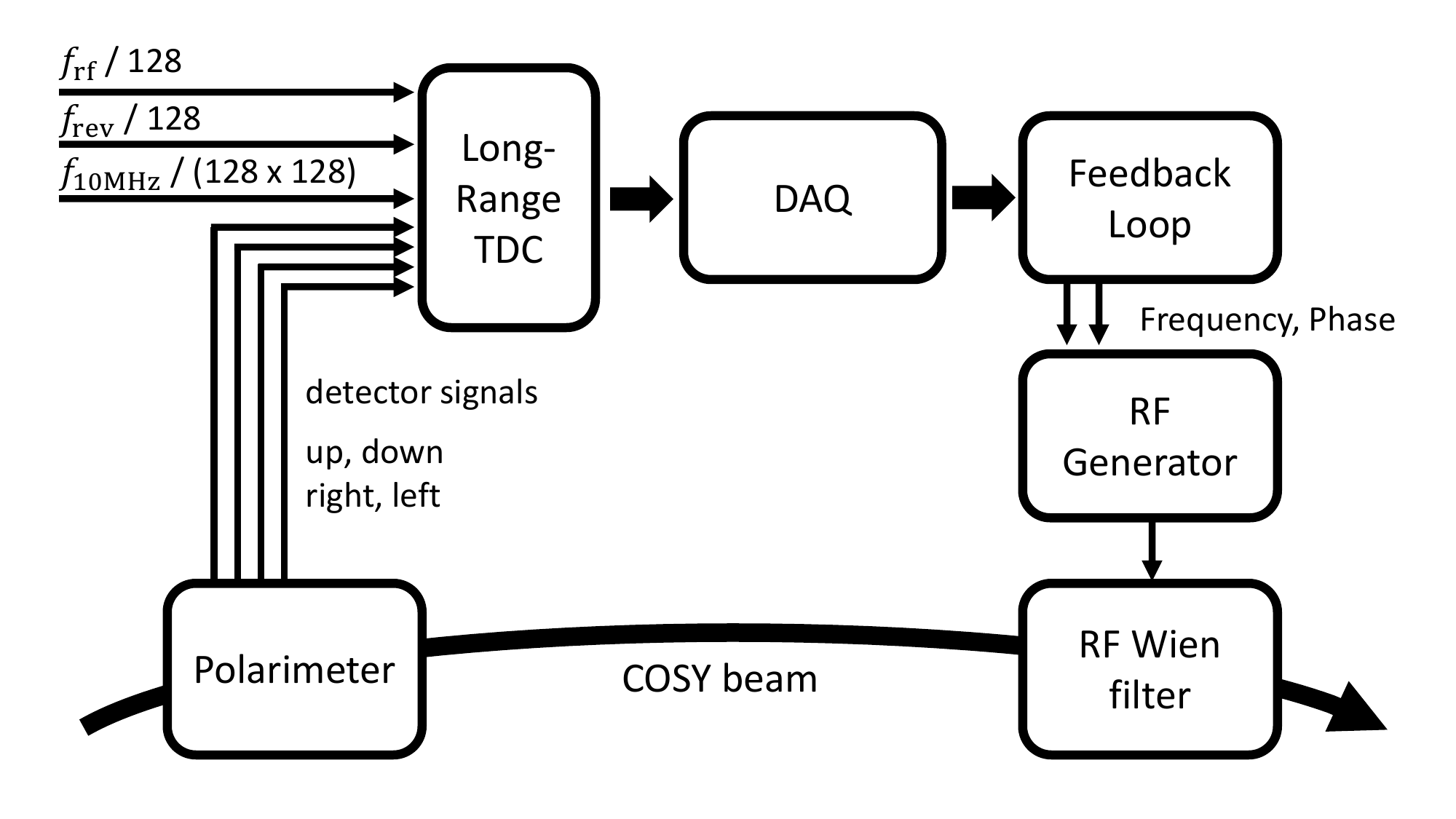}
    \caption{Sketch of the experimental setup. The signals of the four detector quadrants and the pre-scaled frequency signals are fed into a long-range TDC, \ie the same time reference was used for all signals. A 10 MHz reference signal is used for normalization. The DAQ system sends the time stamps into the feedback system, which then adjusts the frequency and phase of the rf generator of the Wien filter.}
    \label{fig:setup}
\end{figure}

Part of the COSY beam is used to permanently monitor the beam polarization by means of the analyzing power of elastic scattering on carbon. For the measurements described in this paper, the WASA Forward Detector\cite{WASA-at-COSY:2004mns} and the Jedi Polarimeter (JePo) have been used, the latter specifically designed for EDM measurements\cite{JEDI:2020fzda}. 
The beam is continuously extracted onto an internal carbon target by applying a white noise excitation in a frequency band around a betatron resonance. Elastically scattered deuterons are detected by detector elements in four quadrants (up, down, left, and right) around the beam pipe. 
The sum of the rates is used to control the amplitude of the excitation to keep the rates constant. The chosen rate level determines the statistical quality of each measurement bin and defines the achievable duration of an accelerator cycle.
The asymmetry of the measured rates in the left-right direction provides insight into the vertical component of the polarization, whereas the asymmetry in the up-down direction is a measure of the oscillating, radial in-plane component. Cycles with unpolarized beam are used to adjust any fake asymmetries from asymmetric detector acceptances. 
The frequency signals $\frf$, $\fc$ and $f_\mathrm{10MHz}$ from the rf Wien filter, the COSY rf cavity and the 10-MHz reference, respectively, are prescaled and received via an optical-fibre distribution system.
Frequency and phase of the spin precession are determined as discussed in Ref.~\cite{JEDI:2015vwa}. The stability of this phase w.r.t. the rf signal from the frequency generator of the Wien filter is continuously monitored and the latter is adjusted if needed. A sketch of the overall scheme of the feedback system is depicted in Fig.\ref{fig:setup}. 

For data acquisition components developed for the WASA detector based on a proprietary optimized backplane bus with LVDS (Low Voltage Differential Signaling) technology \cite{Kleines:2006cy} are used. The crates are equipped with a SIS1100/SIS3100 crate controller, the link to the readout computer is based on the physical layer of Gigabit Ethernet. The communication between readout computers and the phase-lock feedback system is done via TCP/IP over Gigabit Ethernet. 
A GPX-based model is used as a long-range Time-to-Digital Converter (TDC).
The adjustable clock time period is set to \qty{95.59}{ps} based on a single internal clock. For normalization, an external GPS-based reference signal of \qty{10}{MHz} is used. 
The internal clock counter reaches its maximum value at $\qty{6.4}{\mu s}$, at which point an overflow bit is sent to a 20-bit register. This allows a full range up to about \qty{6.7}{s}. 
During decoding, another overflow register is introduced, enabling continuous time-stamping of the registered events over the full measurement cycle. 

The frequency and phase computed by the feedback system are communicated via Ethernet to the frequency generator (Rohde \& Schwarz SMB100A) attached to the rf Wien filter. It is important to note that this generator ensures a continuous phase evolution at a frequency change, \ie no phase jumps. The frequency generator is operational at all times, allowing for the synchronization of its frequency and phase settings to the spin-precession frequency while the rf Wien filter itself is deactivated. Once all preparations have been completed, the amplifiers of the Wien filter are activated, initiating the actual measurement process.

\begin{figure}
    \centering
    \includegraphics[width=\columnwidth]{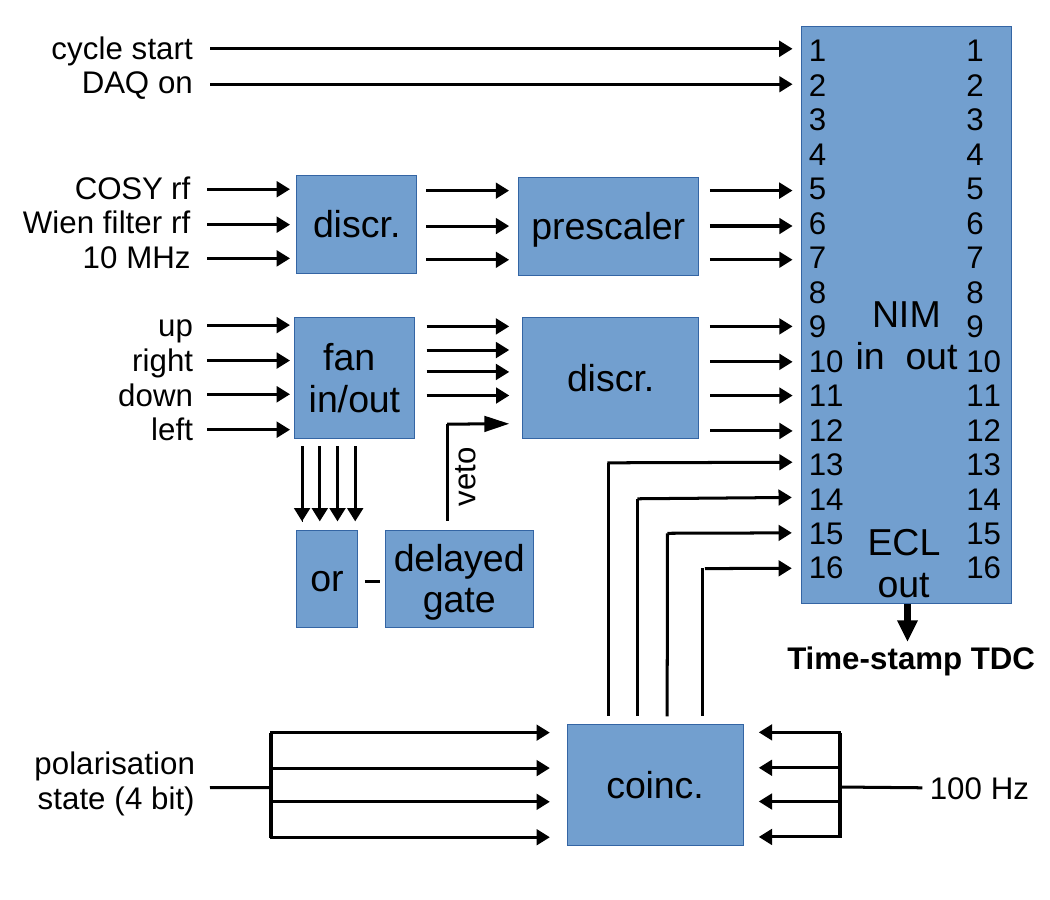}
    \caption{Sketch of the main signal processing. All signals (control signals, frequencies, rates and polarization information) are fed into a single time-stamping TDC.}
    \label{fig:electronics}
\end{figure}

Figure~\ref{fig:electronics} shows the processing of the main signals before entering the time-stamp TDC. The frequency signals are discriminated and prescaled by $128$ (revolution frequency and Wien filter frequency) and $128\times128$ (10 MHz reference signal). A logical OR of the polarimeter signals is used to filter out signal bursts by applying a (typically \qty{1}{$\mu$s} long) veto signal keeping the rate within the specifications of the TDC. The information on the initial polarization state is provided via a 4-bit signal triggered with a \qty{100}{Hz} clock. In addition, other slow control signals (cycle start as time reference, DAQ on) are monitored.

\section{The Phase-Lock Feedback}
\label{sec:phase_lock_feedback}

\begin{figure}
    \centering
    \includegraphics[width=\columnwidth]{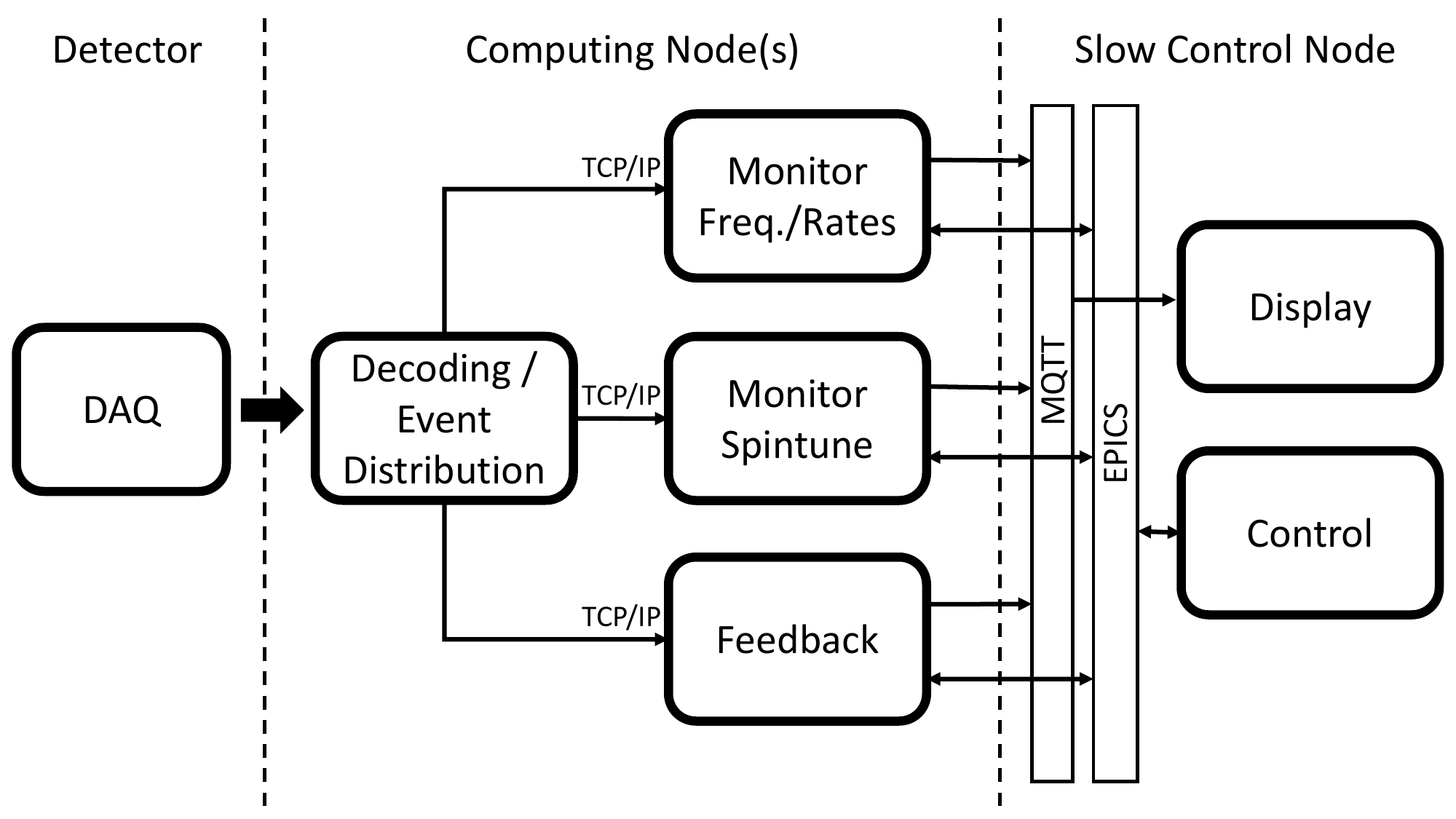}
    \caption{Overall data processing scheme. The event stream from the Data Acquisition is distributed to multiple data processing routines, which use MQTT and EPICS for communication.}
    \label{fig:software}
\end{figure}

The phase-lock loop is implemented as part of a software package developed for monitoring all frequencies, rates, and asymmetries, as well as the spin tune and spin-precession phase. A schematic is shown in Fig.~\ref{fig:software}. The data from the DAQ system are decoded and distributed to several processes. Each of these processes analyzes the data with respect to specific questions. This analysis is done in certain adjustable time intervals $\Delta t$ (typically one second) and the results are sent to a slow control node for display via MQTT (Message Queuing Telemetry Transport, \cite{light2017})\footnote{See: MQTT -- The Standard for IoT Messaging, \url{https:\\mqtt.org}.} and also distributed and archived via EPICS (Experimental Physics and Industrial Control System, \cite{Dalesio:1992fso})\footnote{See: EPICS -- Experimental Physics and Industrial Control System, \url{https:\\epics.anl.gov}.}. The latter is also used to set all process parameters. In the case of more than one bunch in the ring, the system can be configured to monitor each bunch separately. 

The main three processes are:
\paragraph{Monitoring Rates and Frequencies}
This process analyzes the frequencies, their stability, the detector rates, and the left/right asymmetry as a measure of the vertical polarization. In addition, the detector signals are plotted relative to the start of a $\fc$ oscillation period, reflecting the particle distribution in the ring and to unfold the longitudinal bunch extension in the processes b) and c).
\paragraph{Spin Tune Monitoring}
This process analyzes the up/down asymmetry, the spin tune and its phase as described in Ref.~\cite{JEDI:2015vwa}. 
\paragraph{Feedback}
This process does the actual phase-lock feedback and is discussed below.

In each cycle, the measurement with the rf Wien filter was structured as follows: 
\begin{enumerate}
    \item Adjusting the Wien-filter frequency to the resonance frequency \mbox{$\frf=|\nus - 1|\fc$} (\ie $h=-1$), while the Wien filter power supplies are still off
    \item Measuring the relative phase $\phirel$ between the generated Wien filter frequency and the spin precession
    \item Correcting $\frf$ if needed and adjusting $\phirel$ by means of $\phirf$ to the demanded value
    \item Switching on the rf Wien filter power supplies
    \item Maintaining $\phirel$ throughout the measurement by adjusting $\frf$ and $\phirf$
    \item Switching off the rf Wien filter power supplies
\end{enumerate}

The value for $\nus$ used for calculating the initial value of $\frf$ is the averaged, high-precision spin tune $\nus^\mathrm{fixed}$ determined in the previous cycle\cite{JEDI:2015vwa}. The revolution frequency $\fc$ is fixed and defined in the accelerator control system. 

\begin{figure}
    \centering
    \includegraphics[width=\columnwidth]{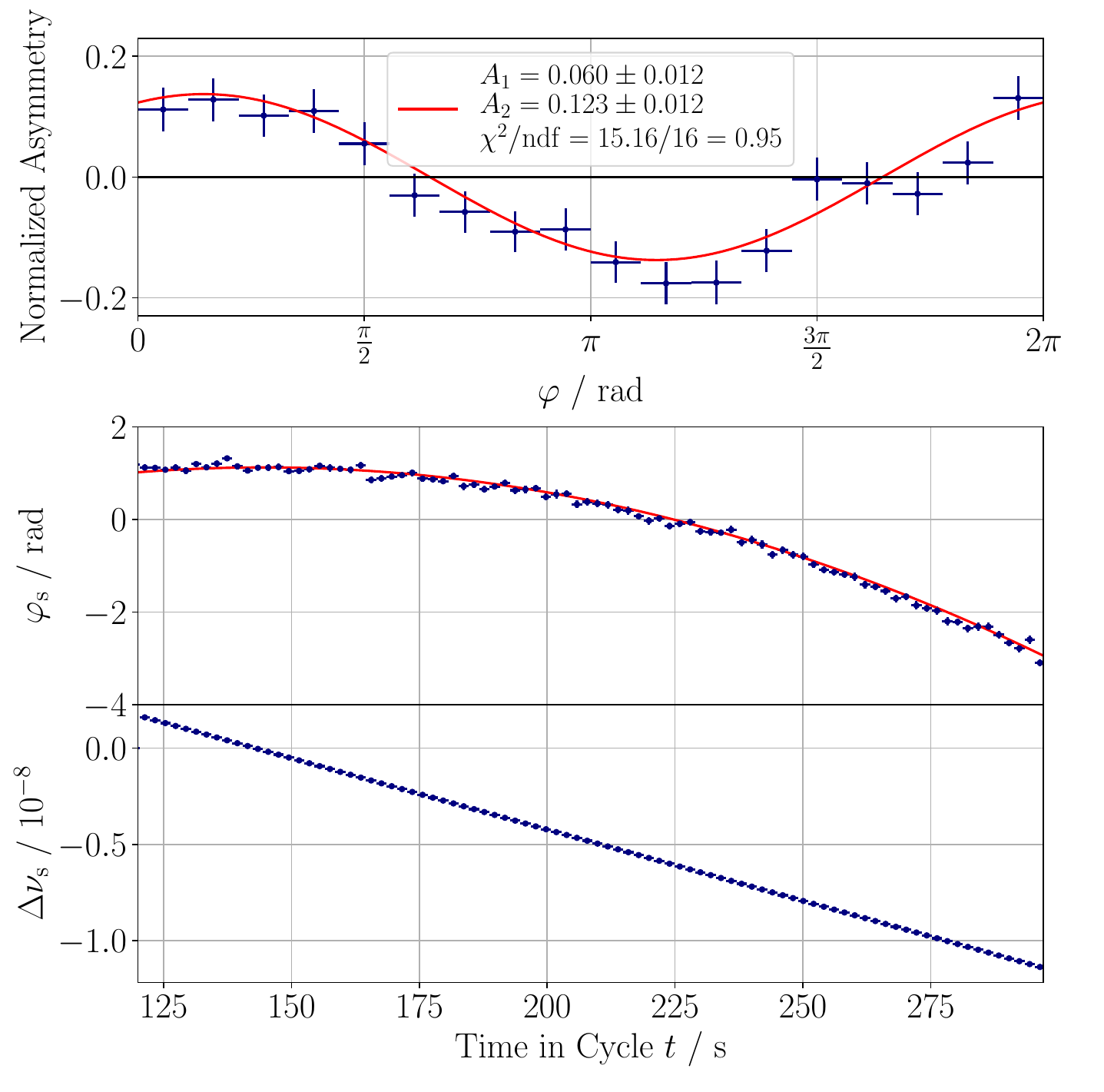}
    \caption{Upper Panel: Normalized asymmetry as a function of the spin-precession phase based on the assumed, fixed spin tune (here $\nus^{\mathrm{fixed}} = \SI{0.1610066737}{}$) of data taken during one $\Delta t = \SI{1}{s}$ period. The distribution is fitted according to Eq. ~(\ref{eq:spin_tune_phase}). In this example the offset of the spin-precession phase is $\varphi_{\mathrm{s}} = \SI{1.12(8)}{rad}$. Lower Panels: Offset of the spin  precession phase $\varphi_s$ and change of spin tune $\Delta \nu_s$ (according to Eq. (\ref{eq:spin_tune_vs_time}), including uncertainties) as a function of time in the cycle. In this example, the change of the spin tune amounts to $\Delta\nus\approx\SI{1e-8}{}$ during a \SI{150}{s} interval. 
    }
    \label{fig:spin_tune_phase}
\end{figure}

The fixed spin tune $\nus^\mathrm{fixed}$ is also used to derive the actual spin tune from the phase walk relative to this assumed spin tune in the current cycle.   
As discussed in \cite{JEDI:2015vwa} this phase analysis 
is more sensitive and faster than a direct bin-by-bin Fourier transform. In the following, the procedure is shortly summarized: 

For every event, the spin-precession phase based on the assumed, fixed spin tune is calculated and mapped to a $4\pi$ interval, \ie two full precessions:
\begin{equation}
\varphi_i = \left( 2\pi\nu_\mathrm{s}^\mathrm{fixed} n \right) \mathrm{mod}\, 4\pi \, .
\label{eq:st_phase}
\end{equation}
These values are accumulated in a histogram such, that events from the up and down detector elements are filled with the weight $1$ and $-1$, respectively. 
The resulting histogram shows the distribution of the measured asymmetry as a function of the spin direction given by the spin-precession phase: an unpolarized, constant background with a sinusoidal oscillation on top. In order to remove the unpolarized background, the two half intervals of the second oscillation are interchanged and subtracted from the first one. After normalization a sinusoidal oscillation between $-1$ and $1$ as a function of $\varphi \in [0, 2\pi)$ remains as shown in Fig.~\ref{fig:spin_tune_phase} (upper panel). This data is fitted according to 
\begin{equation} \label{eq:spin_tune_phase}
    f(\varphi) = A\,\sin(\varphi + \varphi_\mathrm{s}) 
\end{equation}
with the offset of the spin-precession phase $\varphi_\mathrm{s}$ as a fit parameter. (In practice, the fit formula is converted into its linear form $a_1 \sin\varphi + a_2 \cos\varphi$.) 

When the assumed spin tune $\nu_\mathrm{s}^\mathrm{fixed}$ matches the real value, the phase offset $\varphi_\mathrm{s}$ remains constant within the measurement interval $\Delta t$. In reality the real value differs slightly from the assumed value resulting in a phase walk of $ \Delta\varphi_\mathrm{s} = 2\pi\Delta\nu_\mathrm{s}\fc  \Delta t$ over that period.
For typical operation parameters, the deviation from the fixed spin tune is within $\num{1e-8}$, corresponding to a phase change of less than $\qty{50}{mrad}$ per second. 
Compared to the effect originating from the longitudinal extension of the bunch, which is typically $\approx\qty{500}{mrad}$, this still can be considered constant.
By extrapolating the change of $\varphi_\mathrm{s}$ from interval to interval, the change of the spin tune during the cycle can be calculated using 
\begin{equation} \label{eq:spin_tune_vs_time}
    \nus(t) = \nus^\mathrm{fixed} + \frac{1}{2\pi\fc}\frac{\mathrm{d}\varphi_{\mathrm{s}}}{\mathrm{d}t} = \nus^\mathrm{fixed} + \Delta\nus.
\end{equation}
An example is shown in Fig.~\ref{fig:spin_tune_phase} (lower panel).

For the rf Wien filter, the current event time from the polarimeter is used to calculate the phase that the Wien filter frequency signal had at the time $t_i$ of that event
\begin{equation} \label{eq:phi_wf}
    \varphi_{\mathrm{rf},i} = \left[ 2\pi \cdot ( t_i - t_\mathrm{LastRfTag}) \cdot \frf \right] \bmod 2\pi.
\end{equation}
Here, $t_\mathrm{LastRfTag}$ is the time stamp of the last zero crossing from negative to positive of the rf signal. In order to use $\nu_\mathrm{s}^\mathrm{fixed}$ as a reference, the difference between $\varphi_i$ and $\varphi_{\mathrm{rf},i}$ is histogrammed. If sufficiently close to the resonance, this difference can be considered constant. Finally, the event time relative to $\fc$ is used to account for the bunch width to get a more distinct signal. At the end of each time bin, the mean of the distribution is calculated as $\varphi^\mathrm{meas}_\mathrm{rf}$. 

Then the relative phase is the difference between both:
\begin{equation} \label{eq:rel_phase_meas}
    \phirel^\mathrm{meas} = \phis - \varphi^\mathrm{meas}_\mathrm{rf}.
\end{equation}

Once the Wien filter frequency is set to its initial value at the beginning of the measurement cycle, the difference of the relative phase with respect to the demanded value
\begin{eqnarray} \label{eq:phase_correction}
    \Delta\varphi = \phirel^\mathrm{meas} - \phirel^\mathrm{demand}
\end{eqnarray}
is continuously monitored according to the following scheme:

A sequence length $N$ (typically 5-7) is defined via a control parameter and $N$ subsequent values of $\phirel^\mathrm{meas}$, each within a time bin $\Delta t$, are recorded. After the last interval, these values are used to apply a linear fit
\begin{equation}  \label{eq:phase_correction_fit}
    \Delta\varphi(j) = m \cdot \left(j - \frac{N-1}{2}\right) + \Delta\varphi_\mathrm{avg}
\end{equation}
with the bin number $j \in [0, N-1]$. As a fitting result, one gets the slope $m$ and the average offset $\Delta\varphi_\mathrm{avg}$ together with the corresponding statistical uncertainties. 
In addition, the phase difference is extrapolated to bin number $N$, which is the first bin following the measurement sequence. This bin represents the interval during which the calculation and correction occur
\begin{equation}
    \Delta\varphi(N) = \Delta\varphi_\mathrm{avg} + m \cdot \frac{N+1}{2}.
\end{equation}
Both the slope $m$ and the offset of the phase difference, $\Delta\varphi_\mathrm{avg}$ or $\Delta\varphi(N)$, are considered "good" if their deviations from zero fall within two standard deviations. When this condition is met, no correction is applied, otherwise frequency and/or phase are corrected accordingly: 
\begin{eqnarray}
    \frf &\rightarrow& \frf  - \frac{1}{2\pi\Delta t} m \, ,\\ 
    \phirf &\rightarrow& \phirf - \Delta\phirf \, ,
\end{eqnarray}
with $\Delta\phirf$ either $\Delta\varphi_\mathrm{avg}$ or $\Delta\varphi(N)$.
For the slope $m$ the statistical error from the fit is used, for the phase several approaches were considered:
\begin{itemize}
    \item Using the average phase difference $\Delta\varphi_\mathrm{avg}$ as $\Delta\phirf$ and its statistical error in cases without frequency correction, and using the extrapolated phase value $\Delta\varphi(N)$ when a frequency correction is applied. Alternatively, $\Delta\varphi(N)$ could be used in all cases, regardless of a frequency correction.
    \item For the uncertainty of $\Delta\varphi(N)$, either using the uncertainty of $\Delta\varphi_\mathrm{avg}$ directly or propagating the uncertainty of the slope $m$ in addition.
\end{itemize}
For standard data taking, the option to use $\Delta\varphi(N)$ together with the uncertainty of the averaged phase has been selected and is presented here. 

\begin{figure*}
    \centering
    \includegraphics[width=\textwidth]{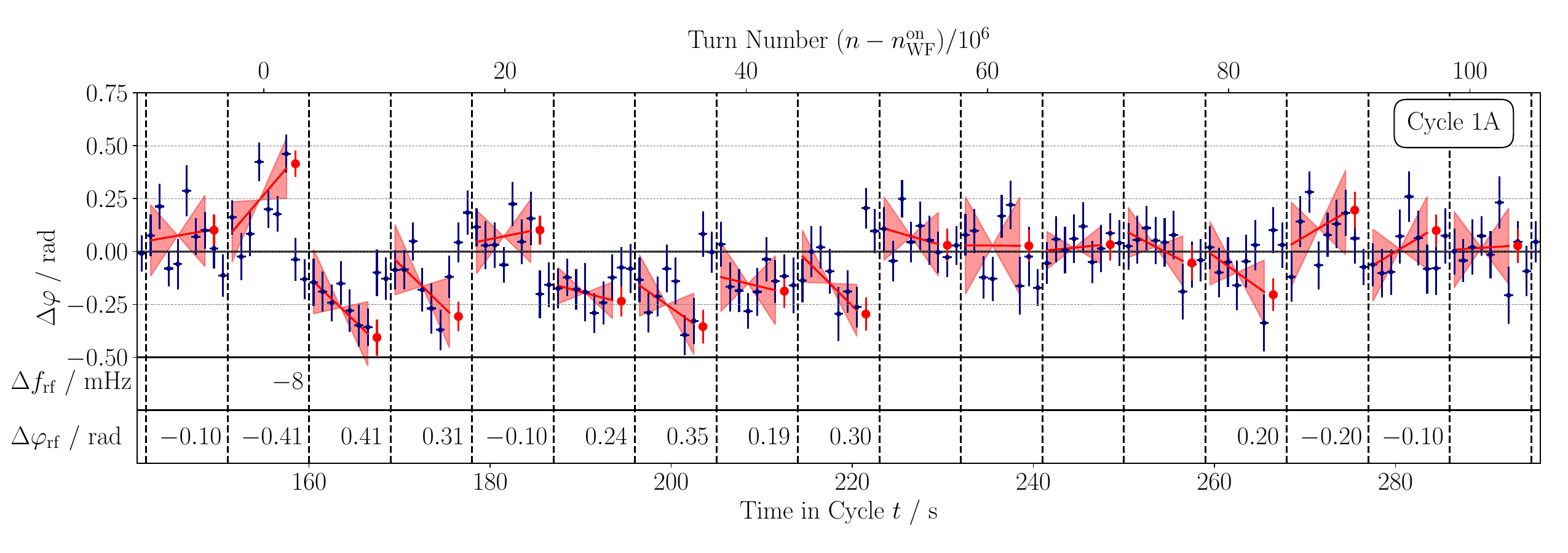}
    \includegraphics[width=\textwidth]{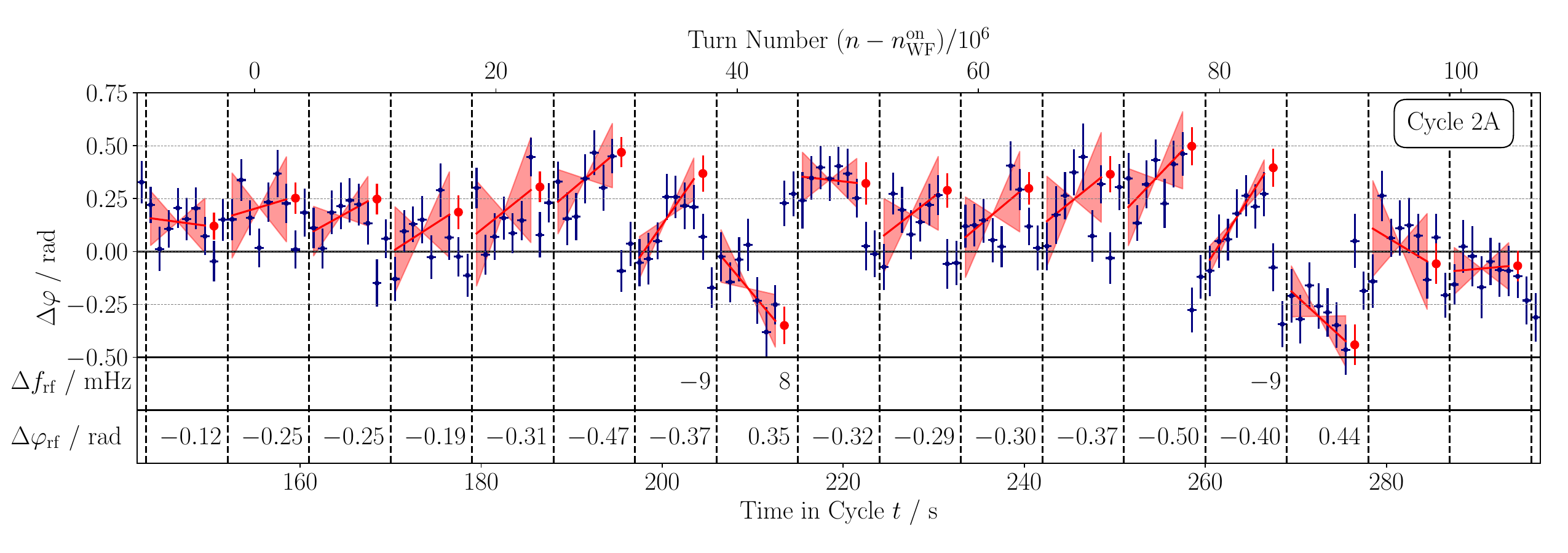}
    \caption{Frequency and phase corrections derived from the measured relative phase for two exemplary cycles labeled 1A and 2A. The upper panel in each plot shows the difference of the relative phase with respect to the demanded value $\Delta \varphi$ as a function of the time in cycle. The time axis is divided in several blocks with 7+2 measurement bins each ($\Delta t = \SI{1}{s}$) indicated by dashed lines. The first 7 values in each block are fitted linearly followed by a correction of frequency and/or phase shown below if the deviation is significant as defined in the text. The red dots show the extrapolated values for the phase difference, and the red lines the fitted slope. The error bars and the shaded area indicate the allowed $2\sigma$ range for phase offset and slope.
    While the applied corrections are indicated only at the time of correction, all changes are additive. 
    The corresponding distributions of $\Delta\varphi$ have a width of $\sigma_{\Delta\varphi}=\SI{0.16}{rad}$ and $\sigma_{\Delta\varphi}=\SI{0.21}{rad}$ centered at $\Delta\varphi = \SI{-0.04}{rad}$ and $\Delta\varphi = \SI{0.10}{rad}$, respectively.
    }
    \label{fig:phase_feeback_action}
\end{figure*}

As all corrections are applied during interval $N$, this bin and the next one are skipped and the collection of samples continues at interval $N+2$.

Fig.~\ref{fig:phase_feeback_action} illustrates the procedure. For two different cycles, several measurement blocks with 7+2 measurement bins are shown together with the fits to the data and the correction to phase and frequency. The resulting distributions of $\Delta\varphi$ have a typical width of $\sigma_{\Delta\varphi}\approx\SI{0.2}{rad}$. This is consistent with the value reported for the first commissioning run, when the revolution frequency was controlled \cite{JEDI:2017bnp}. 

Depending on the variation of the spin tune in the cycle (see Fig.~\ref{fig:spin_tune_phase}), the average relative phase shows a systematic deviation from the demanded value. Therefore, in the data analysis, the measured relative phase has to be used rather than the demanded one.

There are certain limitations and caveats that need to be mentioned:
\begin{enumerate}
    \item The spin tune and its phase are determined from the oscillating radial in-plane polarization. For the analysis, the measured in-plane polarization $\ph$ has to be non-zero taking into account its uncertainty. Hence, the feedback stops working when either the polarization itself becomes too small or the direction of the polarization vector approaches the vertical direction. 
    \item The measured relative phase $\phirel$ differs from the true value in the rf spin rotator due to several factors. These include cable delays and latencies, as well as the additional spin precession that occurs between the positions of the rf spin rotator and the detector. In multi-bunch setups, offsets between the bunches result from the different spin orientations determined by the selection of harmonics and from corrections to account for the particle distribution in the ring. While these offsets remain constant for a given setup and the feedback system provides stabilization of the relative phase, the real phase must be extracted from the data when needed.
    \item Spin oscillations show a certain phase walk compared to the free, in-plane precession (see Ref.~\cite{JEDI:2018txsa} and Fig.~5 and 6 therein). When the phase-lock feedback system is operated on the bunch affected by the rf spin rotator, this phase walk is suppressed and the observed signal is modified. This has to be considered when analyzing the data as discussed in the next section.
\end{enumerate}

\section{Effect of the Phase-Locking on the Spin Rotations in the Wien Filter}

An rf spin rotator typically changes the phase of the spin precession while rotating the spin. By keeping the relative phase constant, active phase locking also affects the primary observable, \ie the vertical spin oscillation. Stabilizing the relative phase and at the same time allowing an rf-induced change of $\varphi_\mathrm{rel}$ is only possible by using a second bunch that remains undisturbed by the rf spin rotator. This was achieved in the second run in 2021 when two bunches were used and the Wien filter for the second bunch was disabled by fast switches \cite{JEDI:2023btw}. Another advantage of this method is that the application of the feedback system is no longer limited by the vanishing in-plane polarization due to the out-of-plane rotation by the spin rotator. 

In Ref.~\cite{JEDI:2018txsa} the phase walk while applying an rf solenoid as spin rotator has been measured and discussed in detail, on resonance as well as off resonance.  
The results have been described by a set of differential equations for the out-of-plane angle $\alpha$ of the spin direction and the relative phase $\phirel$. 
Although these equations were developed under the assumption of an ideal ring, \ie neglecting spin decoherence and other systematic effects, the data could be fitted quite well. 
They are used here to illustrate the two feedback scenarios, as the differences in the resulting phase relations can be easily captured by modifying a single equation.

The two equations from Ref.~\cite{JEDI:2018txsa} are
\begin{eqnarray}
\dfrac{{\rm d}\alpha}{{\rm d}n} &=& \dfrac{k}{2}\cos\phirel\label{eq:alpha}\\
\dfrac{{\rm d}\phirel}{{\rm d}n} &=& \dfrac{k}{2}\left(\sin\phirel \tan\alpha + q\right)\label{eq:phi}
\end{eqnarray}
and the relation to the variables defined in this paper is given by
\begin{eqnarray}
k &=& -\chi_0 \sin\xi = -4\pi\epsilon \label{eq:k}\\
q &=& \dfrac{4\pi\Delta f}{k \fc} \label{eq:q}
\end{eqnarray}
with $k$ defining the strength of the rf device and $q$ being the off-resonance parameter. 

There is an analytic solution given:
\begin{eqnarray}
p_y(n) &=& \sin\alpha(n) \nonumber \\ 
&=& A_1\sin(A_2 + A_3n) - A_4 \label{eq:sinalpha}\,,\\
\cos\phirel(n) &=& \dfrac{A_1 \sqrt{1+q^2} \cos(A_2+A_3n)}{\sqrt{1-\left( A_1\sin(A_2 + A_3n) - A_4 \right)^2}} \label{eq:cosphi}\,,\\
\sin\phirel(n) &=& \dfrac{C + q \sin\alpha(n)}{\cos\alpha(n)} \label{eq:sinphi}\,,
\end{eqnarray}
with the parameters
\begin{eqnarray}
A_1 &=& \dfrac{\sqrt{1+q^2-C^2}}{1+q^2}\,, \label{eq:A1}\\
A_2 &=& \left\{
        \begin{array}{ll}
        \arcsin\left(\dfrac{\sin\alpha(0)+A_4}{A_1}\right) & |\phirel(0)| < \pi/2\,, \\
        \pi - \arcsin\left(\dfrac{\sin\alpha(0)+A_4}{A_1}\right) & |\phirel(0)| > \pi/2\,,
        \end{array}
        \right.  \label{eq:A2}\\\
A_3 &=& \dfrac{k}{2}\sqrt{1+q^2}\,, \label{eq:A3}\\\
A_4 &=& \dfrac{Cq}{1+q^2}\,, \label{eq:A4}\\\
C   &=& \cos\alpha\sin\phirel - q\sin\alpha = \mathrm{const}\,.
\end{eqnarray}
The starting conditions at the time the Wien filter is switched on are given by $\alpha(0)$ and $\phirel(0)$. 
$A_1$ is the oscillation amplitude, $A_2$ is the phase offset of the oscillation matching the start conditions, $A_3$ is the oscillation frequency and $A_4$ is the central value of the oscillation, which is shifted when being off-resonance. $C$ is an auxillary variable defined by Eq.~\ref{eq:sinphi}.
In the following, these equations are used to discuss certain properties of the system
and are compared to data. For the latter, cycles with long spin coherence times have been selected as decoherence is not included in the model. In addition, the active modification of the relative phase contributes to systematic effects (see Fig.~\ref{fig:phase_feeback_action}). The corresponding $\chi^2$ values reflect this. All uncertainties quoted here are purely statistical and should be interpreted as such. For the final EDM data analysis, a more sophisticated and comprehensive model incorporating these systematic effects is employed~\cite{JEDI:2023trd}.

It should be noted that the definition of $\phirel = 0$ here differs from the one used as experimental observable in Sect. ~\ref{sec:phase_lock_feedback}. The measured spin-precession phase based on the up-down asymmetry in the polarimeter is relative to the beam direction. $\phirel$ as used in Eq.~(\ref{eq:phi}) is defined relative to the direction of $\vec c \times \vec m$ as used in Eq.~(\ref{eq:res_strength}).
This adds another contribution to the setup-specific offset $\varphi_\mathrm{off}$, but has no effect on the following discussion.

\subsection{Resonance Width}
\label{sec:bw}

As mentioned before, the necessity to control the generator frequency $\frf$ can be judged using the resonance width $\Gamma$ of the rf spin rotator. Thus, one needs a proper estimate for $\Gamma$. One can utilize the parameter $A_1$ from Eq.~(\ref{eq:A1}), which quantifies the oscillation amplitude. When starting with the spin along the invariant spin axis ($\alpha(0)=\pi/2$ and any $\phirel(0)$), the  parameter $C$ amounts to $q^2$ and with Eq.~(\ref{eq:q}) $A_1$ can be written in the same analytic form as a Lorentz function:
\begin{equation}
    A_1 = \dfrac{1}{1+q^2} = \dfrac{(\Gamma/2)^2}{(\Gamma/2)^2 + (\Delta f)^2}
    \label{eq:bw}
\end{equation}
with $\Gamma = {k f_{\mathrm{rev}}}/{2\pi} = 2\epsilon\fc$. This can be used as a first criterion whether an active control loop is needed or not.

\subsection{Phase Locking with a Single Bunch}
\label{sec:single_bunch}

As discussed in Ref.~\cite{JEDI:2018txsa} and indicated by Eq.~(\ref{eq:phi}) the relative phase $\phirel$ is changing during an rf-induced spin oscillation. Frequency synchronisation with phase locking prevents this change.  
One can consider this effect in the model by replacing Eq.~(\ref{eq:phi}) with  
\begin{eqnarray}
\dfrac{{\rm d}\phirel}{{\rm d}n} &=& 0\,.
\end{eqnarray}
The remaining Eq.~(\ref{eq:alpha}) with a constant phase $\phirel = \phirel(0)$ describes a linear increase in $\alpha$ with a phase dependent slope: 
\begin{equation}
\label{eq:alphadot}
\dfrac{{\rm d}\alpha}{{\rm d}t} = 2\pi\epsilon\fc \cos\phirel\,.
\end{equation}
This results in a spin oscillation of the form
\begin{equation}
\label{eq:lock}
p_y(t) = \sin(\cos\phirel\cdot2\pi\epsilon\fc t) \,.\\
\end{equation}
The oscillation amplitude is always 1 (\ie the spin is fully rotated into the vertical direction independent of the relative phase) and the oscillation frequency is modulated by the phase. As a direct implication, this method can only be used for less than a quarter of a full oscillation, because the feedback stops working when the in-plane component vanishes.

\begin{figure}
    \centering
    \includegraphics[width=\columnwidth]{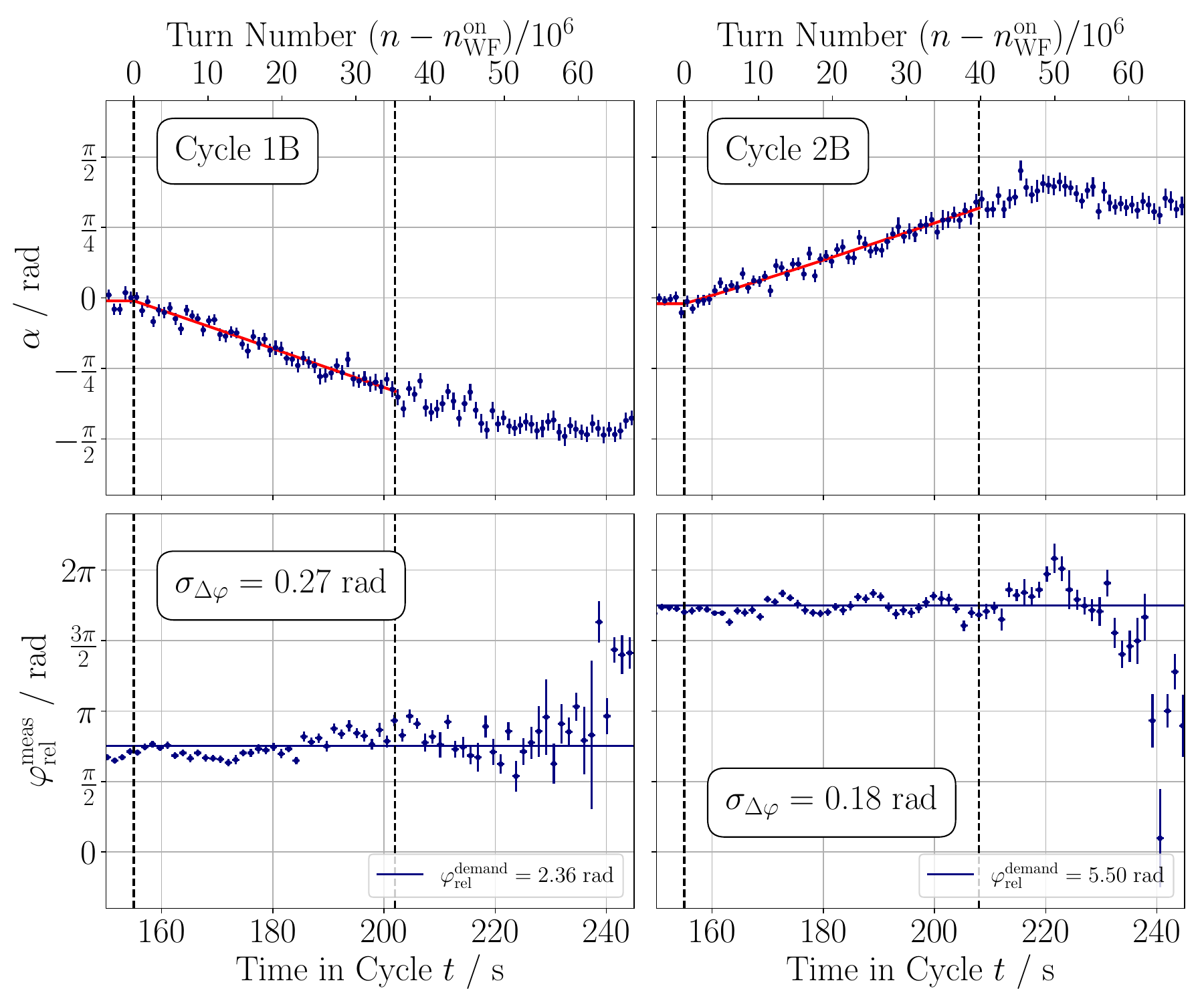}
    \caption{With phase locking: Out-of-plane angle $\alpha$ and relative phase $\phirel$ as a function of time for two different initial values of $\varphi_0$ labeled as cycles 1B and 2B. The first dashed line marks $t_{\mathrm{WF}}^{\mathrm{on}}$. The second dashed line marks the timestamp at which the phase feedback is switched off. For the fit parameters see Table~\ref{tab:scanfix}.
    }
    \label{fig:scanfix}
\end{figure}

\begin{table}
\centering
\caption{Fit parameters from Fig.~\ref{fig:scanfix}}
\label{tab:scanfix}
\begin{tabular*}{\columnwidth}{@{\extracolsep{\fill}}lcc}
& Cycle 1B & Cycle 2B \\
\hline \hline
$\phirel^\mathrm{demand}$ / \si{rad} & 2.36 & 5.50 \\
$\overline\varphi_\mathrm{rel}^\mathrm{meas}$ / \si{rad} & 2.29(1) & 5.43(1) \\
$\sigma_{\Delta\phirel}$ / \si{rad} & 0.27 & 0.18 \\
\hline
$\mathrm{d}\alpha/\mathrm{d}t$ / \si{rad/s} & -0.020(7) & 0.0186(6) \\
$\alpha(0)$ / rad & -0.04(2) & -0.06(2) \\
$\nicefrac{\chi^2}{\mathrm{ndf}}$ & 51/57 & 61/60 \\
\hline \hline
\end{tabular*}%
\end{table}

Fig.~\ref{fig:scanfix} shows data for two different relative phases. The slope is linear until the feedback stops working and the phase is no longer stabilized. The fit results are summarized in Table~\ref{tab:scanfix}. 

\begin{figure}
    \centering
    \includegraphics[width=\columnwidth]{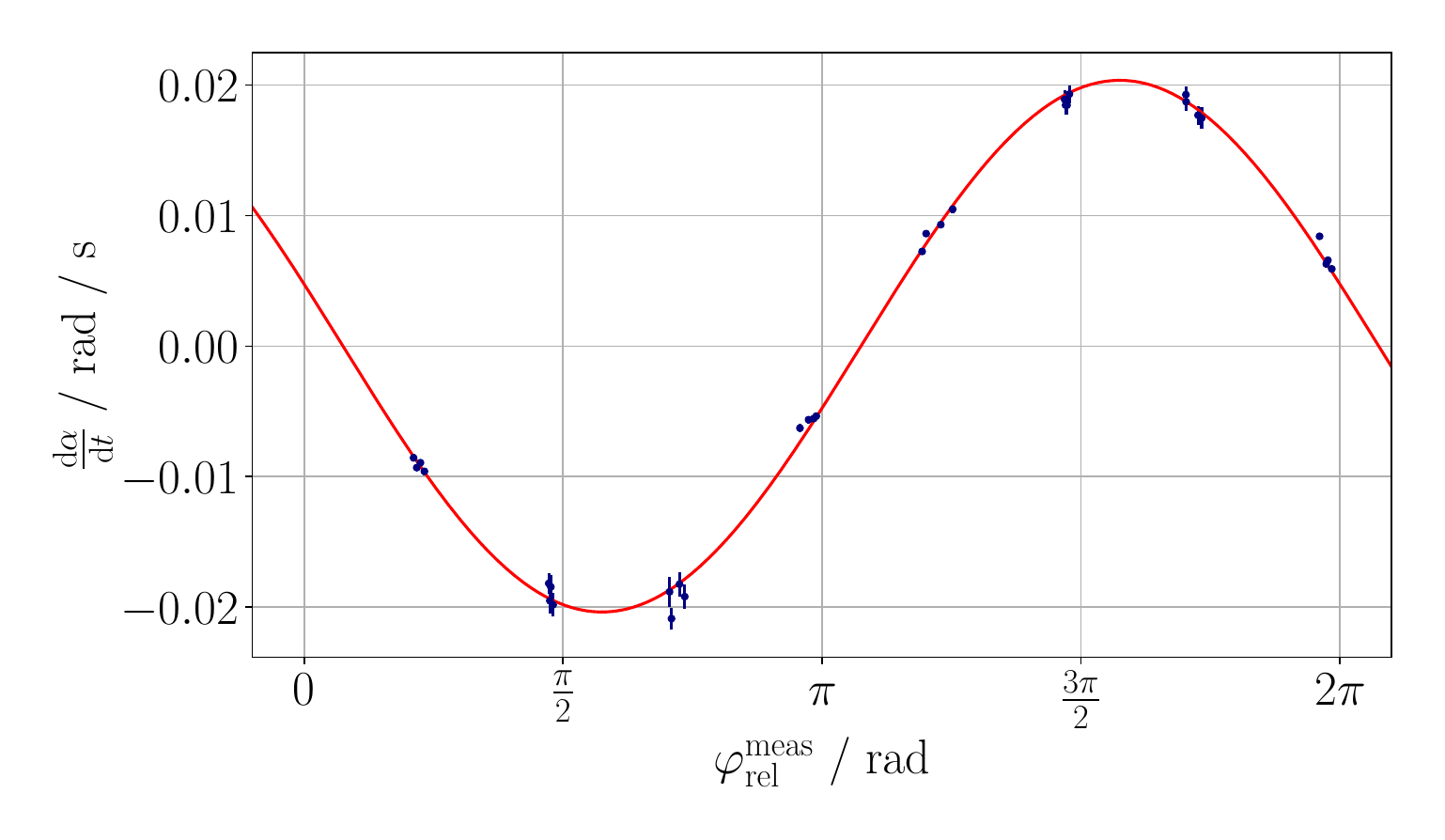}
    \caption{Sinusoidal dependence of the initial slope as a function of the relative phase for the phase-locked case as shown in Fig.~\ref{fig:scanfix}. The fit parameters according to Eqs.~(\ref{eq:alphadot}) and (\ref{eq:phioff}) are given by $\epsilon=\SI{4.32(3)e-9}{}$, $\varphi_{\mathrm{off}}=\SI{-1.335(5)}{rad}$, and $\nicefrac{\chi^2}{\text{ndf}}=\nicefrac {55.13}{30}=1.83$. 
    }
    \label{fig:init_slope_vs_rel_phase}
\end{figure}

With this method, the resonance strength $\epsilon$ can be extracted by varying $\phirel$ from cycle to cycle and fitting the sinusoidal dependence of the slope $\dot\alpha$. This is shown in Fig.~\ref{fig:init_slope_vs_rel_phase}. 
The data are fitted using Eq.~(\ref{eq:alphadot}) with 
\begin{equation}
  \phirel^\mathrm{meas} = \phirel + \varphi_{\mathrm{off}} \label{eq:phioff}
\end{equation}
introducing the additional parameter $\varphi_\mathrm{off}$ to address the unknown, but constant setup-specific offset.
Considering the result of $\varphi_{\mathrm{off}}=\SI{-1.335}{rad}$ for this particular fit, the values for $\phirel$ for the exemplary cycles in Fig.~\ref{fig:scanfix} are $\phirel^\mathrm{1B} = \SI{3.65}{rad}$ and $\phirel^\mathrm{2B} = \SI{0.51}{rad}$, \ie close to the optimal phases of 0 and $\pi$.

\subsection{Phase Locking with Two Bunches}
\label{sec:two_bunches}

Using a second, undisturbed bunch for the phase-lock feedback, the unmodified Eqs.~(\ref{eq:alpha}) and (\ref{eq:phi}) can be applied. For a resonant operation ($q=0$) starting with the spin precessing in the ring plane ($\alpha(0)=0$) one can simplify the equations
(\ref{eq:A1})--(\ref{eq:A4}):
\begin{eqnarray}
A_1 &=& |\cos\phirel(0)| \label{eq:A1_opt}\\
A_2 &=& 0\,\mathrm{or}\,\pi \label{eq:A2_opt}\\
A_3 &=& k/2 \label{eq:A3_opt} \\
A_4 &=& 0 \label{eq:A4_opt}
\end{eqnarray}
and obtains
\begin{equation}
p_y(n) = \sin\alpha(n) = \cos\phirel(0)\sin(k n / 2) \\
\end{equation}
or 
\begin{equation}
\label{eq:free}
p_y(t) = \sin\alpha(t) = \cos\phirel(0)\sin(2\pi\epsilon\fc t) \,.\\
\end{equation}
This is consistent with the expectation: the amplitude of the oscillation depends on the initial relative phase $\phirel(0)$ and the oscillation frequency is fixed at $\omega = 2\pi\varepsilon f_{\mathrm{rev}}$, which can then be used to extract the resonance strength $\epsilon$.

\begin{figure}
    \centering    \includegraphics[width=\columnwidth]{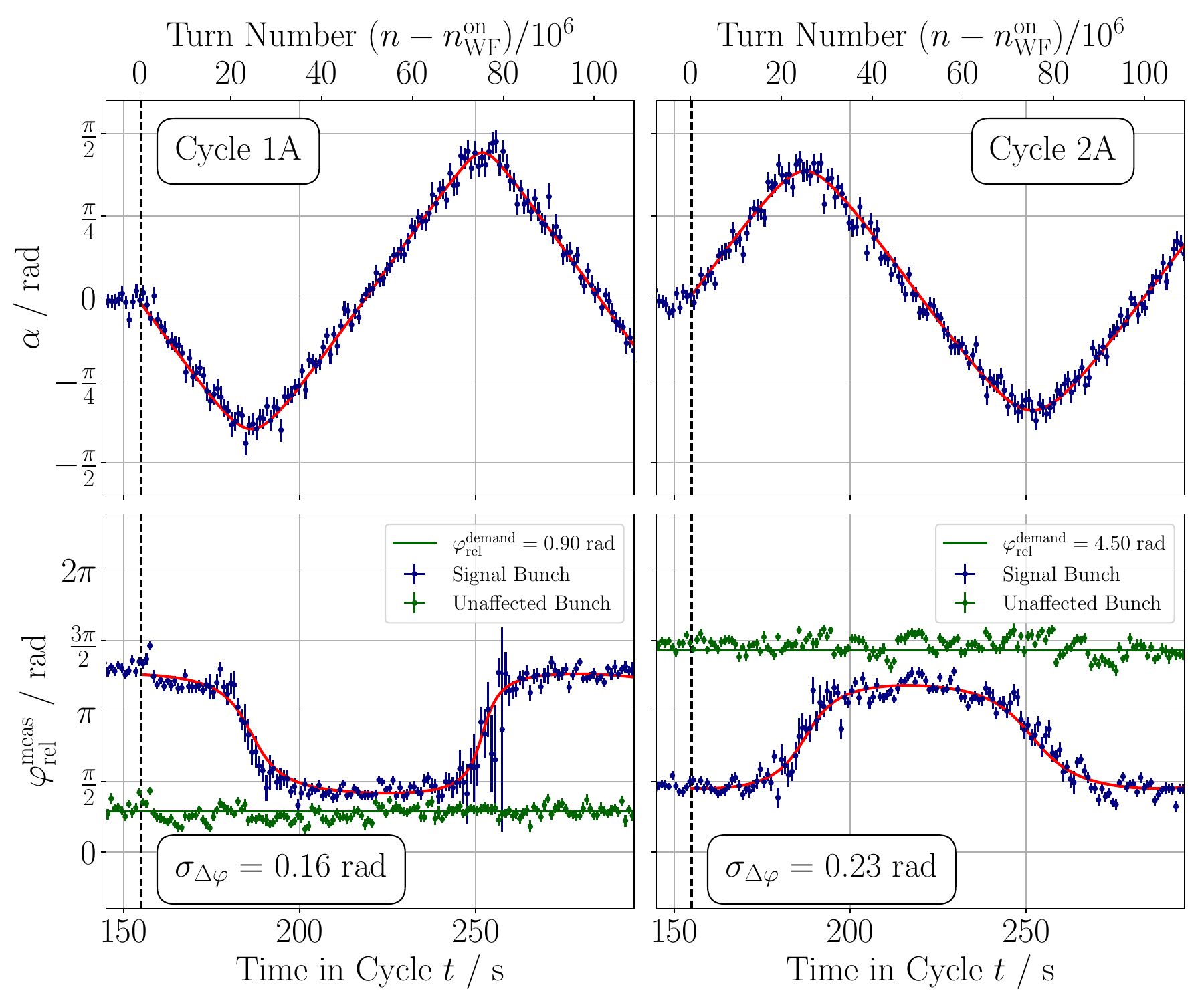}
    \caption{Out-of-plane angle $\alpha$ (top) and relative phase $\phirel^\mathrm{meas}$ (bottom) as a function of time for two initial values of $\phirel^\mathrm{demand}$ 
    (left and right). The dashed line marks the time the Wien filter is switched on $t_{\mathrm{WF}}^{\mathrm{on}}$. The blue data points represent the signal bunch, the green ones the unaffected bunch used by the feedback system. The fit parameters are listed in Table~\ref{tab:scan}.
    }
    \label{fig:scan}
\end{figure}

\begin{table}
\centering
\caption{Fit parameters from Fig.~\ref{fig:scan}}
\label{tab:scan}
\begin{tabular*}{\columnwidth}{@{\extracolsep{\fill}}lcc}
& Cycle 1A & Cycle 2A \\
\hline \hline
Unaffected bunch & & \\
\hline
$\phirel^\mathrm{demand}$ / \si{rad} & 0.90 & 4.50 \\
$\overline\varphi_\mathrm{rel}^\mathrm{meas}$ / \si{rad} & 0.86(1) & 4.60(1) \\
$\sigma_{\Delta\phirel}$ / \si{rad} & 0.16 & 0.21 \\
\hline \hline
Signal bunch & & \\
\hline
$A_1$ & 0.968(2) & 0.908(3) \\
$A_2$ & $\pi + 0.07(1)$ & $0.00(1)$ \\
$A_3$ & \num{6.38(2)e-8} & \num{6.49(3)e-8} \\
$A_4$ & $-0.017(3)$ & $-0.029(4)$ \\
$C$ & $-0.73(2)$ & 0.99(2) \\
$k$ & \num{1.27(1)e-7} & \num{1.29(1)e-7} \\
$q$ & $-0.07(3)$ & $-0.07(2)$ \\
$\alpha(0)$ / rad & $-0.05(3)$ & 0.03(3)\\
$\phirel^{\mathrm{meas}}(0)$ / rad & 3.96(3) & 1.42(1)\\
$\varphi_\mathrm{off}$ & 1.07(1) & 0.99(1) \\
$\phirel(0)$ / rad & 2.89(3) & 0.43(1)\\
$\chi^2$/ndf & 370/277 & 496/277 \\
\hline \hline
\end{tabular*}%
\end{table}

The result of such a measurement is shown in Fig.~\ref{fig:scan}. Here, the out-of-plane angle of the polarization vector and the relative phase are plotted versus the time in cycle for two different initial relative phases using the same two cycles 1A and 2A, which are shown in Fig.~\ref{fig:phase_feeback_action}. Each cycle is fitted with the general Eqs.~(\ref{eq:sinalpha})-(\ref{eq:sinphi}) simultaneously.
The green data points in the phase plot show the unaffected bunch, which is used to monitor and control frequency and phase. The blue data points show the signal bunch.

The corresponding fit results are shown in Table~\ref{tab:scan}. In addition to the variables $A_1$ to $A_4$ and $C$, the values for $k$, $q$ and the start values 
$\alpha(0)$ and $\phirel^{\mathrm{meas}}(0)$ are listed. The values of $A_1$ to $A_4$ are close to the optimal ones in Eqs.~(\ref{eq:A1_opt})-(\ref{eq:A4_opt}) indicating the successful operation of the feedback system. The number for the off-resonance parameter $q$ slightly differs from zero if only the statistical error is considered. However, as mentioned before the fit quality suggests additional systematic uncertainties and larger error margins. Furthermore, its value corresponds to a frequency deviation of $|\Delta f| \approx \SI{5e-4}{Hz}$ (see Eq.~(\ref{eq:q})), which has to be compared to the step size of the frequency generator of $\SI{1e-3}{Hz}$.

In the fits, the offset parameter $\varphi_\mathrm{off}$ amounts to approximately \SI{1}{rad} for this setup. Thus, the actual relative phases for the signal bunch are $\phirel^\mathrm{1A} \approx \qty{2.9}{rad}$ and $\phirel^\mathrm{2A} \approx \qty{0.4}{rad}$. These relative phases are close to the optimal phases, but not exactly. Therefore, the oscillation of the polarization vector does not reach the maximum out-of-plane angles of $\pm\pi/2$ ($A_1 = |\cos\phirel(0)|$).

In general, there is a good agreement between the analytical model and the data presented in Figs.~\ref{fig:scanfix} and \ref{fig:scan}.

The scenarios discussed in the last two subsections \ref{sec:single_bunch} and \ref{sec:two_bunches} overall result in a different evolution of the out-of-plane angle. When comparing Eq.~(\ref{eq:alphadot}) with Eq.~(\ref{eq:free}), one gets the same slope at $t=0$:
\begin{equation}
\frac{\mathrm{d}\alpha}{\mathrm{d}t}^\mathrm{Eq.(\ref*{eq:alphadot})} = \left.\frac{\mathrm{d}\alpha}{\mathrm{d}t}^\mathrm{Eq.(\ref*{eq:free})}\right|_{t=0} = 2\pi\epsilon\fc\ .
\end{equation} 
In general, this can be used to compare the results from the different methods. However, concerning the presented data the results shown in Figs.\ref{fig:scanfix} and \ref{fig:scan} are not directly comparable, because they were taken in two different beam times with two different setups (2018 and 2021), and both the Wien filter power and the setup changed.

\section{Summary \& Conclusion}

This paper reports the successful operation of a phase-lock feedback system that maintains a resonance condition between a \SI{120}{kHz} spin precession in a storage ring and a radio-frequency (rf) Wien filter operated at a harmonic of the spin precession. By continuously measuring the spin precession, the frequency and phase of the Wien filter were dynamically adjusted to sustain a constant, selectable phase difference between the spin precession and the rf signal of the Wien filter.
The performance of the system is quantified by the deviation of the actual phase from the demanded phase, with the variation characterized by a standard deviation of approximately $\sigma_{\Delta\varphi} \approx \qty{0.2}{rad}$. The measured time dependence of the polarization vector and the relative phase is in good agreement with an analytical model.

Feedback systems of this type are mandatory for proposed storage ring experiments using polarized beam in order to search for electric dipole moments \cite{abusaif2021} or axion/ALPs \cite{JEDI:2022hxaa}. The case when the system is combined with the ability to do bunch-selective spin rotations using dedicated bunches for a continuous spin-tune determination, can be of more general interest. It could be used for other applications, that benefit from a precise knowledge of the free spin tune, but do not need long-term stabilization. For example, one can think of fine-tuning fast rf spin-flippers.

\section{Acknowledgments}

We thank the COSY crew for their support in operating the COSY accelerator for the experiment. The work presented here has been carried out in the framework of the JEDI collaboration and was supported by an ERC Advanced Grant from the European Union (Proposal No. 694340: Search for electric dipole moments using storage rings). The work of A.\,Aksentev, A.\,Melnikov and N.N.\,Nikolaev was supported by the Russian Science Foundation (Grant No.\,22-42-04419). This research is part of a project funded by the European Union's Horizon 2020 research and innovation program under grant agreement STRONG-2020-No. 824093.

\bibliography{literature}

\begin{thebibliography}{34}%
\makeatletter
\providecommand \@ifxundefined [1]{%
 \@ifx{#1\undefined}
}%
\providecommand \@ifnum [1]{%
 \ifnum #1\expandafter \@firstoftwo
 \else \expandafter \@secondoftwo
 \fi
}%
\providecommand \@ifx [1]{%
 \ifx #1\expandafter \@firstoftwo
 \else \expandafter \@secondoftwo
 \fi
}%
\providecommand \natexlab [1]{#1}%
\providecommand \enquote  [1]{``#1''}%
\providecommand \bibnamefont  [1]{#1}%
\providecommand \bibfnamefont [1]{#1}%
\providecommand \citenamefont [1]{#1}%
\providecommand \href@noop [0]{\@secondoftwo}%
\providecommand \href [0]{\begingroup \@sanitize@url \@href}%
\providecommand \@href[1]{\@@startlink{#1}\@@href}%
\providecommand \@@href[1]{\endgroup#1\@@endlink}%
\providecommand \@sanitize@url [0]{\catcode `\\12\catcode `\$12\catcode
  `\&12\catcode `\#12\catcode `\^12\catcode `\_12\catcode `\%12\relax}%
\providecommand \@@startlink[1]{}%
\providecommand \@@endlink[0]{}%
\providecommand \url  [0]{\begingroup\@sanitize@url \@url }%
\providecommand \@url [1]{\endgroup\@href {#1}{\urlprefix }}%
\providecommand \urlprefix  [0]{URL }%
\providecommand \Eprint [0]{\href }%
\providecommand \doibase [0]{https://doi.org/}%
\providecommand \selectlanguage [0]{\@gobble}%
\providecommand \bibinfo  [0]{\@secondoftwo}%
\providecommand \bibfield  [0]{\@secondoftwo}%
\providecommand \translation [1]{[#1]}%
\providecommand \BibitemOpen [0]{}%
\providecommand \bibitemStop [0]{}%
\providecommand \bibitemNoStop [0]{.\EOS\space}%
\providecommand \EOS [0]{\spacefactor3000\relax}%
\providecommand \BibitemShut  [1]{\csname bibitem#1\endcsname}%
\let\auto@bib@innerbib\@empty
\bibitem [{\citenamefont {Caussyn}\ \emph {et~al.}(1994)\citenamefont
  {Caussyn}, \citenamefont {Derbenev}, \citenamefont {Ellison}, \citenamefont
  {Lee}, \citenamefont {Rinckel}, \citenamefont {Schwandt}, \citenamefont
  {Sperisen}, \citenamefont {Stephenson}, \citenamefont {{von Przewoski}},
  \citenamefont {Blinov}, \citenamefont {Chu}, \citenamefont {Courant},
  \citenamefont {Crandell}, \citenamefont {Kaufman}, \citenamefont {Krisch},
  \citenamefont {Nurushev}, \citenamefont {Phelps}, \citenamefont {Ratner},
  \citenamefont {Wong},\ and\ \citenamefont {Ohmori}}]{Caussyn:1994aea}%
  \BibitemOpen
  \bibfield  {author} {\bibinfo {author} {\bibfnamefont {D.~D.}\ \bibnamefont
  {Caussyn}}, \bibinfo {author} {\bibfnamefont {{\relax Ya}.~S.}\ \bibnamefont
  {Derbenev}}, \bibinfo {author} {\bibfnamefont {T.~J.~P.}\ \bibnamefont
  {Ellison}}, \bibinfo {author} {\bibfnamefont {S.~Y.}\ \bibnamefont {Lee}},
  \bibinfo {author} {\bibfnamefont {T.}~\bibnamefont {Rinckel}}, \bibinfo
  {author} {\bibfnamefont {P.}~\bibnamefont {Schwandt}}, \bibinfo {author}
  {\bibfnamefont {F.}~\bibnamefont {Sperisen}}, \bibinfo {author}
  {\bibfnamefont {E.~J.}\ \bibnamefont {Stephenson}}, \bibinfo {author}
  {\bibfnamefont {B.}~\bibnamefont {{von Przewoski}}}, \bibinfo {author}
  {\bibfnamefont {B.~B.}\ \bibnamefont {Blinov}}, \bibinfo {author}
  {\bibfnamefont {C.~M.}\ \bibnamefont {Chu}}, \bibinfo {author} {\bibfnamefont
  {E.~D.}\ \bibnamefont {Courant}}, \bibinfo {author} {\bibfnamefont {D.~A.}\
  \bibnamefont {Crandell}}, \bibinfo {author} {\bibfnamefont {W.~A.}\
  \bibnamefont {Kaufman}}, \bibinfo {author} {\bibfnamefont {A.~D.}\
  \bibnamefont {Krisch}}, \bibinfo {author} {\bibfnamefont {T.~S.}\
  \bibnamefont {Nurushev}}, \bibinfo {author} {\bibfnamefont {R.~A.}\
  \bibnamefont {Phelps}}, \bibinfo {author} {\bibfnamefont {L.~G.}\
  \bibnamefont {Ratner}}, \bibinfo {author} {\bibfnamefont {V.~K.}\
  \bibnamefont {Wong}},\ and\ \bibinfo {author} {\bibfnamefont
  {C.}~\bibnamefont {Ohmori}},\ }\bibfield  {title} {\bibinfo {title} {Spin
  {{Flipping}} a {{Stored Polarized Proton Beam}}},\ }\href
  {https://doi.org/10.1103/PhysRevLett.73.2857} {\bibfield  {journal} {\bibinfo
   {journal} {Physical Review Letters}\ }\textbf {\bibinfo {volume} {73}},\
  \bibinfo {pages} {2857} (\bibinfo {year} {1994})}\BibitemShut {NoStop}%
\bibitem [{\citenamefont {Froissart}\ and\ \citenamefont
  {Stora}(1960)}]{Froissart:1960zz}%
  \BibitemOpen
  \bibfield  {author} {\bibinfo {author} {\bibfnamefont {M.}~\bibnamefont
  {Froissart}}\ and\ \bibinfo {author} {\bibfnamefont {R.}~\bibnamefont
  {Stora}},\ }\bibfield  {title} {\bibinfo {title} {Depolarisation d'un
  faisceau de protons polarises dans un synchrotron},\ }\href
  {https://doi.org/10.1016/0029-554X(60)90033-1} {\bibfield  {journal}
  {\bibinfo  {journal} {Nuclear Instruments and Methods}\ }\textbf {\bibinfo
  {volume} {7}},\ \bibinfo {pages} {297} (\bibinfo {year} {1960})}\BibitemShut
  {NoStop}%
\bibitem [{\citenamefont {Blinov}\ \emph {et~al.}(1998)\citenamefont {Blinov},
  \citenamefont {Anferov}, \citenamefont {Derbenev}, \citenamefont {Kageya},
  \citenamefont {Krisch}, \citenamefont {Lorenzon}, \citenamefont {Ratner},
  \citenamefont {Sivers}, \citenamefont {Sourkont}, \citenamefont {Wong},
  \citenamefont {Chu}, \citenamefont {Lee}, \citenamefont {Rinckel},
  \citenamefont {Schwandt}, \citenamefont {Sperisen}, \citenamefont {{von
  Przewoski}},\ and\ \citenamefont {Sato}}]{Blinov:1998ya}%
  \BibitemOpen
  \bibfield  {author} {\bibinfo {author} {\bibfnamefont {B.~B.}\ \bibnamefont
  {Blinov}}, \bibinfo {author} {\bibfnamefont {V.~A.}\ \bibnamefont {Anferov}},
  \bibinfo {author} {\bibfnamefont {{\relax Ya}.~S.}\ \bibnamefont {Derbenev}},
  \bibinfo {author} {\bibfnamefont {T.}~\bibnamefont {Kageya}}, \bibinfo
  {author} {\bibfnamefont {A.~D.}\ \bibnamefont {Krisch}}, \bibinfo {author}
  {\bibfnamefont {W.}~\bibnamefont {Lorenzon}}, \bibinfo {author}
  {\bibfnamefont {L.~G.}\ \bibnamefont {Ratner}}, \bibinfo {author}
  {\bibfnamefont {D.~W.}\ \bibnamefont {Sivers}}, \bibinfo {author}
  {\bibfnamefont {K.~V.}\ \bibnamefont {Sourkont}}, \bibinfo {author}
  {\bibfnamefont {V.~K.}\ \bibnamefont {Wong}}, \bibinfo {author}
  {\bibfnamefont {C.~M.}\ \bibnamefont {Chu}}, \bibinfo {author} {\bibfnamefont
  {S.~Y.}\ \bibnamefont {Lee}}, \bibinfo {author} {\bibfnamefont
  {T.}~\bibnamefont {Rinckel}}, \bibinfo {author} {\bibfnamefont
  {P.}~\bibnamefont {Schwandt}}, \bibinfo {author} {\bibfnamefont
  {F.}~\bibnamefont {Sperisen}}, \bibinfo {author} {\bibfnamefont
  {B.}~\bibnamefont {{von Przewoski}}},\ and\ \bibinfo {author} {\bibfnamefont
  {H.}~\bibnamefont {Sato}},\ }\bibfield  {title} {\bibinfo {title} {Spin
  {{Flipping}} in the {{Presence}} of a {{Full Siberian Snake}}},\ }\href
  {https://doi.org/10.1103/PhysRevLett.81.2906} {\bibfield  {journal} {\bibinfo
   {journal} {Physical Review Letters}\ }\textbf {\bibinfo {volume} {81}},\
  \bibinfo {pages} {2906} (\bibinfo {year} {1998})}\BibitemShut {NoStop}%
\bibitem [{\citenamefont {Anferov}\ \emph {et~al.}(1999)\citenamefont {Anferov}
  \emph {et~al.}}]{Anferov:1998wr}%
  \BibitemOpen
  \bibfield  {author} {\bibinfo {author} {\bibfnamefont {V.~A.}\ \bibnamefont
  {Anferov}} \emph {et~al.},\ }\bibfield  {title} {\bibinfo {title} {Spin
  flipping with a full {{Siberian}} snake in a cooler ring},\ }in\ \href
  {https://doi.org/10.1142/3952} {\emph {\bibinfo {booktitle} {13th
  {{International Symposium}} on {{High-Energy Spin Physics}} ({{SPIN}} 98)}}}\
  (\bibinfo  {publisher} {World Scientific},\ \bibinfo {year} {1999})\ pp.\
  \bibinfo {pages} {503--505}\BibitemShut {NoStop}%
\bibitem [{\citenamefont {Morozov}\ \emph {et~al.}(2004)\citenamefont
  {Morozov}, \citenamefont {Krisch}, \citenamefont {Leonova}, \citenamefont
  {Raymond}, \citenamefont {Wong}, \citenamefont {Yonehara}, \citenamefont
  {Bechstedt}, \citenamefont {Gebel}, \citenamefont {Lehrach}, \citenamefont
  {Lorentz}, \citenamefont {Maier}, \citenamefont {Prasuhn}, \citenamefont
  {Schnase}, \citenamefont {Stockhorst}, \citenamefont {Eversheim},
  \citenamefont {Hinterberger}, \citenamefont {Rohdje{\ss}},\ and\
  \citenamefont {Ulbrich}}]{Morozov:2004rb}%
  \BibitemOpen
  \bibfield  {author} {\bibinfo {author} {\bibfnamefont {V.~S.}\ \bibnamefont
  {Morozov}}, \bibinfo {author} {\bibfnamefont {A.~D.}\ \bibnamefont {Krisch}},
  \bibinfo {author} {\bibfnamefont {M.~A.}\ \bibnamefont {Leonova}}, \bibinfo
  {author} {\bibfnamefont {R.~S.}\ \bibnamefont {Raymond}}, \bibinfo {author}
  {\bibfnamefont {V.~K.}\ \bibnamefont {Wong}}, \bibinfo {author}
  {\bibfnamefont {K.}~\bibnamefont {Yonehara}}, \bibinfo {author}
  {\bibfnamefont {U.}~\bibnamefont {Bechstedt}}, \bibinfo {author}
  {\bibfnamefont {R.}~\bibnamefont {Gebel}}, \bibinfo {author} {\bibfnamefont
  {A.}~\bibnamefont {Lehrach}}, \bibinfo {author} {\bibfnamefont
  {B.}~\bibnamefont {Lorentz}}, \bibinfo {author} {\bibfnamefont
  {R.}~\bibnamefont {Maier}}, \bibinfo {author} {\bibfnamefont
  {D.}~\bibnamefont {Prasuhn}}, \bibinfo {author} {\bibfnamefont
  {A.}~\bibnamefont {Schnase}}, \bibinfo {author} {\bibfnamefont
  {H.}~\bibnamefont {Stockhorst}}, \bibinfo {author} {\bibfnamefont
  {D.}~\bibnamefont {Eversheim}}, \bibinfo {author} {\bibfnamefont
  {F.}~\bibnamefont {Hinterberger}}, \bibinfo {author} {\bibfnamefont
  {H.}~\bibnamefont {Rohdje{\ss}}},\ and\ \bibinfo {author} {\bibfnamefont
  {K.}~\bibnamefont {Ulbrich}},\ }\bibfield  {title} {\bibinfo {title} {Spin
  manipulation of 1.94 {GeV/c} polarized protons stored in the {{COSY}} cooler
  synchrotron},\ }\href {https://doi.org/10.1103/PhysRevSTAB.7.024002}
  {\bibfield  {journal} {\bibinfo  {journal} {Physical Review Special Topics -
  Accelerators and Beams}\ }\textbf {\bibinfo {volume} {7}},\ \bibinfo {pages}
  {024002} (\bibinfo {year} {2004})}\BibitemShut {NoStop}%
\bibitem [{\citenamefont {Morozov}\ \emph {et~al.}(2001)\citenamefont
  {Morozov}, \citenamefont {Anferov}, \citenamefont {Blinov}, \citenamefont
  {Krisch}, \citenamefont {Lorenzon}, \citenamefont {Peters}, \citenamefont
  {Yonehara}, \citenamefont {Farkhonden}, \citenamefont {Franklin},
  \citenamefont {Jacobs}, \citenamefont {Kolster}, \citenamefont {Sirca},
  \citenamefont {Smith}, \citenamefont {Tsentalovich}, \citenamefont {Vieregg},
  \citenamefont {Zwart},\ and\ \citenamefont {Six}}]{Morozov:2001ne}%
  \BibitemOpen
  \bibfield  {author} {\bibinfo {author} {\bibfnamefont {V.~S.}\ \bibnamefont
  {Morozov}}, \bibinfo {author} {\bibfnamefont {V.~A.}\ \bibnamefont
  {Anferov}}, \bibinfo {author} {\bibfnamefont {B.~B.}\ \bibnamefont {Blinov}},
  \bibinfo {author} {\bibfnamefont {A.~D.}\ \bibnamefont {Krisch}}, \bibinfo
  {author} {\bibfnamefont {W.}~\bibnamefont {Lorenzon}}, \bibinfo {author}
  {\bibfnamefont {C.~C.}\ \bibnamefont {Peters}}, \bibinfo {author}
  {\bibfnamefont {K.}~\bibnamefont {Yonehara}}, \bibinfo {author}
  {\bibfnamefont {M.}~\bibnamefont {Farkhonden}}, \bibinfo {author}
  {\bibfnamefont {W.~A.}\ \bibnamefont {Franklin}}, \bibinfo {author}
  {\bibfnamefont {K.~D.}\ \bibnamefont {Jacobs}}, \bibinfo {author}
  {\bibfnamefont {H.}~\bibnamefont {Kolster}}, \bibinfo {author} {\bibfnamefont
  {S.}~\bibnamefont {Sirca}}, \bibinfo {author} {\bibfnamefont
  {T.}~\bibnamefont {Smith}}, \bibinfo {author} {\bibfnamefont
  {E.}~\bibnamefont {Tsentalovich}}, \bibinfo {author} {\bibfnamefont
  {J.}~\bibnamefont {Vieregg}}, \bibinfo {author} {\bibfnamefont {G.~T.}\
  \bibnamefont {Zwart}},\ and\ \bibinfo {author} {\bibfnamefont
  {E.}~\bibnamefont {Six}},\ }\bibfield  {title} {\bibinfo {title}
  {Spin-flipping polarized electrons},\ }\href
  {https://doi.org/10.1103/PhysRevSTAB.4.104002} {\bibfield  {journal}
  {\bibinfo  {journal} {Physical Review Special Topics - Accelerators and
  Beams}\ }\textbf {\bibinfo {volume} {4}},\ \bibinfo {pages} {104002}
  (\bibinfo {year} {2001})}\BibitemShut {NoStop}%
\bibitem [{\citenamefont {Huang}\ \emph {et~al.}(2018)\citenamefont {Huang},
  \citenamefont {Kewisch}, \citenamefont {Liu}, \citenamefont {Marusic},
  \citenamefont {Meng}, \citenamefont {M{\'e}ot}, \citenamefont {Oddo},
  \citenamefont {Ptitsyn}, \citenamefont {Ranjbar},\ and\ \citenamefont
  {Roser}}]{Huang:2018jtf}%
  \BibitemOpen
  \bibfield  {author} {\bibinfo {author} {\bibfnamefont {H.}~\bibnamefont
  {Huang}}, \bibinfo {author} {\bibfnamefont {J.}~\bibnamefont {Kewisch}},
  \bibinfo {author} {\bibfnamefont {C.}~\bibnamefont {Liu}}, \bibinfo {author}
  {\bibfnamefont {A.}~\bibnamefont {Marusic}}, \bibinfo {author} {\bibfnamefont
  {W.}~\bibnamefont {Meng}}, \bibinfo {author} {\bibfnamefont {F.}~\bibnamefont
  {M{\'e}ot}}, \bibinfo {author} {\bibfnamefont {P.}~\bibnamefont {Oddo}},
  \bibinfo {author} {\bibfnamefont {V.}~\bibnamefont {Ptitsyn}}, \bibinfo
  {author} {\bibfnamefont {V.}~\bibnamefont {Ranjbar}},\ and\ \bibinfo {author}
  {\bibfnamefont {T.}~\bibnamefont {Roser}},\ }\bibfield  {title} {\bibinfo
  {title} {High {{Spin-Flip Efficiency}} at 255 {{GeV}} for {{Polarized
  Protons}} in a {{Ring With Two Full Siberian Snakes}}},\ }\href
  {https://doi.org/10.1103/PhysRevLett.120.264804} {\bibfield  {journal}
  {\bibinfo  {journal} {Physical Review Letters}\ }\textbf {\bibinfo {volume}
  {120}},\ \bibinfo {pages} {264804} (\bibinfo {year} {2018})}\BibitemShut
  {NoStop}%
\bibitem [{\citenamefont {Huang}\ \emph {et~al.}(2019)\citenamefont {Huang},
  \citenamefont {Kewisch}, \citenamefont {Liu}, \citenamefont {Marusic},
  \citenamefont {Meng}, \citenamefont {M{\'e}ot}, \citenamefont {Oddo},
  \citenamefont {Ptitsyn}, \citenamefont {Ranjbar}, \citenamefont {Roser},\
  and\ \citenamefont {Schmidke}}]{Huang:2019mce}%
  \BibitemOpen
  \bibfield  {author} {\bibinfo {author} {\bibfnamefont {H.}~\bibnamefont
  {Huang}}, \bibinfo {author} {\bibfnamefont {J.}~\bibnamefont {Kewisch}},
  \bibinfo {author} {\bibfnamefont {C.}~\bibnamefont {Liu}}, \bibinfo {author}
  {\bibfnamefont {A.}~\bibnamefont {Marusic}}, \bibinfo {author} {\bibfnamefont
  {W.}~\bibnamefont {Meng}}, \bibinfo {author} {\bibfnamefont {F.}~\bibnamefont
  {M{\'e}ot}}, \bibinfo {author} {\bibfnamefont {P.}~\bibnamefont {Oddo}},
  \bibinfo {author} {\bibfnamefont {V.}~\bibnamefont {Ptitsyn}}, \bibinfo
  {author} {\bibfnamefont {V.}~\bibnamefont {Ranjbar}}, \bibinfo {author}
  {\bibfnamefont {T.}~\bibnamefont {Roser}},\ and\ \bibinfo {author}
  {\bibfnamefont {W.~B.}\ \bibnamefont {Schmidke}},\ }\bibfield  {title}
  {\bibinfo {title} {Measurement of the {{Spin Tune Using}} the {{Coherent Spin
  Motion}} of {{Polarized Protons}} in a {{Storage Ring}}},\ }\href
  {https://doi.org/10.1103/PhysRevLett.122.204803} {\bibfield  {journal}
  {\bibinfo  {journal} {Physical Review Letters}\ }\textbf {\bibinfo {volume}
  {122}},\ \bibinfo {pages} {204803} (\bibinfo {year} {2019})}\BibitemShut
  {NoStop}%
\bibitem [{\citenamefont {Bai}\ \emph {et~al.}(1998)\citenamefont {Bai},
  \citenamefont {Ahrens}, \citenamefont {Alessi}, \citenamefont {Brown},
  \citenamefont {Bunce}, \citenamefont {Cameron}, \citenamefont {Chu},
  \citenamefont {Glenn}, \citenamefont {Huang}, \citenamefont {Kponou},
  \citenamefont {Krueger}, \citenamefont {Lamble}, \citenamefont {Luccio},
  \citenamefont {Makdisi}, \citenamefont {Lee}, \citenamefont {Okamura},
  \citenamefont {Ratner}, \citenamefont {Reece}, \citenamefont {Roser},
  \citenamefont {Spinka}, \citenamefont {Syphers}, \citenamefont {Tsoupas},
  \citenamefont {Underwood}, \citenamefont {{van Asselt}}, \citenamefont
  {Williams},\ and\ \citenamefont {Yokosawa}}]{Bai:1998gj}%
  \BibitemOpen
  \bibfield  {author} {\bibinfo {author} {\bibfnamefont {M.}~\bibnamefont
  {Bai}}, \bibinfo {author} {\bibfnamefont {L.}~\bibnamefont {Ahrens}},
  \bibinfo {author} {\bibfnamefont {J.}~\bibnamefont {Alessi}}, \bibinfo
  {author} {\bibfnamefont {K.}~\bibnamefont {Brown}}, \bibinfo {author}
  {\bibfnamefont {G.}~\bibnamefont {Bunce}}, \bibinfo {author} {\bibfnamefont
  {P.}~\bibnamefont {Cameron}}, \bibinfo {author} {\bibfnamefont {C.~M.}\
  \bibnamefont {Chu}}, \bibinfo {author} {\bibfnamefont {J.~W.}\ \bibnamefont
  {Glenn}}, \bibinfo {author} {\bibfnamefont {H.}~\bibnamefont {Huang}},
  \bibinfo {author} {\bibfnamefont {A.~E.}\ \bibnamefont {Kponou}}, \bibinfo
  {author} {\bibfnamefont {K.}~\bibnamefont {Krueger}}, \bibinfo {author}
  {\bibfnamefont {W.}~\bibnamefont {Lamble}}, \bibinfo {author} {\bibfnamefont
  {A.}~\bibnamefont {Luccio}}, \bibinfo {author} {\bibfnamefont {Y.~I.}\
  \bibnamefont {Makdisi}}, \bibinfo {author} {\bibfnamefont {S.~Y.}\
  \bibnamefont {Lee}}, \bibinfo {author} {\bibfnamefont {M.}~\bibnamefont
  {Okamura}}, \bibinfo {author} {\bibfnamefont {L.}~\bibnamefont {Ratner}},
  \bibinfo {author} {\bibfnamefont {K.}~\bibnamefont {Reece}}, \bibinfo
  {author} {\bibfnamefont {T.}~\bibnamefont {Roser}}, \bibinfo {author}
  {\bibfnamefont {H.}~\bibnamefont {Spinka}}, \bibinfo {author} {\bibfnamefont
  {M.~J.}\ \bibnamefont {Syphers}}, \bibinfo {author} {\bibfnamefont
  {N.}~\bibnamefont {Tsoupas}}, \bibinfo {author} {\bibfnamefont {D.~G.}\
  \bibnamefont {Underwood}}, \bibinfo {author} {\bibfnamefont {W.}~\bibnamefont
  {{van Asselt}}}, \bibinfo {author} {\bibfnamefont {N.}~\bibnamefont
  {Williams}},\ and\ \bibinfo {author} {\bibfnamefont {A.}~\bibnamefont
  {Yokosawa}},\ }\bibfield  {title} {\bibinfo {title} {Overcoming {{Intrinsic
  Spin Resonances}} with an rf {{Dipole}}},\ }\href
  {https://doi.org/10.1103/PhysRevLett.80.4673} {\bibfield  {journal} {\bibinfo
   {journal} {Physical Review Letters}\ }\textbf {\bibinfo {volume} {80}},\
  \bibinfo {pages} {4673} (\bibinfo {year} {1998})}\BibitemShut {NoStop}%
\bibitem [{\citenamefont {Slim}\ \emph {et~al.}(2016)\citenamefont {Slim},
  \citenamefont {Gebel}, \citenamefont {Heberling}, \citenamefont {Hinder},
  \citenamefont {H{\"o}lscher}, \citenamefont {Lehrach}, \citenamefont
  {Lorentz}, \citenamefont {Mey}, \citenamefont {Nass}, \citenamefont
  {Rathmann}, \citenamefont {Reifferscheidt}, \citenamefont {Soltner},
  \citenamefont {Straatmann}, \citenamefont {Trinkel},\ and\ \citenamefont
  {Wolters}}]{Slim:2016pim}%
  \BibitemOpen
  \bibfield  {author} {\bibinfo {author} {\bibfnamefont {J.}~\bibnamefont
  {Slim}}, \bibinfo {author} {\bibfnamefont {R.}~\bibnamefont {Gebel}},
  \bibinfo {author} {\bibfnamefont {D.}~\bibnamefont {Heberling}}, \bibinfo
  {author} {\bibfnamefont {F.}~\bibnamefont {Hinder}}, \bibinfo {author}
  {\bibfnamefont {D.}~\bibnamefont {H{\"o}lscher}}, \bibinfo {author}
  {\bibfnamefont {A.}~\bibnamefont {Lehrach}}, \bibinfo {author} {\bibfnamefont
  {B.}~\bibnamefont {Lorentz}}, \bibinfo {author} {\bibfnamefont
  {S.}~\bibnamefont {Mey}}, \bibinfo {author} {\bibfnamefont {A.}~\bibnamefont
  {Nass}}, \bibinfo {author} {\bibfnamefont {F.}~\bibnamefont {Rathmann}},
  \bibinfo {author} {\bibfnamefont {L.}~\bibnamefont {Reifferscheidt}},
  \bibinfo {author} {\bibfnamefont {H.}~\bibnamefont {Soltner}}, \bibinfo
  {author} {\bibfnamefont {H.}~\bibnamefont {Straatmann}}, \bibinfo {author}
  {\bibfnamefont {F.}~\bibnamefont {Trinkel}},\ and\ \bibinfo {author}
  {\bibfnamefont {J.}~\bibnamefont {Wolters}},\ }\bibfield  {title} {\bibinfo
  {title} {Electromagnetic {{Simulation}} and {{Design}} of a {{Novel Waveguide
  RF Wien Filter}} for {{Electric Dipole Moment Measurements}} of {{Protons}}
  and {{Deuterons}}},\ }\href {https://doi.org/10.1016/j.nima.2016.05.012}
  {\bibfield  {journal} {\bibinfo  {journal} {Nuclear Instruments and Methods
  in Physics Research Section A: Accelerators, Spectrometers, Detectors and
  Associated Equipment}\ }\textbf {\bibinfo {volume} {828}},\ \bibinfo {pages}
  {116} (\bibinfo {year} {2016})}\BibitemShut {NoStop}%
\bibitem [{\citenamefont {Slim}\ \emph {et~al.}(2020)\citenamefont {Slim},
  \citenamefont {Nass}, \citenamefont {Rathmann}, \citenamefont {Soltner},
  \citenamefont {Tagliente},\ and\ \citenamefont {Heberling}}]{Slim:2020ufk}%
  \BibitemOpen
  \bibfield  {author} {\bibinfo {author} {\bibfnamefont {J.}~\bibnamefont
  {Slim}}, \bibinfo {author} {\bibfnamefont {A.}~\bibnamefont {Nass}}, \bibinfo
  {author} {\bibfnamefont {F.}~\bibnamefont {Rathmann}}, \bibinfo {author}
  {\bibfnamefont {H.}~\bibnamefont {Soltner}}, \bibinfo {author} {\bibfnamefont
  {G.}~\bibnamefont {Tagliente}},\ and\ \bibinfo {author} {\bibfnamefont
  {D.}~\bibnamefont {Heberling}},\ }\bibfield  {title} {\bibinfo {title} {The
  driving circuit of the waveguide {{RF Wien}} filter for the deuteron {{EDM}}
  precursor experiment at {{COSY}}},\ }\href
  {https://doi.org/10.1088/1748-0221/15/03/P03021} {\bibfield  {journal}
  {\bibinfo  {journal} {Journal of Instrumentation}\ }\textbf {\bibinfo
  {volume} {15}}\bibinfo  {number} { (03)},\ \bibinfo {pages}
  {P03021}}\BibitemShut {NoStop}%
\bibitem [{\citenamefont {Navas}\ \emph {et~al.}(2024)\citenamefont {Navas}
  \emph {et~al.}}]{ParticleDataGroup:2024cfk}%
  \BibitemOpen
\bibfield  {number} {  }\bibfield  {author} {\bibinfo {author} {\bibfnamefont
  {S.}~\bibnamefont {Navas}} \emph {et~al.} (\bibinfo {collaboration} {Particle
  Data Group}),\ }\bibfield  {title} {\bibinfo {title} {{Review of particle
  physics}},\ }\href {https://doi.org/10.1103/PhysRevD.110.030001} {\bibfield
  {journal} {\bibinfo  {journal} {Phys. Rev. D}\ }\textbf {\bibinfo {volume}
  {110}},\ \bibinfo {pages} {030001} (\bibinfo {year} {2024})}\BibitemShut
  {NoStop}%
\bibitem [{\citenamefont {Bargmann}\ \emph {et~al.}(1959)\citenamefont
  {Bargmann}, \citenamefont {Michel},\ and\ \citenamefont
  {Telegdi}}]{Bargmann:1959gza}%
  \BibitemOpen
  \bibfield  {author} {\bibinfo {author} {\bibfnamefont {V.}~\bibnamefont
  {Bargmann}}, \bibinfo {author} {\bibfnamefont {L.}~\bibnamefont {Michel}},\
  and\ \bibinfo {author} {\bibfnamefont {V.~L.}\ \bibnamefont {Telegdi}},\
  }\bibfield  {title} {\bibinfo {title} {Precession of the {{Polarization}} of
  {{Particles Moving}} in a {{Homogeneous Electromagnetic Field}}},\ }\href
  {https://doi.org/10.1103/PhysRevLett.2.435} {\bibfield  {journal} {\bibinfo
  {journal} {Physical Review Letters}\ }\textbf {\bibinfo {volume} {2}},\
  \bibinfo {pages} {435} (\bibinfo {year} {1959})}\BibitemShut {NoStop}%
\bibitem [{\citenamefont {Fukuyama}\ and\ \citenamefont
  {Silenko}(2013)}]{Fukuyama:2013ioa}%
  \BibitemOpen
  \bibfield  {author} {\bibinfo {author} {\bibfnamefont {T.}~\bibnamefont
  {Fukuyama}}\ and\ \bibinfo {author} {\bibfnamefont {A.~J.}\ \bibnamefont
  {Silenko}},\ }\bibfield  {title} {\bibinfo {title} {Derivation of
  {{Generalized Thomas}}--{{Bargmann}}--{{Michel}}--{{Telegdi Equation}} for a
  {{Particle}} with {{Electric Dipole Moment}}},\ }\href
  {https://doi.org/10.1142/S0217751X13501479} {\bibfield  {journal} {\bibinfo
  {journal} {International Journal of Modern Physics A}\ }\textbf {\bibinfo
  {volume} {28}},\ \bibinfo {pages} {1350147} (\bibinfo {year}
  {2013})}\BibitemShut {NoStop}%
\bibitem [{\citenamefont {Rathmann}\ \emph {et~al.}(2013)\citenamefont
  {Rathmann}, \citenamefont {Saleev},\ and\ \citenamefont
  {Nikolaev}}]{Rathmann:2013rqa}%
  \BibitemOpen
  \bibfield  {author} {\bibinfo {author} {\bibfnamefont {F.}~\bibnamefont
  {Rathmann}}, \bibinfo {author} {\bibfnamefont {A.}~\bibnamefont {Saleev}},\
  and\ \bibinfo {author} {\bibfnamefont {N.~N.}\ \bibnamefont {Nikolaev}},\
  }\bibfield  {title} {\bibinfo {title} {The search for electric dipole moments
  of light ions in storage rings},\ }\href
  {https://doi.org/10.1088/1742-6596/447/1/012011} {\bibfield  {journal}
  {\bibinfo  {journal} {Journal of Physics: Conference Series}\ }\textbf
  {\bibinfo {volume} {447}},\ \bibinfo {pages} {012011} (\bibinfo {year}
  {2013})}\BibitemShut {NoStop}%
\bibitem [{\citenamefont {Morse}\ \emph {et~al.}(2013)\citenamefont {Morse},
  \citenamefont {Orlov},\ and\ \citenamefont {Semertzidis}}]{Morse:2013hoa}%
  \BibitemOpen
  \bibfield  {author} {\bibinfo {author} {\bibfnamefont {W.~M.}\ \bibnamefont
  {Morse}}, \bibinfo {author} {\bibfnamefont {Y.~F.}\ \bibnamefont {Orlov}},\
  and\ \bibinfo {author} {\bibfnamefont {Y.~K.}\ \bibnamefont {Semertzidis}},\
  }\bibfield  {title} {\bibinfo {title} {Rf {{Wien}} filter in an electric
  dipole moment storage ring: {{The}} ``partially frozen spin'' effect},\
  }\href {https://doi.org/10.1103/PhysRevSTAB.16.114001} {\bibfield  {journal}
  {\bibinfo  {journal} {Physical Review Special Topics - Accelerators and
  Beams}\ }\textbf {\bibinfo {volume} {16}},\ \bibinfo {pages} {114001}
  (\bibinfo {year} {2013})}\BibitemShut {NoStop}%
\bibitem [{\citenamefont {Rathmann}\ \emph {et~al.}(2020)\citenamefont
  {Rathmann}, \citenamefont {Nikolaev},\ and\ \citenamefont
  {Slim}}]{Rathmann:2019lwi}%
  \BibitemOpen
  \bibfield  {author} {\bibinfo {author} {\bibfnamefont {F.}~\bibnamefont
  {Rathmann}}, \bibinfo {author} {\bibfnamefont {N.~N.}\ \bibnamefont
  {Nikolaev}},\ and\ \bibinfo {author} {\bibfnamefont {J.}~\bibnamefont
  {Slim}},\ }\bibfield  {title} {\bibinfo {title} {Spin dynamics investigations
  for the electric dipole moment experiment},\ }\href
  {https://doi.org/10.1103/PhysRevAccelBeams.23.024601} {\bibfield  {journal}
  {\bibinfo  {journal} {Physical Review Accelerators and Beams}\ }\textbf
  {\bibinfo {volume} {23}},\ \bibinfo {pages} {024601} (\bibinfo {year}
  {2020})}\BibitemShut {NoStop}%
\bibitem [{\citenamefont {Hempelmann}\ \emph {et~al.}(2017)\citenamefont
  {Hempelmann}, \citenamefont {Hejny}, \citenamefont {Pretz}, \citenamefont
  {Stephenson}, \citenamefont {Augustyniak}, \citenamefont {Bagdasarian},
  \citenamefont {Bai}, \citenamefont {Barion}, \citenamefont {Berz},
  \citenamefont {Chekmenev}, \citenamefont {Ciullo}, \citenamefont {Dymov},
  \citenamefont {Etzkorn}, \citenamefont {Eversmann}, \citenamefont {Gaisser},
  \citenamefont {Gebel}, \citenamefont {Grigoryev}, \citenamefont {Grzonka},
  \citenamefont {Guidoboni}, \citenamefont {Hanraths}, \citenamefont
  {Heberling}, \citenamefont {Hetzel}, \citenamefont {Hinder}, \citenamefont
  {Kacharava}, \citenamefont {Kamerdzhiev}, \citenamefont {Keshelashvili},
  \citenamefont {Koop}, \citenamefont {Kulikov}, \citenamefont {Lehrach},
  \citenamefont {Lenisa}, \citenamefont {Lomidze}, \citenamefont {Lorentz},
  \citenamefont {Maanen}, \citenamefont {Macharashvili}, \citenamefont
  {Magiera}, \citenamefont {Mchedlishvili}, \citenamefont {Mey}, \citenamefont
  {M{\"u}ller}, \citenamefont {Nass}, \citenamefont {Nikolaev}, \citenamefont
  {Pesce}, \citenamefont {Prasuhn}, \citenamefont {Rathmann}, \citenamefont
  {Rosenthal}, \citenamefont {Saleev}, \citenamefont {Schmidt}, \citenamefont
  {Semertzidis}, \citenamefont {Shmakova}, \citenamefont {Silenko},
  \citenamefont {Slim}, \citenamefont {Soltner}, \citenamefont {Stahl},
  \citenamefont {Stassen}, \citenamefont {Stockhorst}, \citenamefont
  {Str{\"o}her}, \citenamefont {Tabidze}, \citenamefont {Tagliente},
  \citenamefont {Talman}, \citenamefont {Th{\"o}rngren~Engblom}, \citenamefont
  {Trinkel}, \citenamefont {Uzikov}, \citenamefont {Valdau}, \citenamefont
  {Valetov}, \citenamefont {Vassiliev}, \citenamefont {Weidemann},
  \citenamefont {Wro{\'n}ska}, \citenamefont {W{\"u}stner}, \citenamefont
  {Zupra{\'n}ski},\ and\ \citenamefont {{\.Z}urek}}]{JEDI:2017bnp}%
  \BibitemOpen
  \bibfield  {author} {\bibinfo {author} {\bibfnamefont {N.}~\bibnamefont
  {Hempelmann}}, \bibinfo {author} {\bibfnamefont {V.}~\bibnamefont {Hejny}},
  \bibinfo {author} {\bibfnamefont {J.}~\bibnamefont {Pretz}}, \bibinfo
  {author} {\bibfnamefont {E.}~\bibnamefont {Stephenson}}, \bibinfo {author}
  {\bibfnamefont {W.}~\bibnamefont {Augustyniak}}, \bibinfo {author}
  {\bibfnamefont {Z.}~\bibnamefont {Bagdasarian}}, \bibinfo {author}
  {\bibfnamefont {M.}~\bibnamefont {Bai}}, \bibinfo {author} {\bibfnamefont
  {L.}~\bibnamefont {Barion}}, \bibinfo {author} {\bibfnamefont
  {M.}~\bibnamefont {Berz}}, \bibinfo {author} {\bibfnamefont {S.}~\bibnamefont
  {Chekmenev}}, \bibinfo {author} {\bibfnamefont {G.}~\bibnamefont {Ciullo}},
  \bibinfo {author} {\bibfnamefont {S.}~\bibnamefont {Dymov}}, \bibinfo
  {author} {\bibfnamefont {F.-J.}\ \bibnamefont {Etzkorn}}, \bibinfo {author}
  {\bibfnamefont {D.}~\bibnamefont {Eversmann}}, \bibinfo {author}
  {\bibfnamefont {M.}~\bibnamefont {Gaisser}}, \bibinfo {author} {\bibfnamefont
  {R.}~\bibnamefont {Gebel}}, \bibinfo {author} {\bibfnamefont
  {K.}~\bibnamefont {Grigoryev}}, \bibinfo {author} {\bibfnamefont
  {D.}~\bibnamefont {Grzonka}}, \bibinfo {author} {\bibfnamefont
  {G.}~\bibnamefont {Guidoboni}}, \bibinfo {author} {\bibfnamefont
  {T.}~\bibnamefont {Hanraths}}, \bibinfo {author} {\bibfnamefont
  {D.}~\bibnamefont {Heberling}}, \bibinfo {author} {\bibfnamefont
  {J.}~\bibnamefont {Hetzel}}, \bibinfo {author} {\bibfnamefont
  {F.}~\bibnamefont {Hinder}}, \bibinfo {author} {\bibfnamefont
  {A.}~\bibnamefont {Kacharava}}, \bibinfo {author} {\bibfnamefont
  {V.}~\bibnamefont {Kamerdzhiev}}, \bibinfo {author} {\bibfnamefont
  {I.}~\bibnamefont {Keshelashvili}}, \bibinfo {author} {\bibfnamefont
  {I.}~\bibnamefont {Koop}}, \bibinfo {author} {\bibfnamefont {A.}~\bibnamefont
  {Kulikov}}, \bibinfo {author} {\bibfnamefont {A.}~\bibnamefont {Lehrach}},
  \bibinfo {author} {\bibfnamefont {P.}~\bibnamefont {Lenisa}}, \bibinfo
  {author} {\bibfnamefont {N.}~\bibnamefont {Lomidze}}, \bibinfo {author}
  {\bibfnamefont {B.}~\bibnamefont {Lorentz}}, \bibinfo {author} {\bibfnamefont
  {P.}~\bibnamefont {Maanen}}, \bibinfo {author} {\bibfnamefont
  {G.}~\bibnamefont {Macharashvili}}, \bibinfo {author} {\bibfnamefont
  {A.}~\bibnamefont {Magiera}}, \bibinfo {author} {\bibfnamefont
  {D.}~\bibnamefont {Mchedlishvili}}, \bibinfo {author} {\bibfnamefont
  {S.}~\bibnamefont {Mey}}, \bibinfo {author} {\bibfnamefont {F.}~\bibnamefont
  {M{\"u}ller}}, \bibinfo {author} {\bibfnamefont {A.}~\bibnamefont {Nass}},
  \bibinfo {author} {\bibfnamefont {N.~N.}\ \bibnamefont {Nikolaev}}, \bibinfo
  {author} {\bibfnamefont {A.}~\bibnamefont {Pesce}}, \bibinfo {author}
  {\bibfnamefont {D.}~\bibnamefont {Prasuhn}}, \bibinfo {author} {\bibfnamefont
  {F.}~\bibnamefont {Rathmann}}, \bibinfo {author} {\bibfnamefont
  {M.}~\bibnamefont {Rosenthal}}, \bibinfo {author} {\bibfnamefont
  {A.}~\bibnamefont {Saleev}}, \bibinfo {author} {\bibfnamefont
  {V.}~\bibnamefont {Schmidt}}, \bibinfo {author} {\bibfnamefont
  {Y.}~\bibnamefont {Semertzidis}}, \bibinfo {author} {\bibfnamefont
  {V.}~\bibnamefont {Shmakova}}, \bibinfo {author} {\bibfnamefont
  {A.}~\bibnamefont {Silenko}}, \bibinfo {author} {\bibfnamefont
  {J.}~\bibnamefont {Slim}}, \bibinfo {author} {\bibfnamefont {H.}~\bibnamefont
  {Soltner}}, \bibinfo {author} {\bibfnamefont {A.}~\bibnamefont {Stahl}},
  \bibinfo {author} {\bibfnamefont {R.}~\bibnamefont {Stassen}}, \bibinfo
  {author} {\bibfnamefont {H.}~\bibnamefont {Stockhorst}}, \bibinfo {author}
  {\bibfnamefont {H.}~\bibnamefont {Str{\"o}her}}, \bibinfo {author}
  {\bibfnamefont {M.}~\bibnamefont {Tabidze}}, \bibinfo {author} {\bibfnamefont
  {G.}~\bibnamefont {Tagliente}}, \bibinfo {author} {\bibfnamefont
  {R.}~\bibnamefont {Talman}}, \bibinfo {author} {\bibfnamefont
  {P.}~\bibnamefont {Th{\"o}rngren~Engblom}}, \bibinfo {author} {\bibfnamefont
  {F.}~\bibnamefont {Trinkel}}, \bibinfo {author} {\bibfnamefont {{\relax
  Yu}.}~\bibnamefont {Uzikov}}, \bibinfo {author} {\bibfnamefont {{\relax
  Yu}.}~\bibnamefont {Valdau}}, \bibinfo {author} {\bibfnamefont
  {E.}~\bibnamefont {Valetov}}, \bibinfo {author} {\bibfnamefont
  {A.}~\bibnamefont {Vassiliev}}, \bibinfo {author} {\bibfnamefont
  {C.}~\bibnamefont {Weidemann}}, \bibinfo {author} {\bibfnamefont
  {A.}~\bibnamefont {Wro{\'n}ska}}, \bibinfo {author} {\bibfnamefont
  {P.}~\bibnamefont {W{\"u}stner}}, \bibinfo {author} {\bibfnamefont
  {P.}~\bibnamefont {Zupra{\'n}ski}},\ and\ \bibinfo {author} {\bibfnamefont
  {M.}~\bibnamefont {{\.Z}urek}},\ }\bibfield  {title} {\bibinfo {title} {Phase
  {{Locking}} the {{Spin Precession}} in a {{Storage Ring}}},\ }\href
  {https://doi.org/10.1103/PhysRevLett.119.014801} {\bibfield  {journal}
  {\bibinfo  {journal} {Physical Review Letters}\ }\textbf {\bibinfo {volume}
  {119}},\ \bibinfo {pages} {014801} (\bibinfo {year} {2017})}\BibitemShut
  {NoStop}%
\bibitem [{\citenamefont {Slim}\ \emph {et~al.}(2023)\citenamefont {Slim},
  \citenamefont {Rathmann}, \citenamefont {Andres}, \citenamefont {Hejny},
  \citenamefont {Nass}, \citenamefont {Kacharava}, \citenamefont {Lenisa},
  \citenamefont {Nikolaev}, \citenamefont {Pretz}, \citenamefont {Saleev},
  \citenamefont {Shmakova}, \citenamefont {Soltner}, \citenamefont {Abusaif},
  \citenamefont {Aggarwal}, \citenamefont {Aksentev}, \citenamefont {Alberdi},
  \citenamefont {Barion}, \citenamefont {Bekman}, \citenamefont {Bey{\ss}},
  \citenamefont {B{\"o}hme}, \citenamefont {Breitkreutz}, \citenamefont
  {Canale}, \citenamefont {Ciullo}, \citenamefont {Dymov}, \citenamefont
  {Fr{\"o}hlich}, \citenamefont {Gebel}, \citenamefont {Gaisser}, \citenamefont
  {Grigoryev}, \citenamefont {Grzonka}, \citenamefont {Hetzel}, \citenamefont
  {Javakhishvili}, \citenamefont {Kamerdzhiev}, \citenamefont {Karanth},
  \citenamefont {Keshelashvili}, \citenamefont {Kononov}, \citenamefont
  {Laihem}, \citenamefont {Lehrach}, \citenamefont {Lomidze}, \citenamefont
  {Lorentz}, \citenamefont {Macharashvili}, \citenamefont {Magiera},
  \citenamefont {Mchedlishvili}, \citenamefont {Melnikov}, \citenamefont
  {M{\"u}ller}, \citenamefont {Pesce}, \citenamefont {Poncza}, \citenamefont
  {Prasuhn}, \citenamefont {Shergelashvili}, \citenamefont {Shurkhno},
  \citenamefont {Siddique}, \citenamefont {Silenko}, \citenamefont {Stassen},
  \citenamefont {Stephenson}, \citenamefont {Str{\"o}her}, \citenamefont
  {Tabidze}, \citenamefont {Tagliente}, \citenamefont {Valdau}, \citenamefont
  {Vitz}, \citenamefont {Wagner}, \citenamefont {Wirzba}, \citenamefont
  {Wro{\'n}ska}, \citenamefont {W{\"u}stner},\ and\ \citenamefont
  {{\.Z}urek}}]{JEDI:2023btw}%
  \BibitemOpen
  \bibfield  {author} {\bibinfo {author} {\bibfnamefont {J.}~\bibnamefont
  {Slim}}, \bibinfo {author} {\bibfnamefont {F.}~\bibnamefont {Rathmann}},
  \bibinfo {author} {\bibfnamefont {A.}~\bibnamefont {Andres}}, \bibinfo
  {author} {\bibfnamefont {V.}~\bibnamefont {Hejny}}, \bibinfo {author}
  {\bibfnamefont {A.}~\bibnamefont {Nass}}, \bibinfo {author} {\bibfnamefont
  {A.}~\bibnamefont {Kacharava}}, \bibinfo {author} {\bibfnamefont
  {P.}~\bibnamefont {Lenisa}}, \bibinfo {author} {\bibfnamefont {N.~N.}\
  \bibnamefont {Nikolaev}}, \bibinfo {author} {\bibfnamefont {J.}~\bibnamefont
  {Pretz}}, \bibinfo {author} {\bibfnamefont {A.}~\bibnamefont {Saleev}},
  \bibinfo {author} {\bibfnamefont {V.}~\bibnamefont {Shmakova}}, \bibinfo
  {author} {\bibfnamefont {H.}~\bibnamefont {Soltner}}, \bibinfo {author}
  {\bibfnamefont {F.}~\bibnamefont {Abusaif}}, \bibinfo {author} {\bibfnamefont
  {A.}~\bibnamefont {Aggarwal}}, \bibinfo {author} {\bibfnamefont
  {A.}~\bibnamefont {Aksentev}}, \bibinfo {author} {\bibfnamefont
  {B.}~\bibnamefont {Alberdi}}, \bibinfo {author} {\bibfnamefont
  {L.}~\bibnamefont {Barion}}, \bibinfo {author} {\bibfnamefont
  {I.}~\bibnamefont {Bekman}}, \bibinfo {author} {\bibfnamefont
  {M.}~\bibnamefont {Bey{\ss}}}, \bibinfo {author} {\bibfnamefont
  {C.}~\bibnamefont {B{\"o}hme}}, \bibinfo {author} {\bibfnamefont
  {B.}~\bibnamefont {Breitkreutz}}, \bibinfo {author} {\bibfnamefont
  {N.}~\bibnamefont {Canale}}, \bibinfo {author} {\bibfnamefont
  {G.}~\bibnamefont {Ciullo}}, \bibinfo {author} {\bibfnamefont
  {S.}~\bibnamefont {Dymov}}, \bibinfo {author} {\bibfnamefont {N.-O.}\
  \bibnamefont {Fr{\"o}hlich}}, \bibinfo {author} {\bibfnamefont
  {R.}~\bibnamefont {Gebel}}, \bibinfo {author} {\bibfnamefont
  {M.}~\bibnamefont {Gaisser}}, \bibinfo {author} {\bibfnamefont
  {K.}~\bibnamefont {Grigoryev}}, \bibinfo {author} {\bibfnamefont
  {D.}~\bibnamefont {Grzonka}}, \bibinfo {author} {\bibfnamefont
  {J.}~\bibnamefont {Hetzel}}, \bibinfo {author} {\bibfnamefont
  {O.}~\bibnamefont {Javakhishvili}}, \bibinfo {author} {\bibfnamefont
  {V.}~\bibnamefont {Kamerdzhiev}}, \bibinfo {author} {\bibfnamefont
  {S.}~\bibnamefont {Karanth}}, \bibinfo {author} {\bibfnamefont
  {I.}~\bibnamefont {Keshelashvili}}, \bibinfo {author} {\bibfnamefont
  {A.}~\bibnamefont {Kononov}}, \bibinfo {author} {\bibfnamefont
  {K.}~\bibnamefont {Laihem}}, \bibinfo {author} {\bibfnamefont
  {A.}~\bibnamefont {Lehrach}}, \bibinfo {author} {\bibfnamefont
  {N.}~\bibnamefont {Lomidze}}, \bibinfo {author} {\bibfnamefont
  {B.}~\bibnamefont {Lorentz}}, \bibinfo {author} {\bibfnamefont
  {G.}~\bibnamefont {Macharashvili}}, \bibinfo {author} {\bibfnamefont
  {A.}~\bibnamefont {Magiera}}, \bibinfo {author} {\bibfnamefont
  {D.}~\bibnamefont {Mchedlishvili}}, \bibinfo {author} {\bibfnamefont
  {A.}~\bibnamefont {Melnikov}}, \bibinfo {author} {\bibfnamefont
  {F.}~\bibnamefont {M{\"u}ller}}, \bibinfo {author} {\bibfnamefont
  {A.}~\bibnamefont {Pesce}}, \bibinfo {author} {\bibfnamefont
  {V.}~\bibnamefont {Poncza}}, \bibinfo {author} {\bibfnamefont
  {D.}~\bibnamefont {Prasuhn}}, \bibinfo {author} {\bibfnamefont
  {D.}~\bibnamefont {Shergelashvili}}, \bibinfo {author} {\bibfnamefont
  {N.}~\bibnamefont {Shurkhno}}, \bibinfo {author} {\bibfnamefont
  {S.}~\bibnamefont {Siddique}}, \bibinfo {author} {\bibfnamefont
  {A.}~\bibnamefont {Silenko}}, \bibinfo {author} {\bibfnamefont
  {S.}~\bibnamefont {Stassen}}, \bibinfo {author} {\bibfnamefont {E.~J.}\
  \bibnamefont {Stephenson}}, \bibinfo {author} {\bibfnamefont
  {H.}~\bibnamefont {Str{\"o}her}}, \bibinfo {author} {\bibfnamefont
  {M.}~\bibnamefont {Tabidze}}, \bibinfo {author} {\bibfnamefont
  {G.}~\bibnamefont {Tagliente}}, \bibinfo {author} {\bibfnamefont
  {Y.}~\bibnamefont {Valdau}}, \bibinfo {author} {\bibfnamefont
  {M.}~\bibnamefont {Vitz}}, \bibinfo {author} {\bibfnamefont {T.}~\bibnamefont
  {Wagner}}, \bibinfo {author} {\bibfnamefont {A.}~\bibnamefont {Wirzba}},
  \bibinfo {author} {\bibfnamefont {A.}~\bibnamefont {Wro{\'n}ska}}, \bibinfo
  {author} {\bibfnamefont {P.}~\bibnamefont {W{\"u}stner}},\ and\ \bibinfo
  {author} {\bibfnamefont {M.}~\bibnamefont {{\.Z}urek}},\ }\href
  {https://doi.org/10.48550/arXiv.2309.06561} {\bibinfo {title} {Pilot bunch
  and co-magnetometry of polarized particles stored in a ring}} (\bibinfo
  {year} {2023}),\ \bibinfo {note} {accepted for publication in Phys. Rev.
  Research},\ \Eprint {https://arxiv.org/abs/2309.06561} {arXiv:2309.06561
  [hep-ex, physics:physics]} \BibitemShut {NoStop}%
\bibitem [{\citenamefont {Bagdasarian}\ \emph {et~al.}(2014)\citenamefont
  {Bagdasarian}, \citenamefont {Bertelli}, \citenamefont {Chiladze},
  \citenamefont {Ciullo}, \citenamefont {Dietrich}, \citenamefont {Dymov},
  \citenamefont {Eversmann}, \citenamefont {Fanourakis}, \citenamefont
  {Gaisser}, \citenamefont {Gebel}, \citenamefont {Gou}, \citenamefont
  {Guidoboni}, \citenamefont {Hejny}, \citenamefont {Kacharava}, \citenamefont
  {Kamerdzhiev}, \citenamefont {Lehrach}, \citenamefont {Lenisa}, \citenamefont
  {Lorentz}, \citenamefont {Magallanes}, \citenamefont {Maier}, \citenamefont
  {Mchedlishvili}, \citenamefont {Morse}, \citenamefont {Nass}, \citenamefont
  {Oellers}, \citenamefont {Pesce}, \citenamefont {Prasuhn}, \citenamefont
  {Pretz}, \citenamefont {Rathmann}, \citenamefont {Shmakova}, \citenamefont
  {Semertzidis}, \citenamefont {Stephenson}, \citenamefont {Stockhorst},
  \citenamefont {Str{\"o}her}, \citenamefont {Talman}, \citenamefont
  {Th{\"o}rngren~Engblom}, \citenamefont {Valdau}, \citenamefont {Weidemann},\
  and\ \citenamefont {W{\"u}stner}}]{Bagdasarian:2014ega}%
  \BibitemOpen
  \bibfield  {author} {\bibinfo {author} {\bibfnamefont {Z.}~\bibnamefont
  {Bagdasarian}}, \bibinfo {author} {\bibfnamefont {S.}~\bibnamefont
  {Bertelli}}, \bibinfo {author} {\bibfnamefont {D.}~\bibnamefont {Chiladze}},
  \bibinfo {author} {\bibfnamefont {G.}~\bibnamefont {Ciullo}}, \bibinfo
  {author} {\bibfnamefont {J.}~\bibnamefont {Dietrich}}, \bibinfo {author}
  {\bibfnamefont {S.}~\bibnamefont {Dymov}}, \bibinfo {author} {\bibfnamefont
  {D.}~\bibnamefont {Eversmann}}, \bibinfo {author} {\bibfnamefont
  {G.}~\bibnamefont {Fanourakis}}, \bibinfo {author} {\bibfnamefont
  {M.}~\bibnamefont {Gaisser}}, \bibinfo {author} {\bibfnamefont
  {R.}~\bibnamefont {Gebel}}, \bibinfo {author} {\bibfnamefont
  {B.}~\bibnamefont {Gou}}, \bibinfo {author} {\bibfnamefont {G.}~\bibnamefont
  {Guidoboni}}, \bibinfo {author} {\bibfnamefont {V.}~\bibnamefont {Hejny}},
  \bibinfo {author} {\bibfnamefont {A.}~\bibnamefont {Kacharava}}, \bibinfo
  {author} {\bibfnamefont {V.}~\bibnamefont {Kamerdzhiev}}, \bibinfo {author}
  {\bibfnamefont {A.}~\bibnamefont {Lehrach}}, \bibinfo {author} {\bibfnamefont
  {P.}~\bibnamefont {Lenisa}}, \bibinfo {author} {\bibfnamefont
  {B.}~\bibnamefont {Lorentz}}, \bibinfo {author} {\bibfnamefont
  {L.}~\bibnamefont {Magallanes}}, \bibinfo {author} {\bibfnamefont
  {R.}~\bibnamefont {Maier}}, \bibinfo {author} {\bibfnamefont
  {D.}~\bibnamefont {Mchedlishvili}}, \bibinfo {author} {\bibfnamefont {W.~M.}\
  \bibnamefont {Morse}}, \bibinfo {author} {\bibfnamefont {A.}~\bibnamefont
  {Nass}}, \bibinfo {author} {\bibfnamefont {D.}~\bibnamefont {Oellers}},
  \bibinfo {author} {\bibfnamefont {A.}~\bibnamefont {Pesce}}, \bibinfo
  {author} {\bibfnamefont {D.}~\bibnamefont {Prasuhn}}, \bibinfo {author}
  {\bibfnamefont {J.}~\bibnamefont {Pretz}}, \bibinfo {author} {\bibfnamefont
  {F.}~\bibnamefont {Rathmann}}, \bibinfo {author} {\bibfnamefont
  {V.}~\bibnamefont {Shmakova}}, \bibinfo {author} {\bibfnamefont {Y.~K.}\
  \bibnamefont {Semertzidis}}, \bibinfo {author} {\bibfnamefont {E.~J.}\
  \bibnamefont {Stephenson}}, \bibinfo {author} {\bibfnamefont
  {H.}~\bibnamefont {Stockhorst}}, \bibinfo {author} {\bibfnamefont
  {H.}~\bibnamefont {Str{\"o}her}}, \bibinfo {author} {\bibfnamefont
  {R.}~\bibnamefont {Talman}}, \bibinfo {author} {\bibfnamefont
  {P.}~\bibnamefont {Th{\"o}rngren~Engblom}}, \bibinfo {author} {\bibfnamefont
  {{\relax Yu}.}~\bibnamefont {Valdau}}, \bibinfo {author} {\bibfnamefont
  {C.}~\bibnamefont {Weidemann}},\ and\ \bibinfo {author} {\bibfnamefont
  {P.}~\bibnamefont {W{\"u}stner}},\ }\bibfield  {title} {\bibinfo {title}
  {Measuring the polarization of a rapidly precessing deuteron beam},\ }\href
  {https://doi.org/10.1103/PhysRevSTAB.17.052803} {\bibfield  {journal}
  {\bibinfo  {journal} {Physical Review Special Topics - Accelerators and
  Beams}\ }\textbf {\bibinfo {volume} {17}},\ \bibinfo {pages} {052803}
  (\bibinfo {year} {2014})}\BibitemShut {NoStop}%
\bibitem [{\citenamefont {Eversmann}\ \emph {et~al.}(2015)\citenamefont
  {Eversmann}, \citenamefont {Hejny}, \citenamefont {Hinder}, \citenamefont
  {Kacharava}, \citenamefont {Pretz}, \citenamefont {Rathmann}, \citenamefont
  {Rosenthal}, \citenamefont {Trinkel}, \citenamefont {Andrianov},
  \citenamefont {Augustyniak}, \citenamefont {Bagdasarian}, \citenamefont
  {Bai}, \citenamefont {Bernreuther}, \citenamefont {Bertelli}, \citenamefont
  {Berz}, \citenamefont {Bsaisou}, \citenamefont {Chekmenev}, \citenamefont
  {Chiladze}, \citenamefont {Ciullo}, \citenamefont {Contalbrigo},
  \citenamefont {{de Vries}}, \citenamefont {Dymov}, \citenamefont {Engels},
  \citenamefont {Esser}, \citenamefont {Felden}, \citenamefont {Gaisser},
  \citenamefont {Gebel}, \citenamefont {Gl{\"u}ckler}, \citenamefont
  {Goldenbaum}, \citenamefont {Grigoryev}, \citenamefont {Grzonka},
  \citenamefont {Guidoboni}, \citenamefont {Hanhart}, \citenamefont
  {Heberling}, \citenamefont {Hempelmann}, \citenamefont {Hetzel},
  \citenamefont {Hipple}, \citenamefont {H{\"o}lscher}, \citenamefont {Ivanov},
  \citenamefont {Kamerdzhiev}, \citenamefont {Kamys}, \citenamefont
  {Keshelashvili}, \citenamefont {Khoukaz}, \citenamefont {Koop}, \citenamefont
  {Krause}, \citenamefont {Krewald}, \citenamefont {Kulikov}, \citenamefont
  {Lehrach}, \citenamefont {Lenisa}, \citenamefont {Lomidze}, \citenamefont
  {Lorentz}, \citenamefont {Maanen}, \citenamefont {Macharashvili},
  \citenamefont {Magiera}, \citenamefont {Maier}, \citenamefont {Makino},
  \citenamefont {Maria{\'n}ski}, \citenamefont {Mchedlishvili}, \citenamefont
  {Mei{\ss}ner}, \citenamefont {Mey}, \citenamefont {Nass}, \citenamefont
  {Natour}, \citenamefont {Nikolaev}, \citenamefont {Nioradze}, \citenamefont
  {Nogga}, \citenamefont {Nowakowski}, \citenamefont {Pesce}, \citenamefont
  {Prasuhn}, \citenamefont {Ritman}, \citenamefont {Rudy}, \citenamefont
  {Saleev}, \citenamefont {Semertzidis}, \citenamefont {Senichev},
  \citenamefont {Shmakova}, \citenamefont {Silenko}, \citenamefont {Slim},
  \citenamefont {Soltner}, \citenamefont {Stahl}, \citenamefont {Stassen},
  \citenamefont {Statera}, \citenamefont {Stephenson}, \citenamefont
  {Stockhorst}, \citenamefont {Straatmann}, \citenamefont {Str{\"o}her},
  \citenamefont {Tabidze}, \citenamefont {Talman}, \citenamefont
  {Th{\"o}rngren~Engblom}, \citenamefont {Trzci{\'n}ski}, \citenamefont
  {Uzikov}, \citenamefont {Valdau}, \citenamefont {Valetov}, \citenamefont
  {Vassiliev}, \citenamefont {Weidemann}, \citenamefont {Wilkin}, \citenamefont
  {Wirzba}, \citenamefont {Wro{\'n}ska}, \citenamefont {W{\"u}stner},
  \citenamefont {Zakrzewska}, \citenamefont {Zupra{\'n}ski},\ and\
  \citenamefont {Zyuzin}}]{JEDI:2015vwa}%
  \BibitemOpen
  \bibfield  {author} {\bibinfo {author} {\bibfnamefont {D.}~\bibnamefont
  {Eversmann}}, \bibinfo {author} {\bibfnamefont {V.}~\bibnamefont {Hejny}},
  \bibinfo {author} {\bibfnamefont {F.}~\bibnamefont {Hinder}}, \bibinfo
  {author} {\bibfnamefont {A.}~\bibnamefont {Kacharava}}, \bibinfo {author}
  {\bibfnamefont {J.}~\bibnamefont {Pretz}}, \bibinfo {author} {\bibfnamefont
  {F.}~\bibnamefont {Rathmann}}, \bibinfo {author} {\bibfnamefont
  {M.}~\bibnamefont {Rosenthal}}, \bibinfo {author} {\bibfnamefont
  {F.}~\bibnamefont {Trinkel}}, \bibinfo {author} {\bibfnamefont
  {S.}~\bibnamefont {Andrianov}}, \bibinfo {author} {\bibfnamefont
  {W.}~\bibnamefont {Augustyniak}}, \bibinfo {author} {\bibfnamefont
  {Z.}~\bibnamefont {Bagdasarian}}, \bibinfo {author} {\bibfnamefont
  {M.}~\bibnamefont {Bai}}, \bibinfo {author} {\bibfnamefont {W.}~\bibnamefont
  {Bernreuther}}, \bibinfo {author} {\bibfnamefont {S.}~\bibnamefont
  {Bertelli}}, \bibinfo {author} {\bibfnamefont {M.}~\bibnamefont {Berz}},
  \bibinfo {author} {\bibfnamefont {J.}~\bibnamefont {Bsaisou}}, \bibinfo
  {author} {\bibfnamefont {S.}~\bibnamefont {Chekmenev}}, \bibinfo {author}
  {\bibfnamefont {D.}~\bibnamefont {Chiladze}}, \bibinfo {author}
  {\bibfnamefont {G.}~\bibnamefont {Ciullo}}, \bibinfo {author} {\bibfnamefont
  {M.}~\bibnamefont {Contalbrigo}}, \bibinfo {author} {\bibfnamefont
  {J.}~\bibnamefont {{de Vries}}}, \bibinfo {author} {\bibfnamefont
  {S.}~\bibnamefont {Dymov}}, \bibinfo {author} {\bibfnamefont
  {R.}~\bibnamefont {Engels}}, \bibinfo {author} {\bibfnamefont {F.~M.}\
  \bibnamefont {Esser}}, \bibinfo {author} {\bibfnamefont {O.}~\bibnamefont
  {Felden}}, \bibinfo {author} {\bibfnamefont {M.}~\bibnamefont {Gaisser}},
  \bibinfo {author} {\bibfnamefont {R.}~\bibnamefont {Gebel}}, \bibinfo
  {author} {\bibfnamefont {H.}~\bibnamefont {Gl{\"u}ckler}}, \bibinfo {author}
  {\bibfnamefont {F.}~\bibnamefont {Goldenbaum}}, \bibinfo {author}
  {\bibfnamefont {K.}~\bibnamefont {Grigoryev}}, \bibinfo {author}
  {\bibfnamefont {D.}~\bibnamefont {Grzonka}}, \bibinfo {author} {\bibfnamefont
  {G.}~\bibnamefont {Guidoboni}}, \bibinfo {author} {\bibfnamefont
  {C.}~\bibnamefont {Hanhart}}, \bibinfo {author} {\bibfnamefont
  {D.}~\bibnamefont {Heberling}}, \bibinfo {author} {\bibfnamefont
  {N.}~\bibnamefont {Hempelmann}}, \bibinfo {author} {\bibfnamefont
  {J.}~\bibnamefont {Hetzel}}, \bibinfo {author} {\bibfnamefont
  {R.}~\bibnamefont {Hipple}}, \bibinfo {author} {\bibfnamefont
  {D.}~\bibnamefont {H{\"o}lscher}}, \bibinfo {author} {\bibfnamefont
  {A.}~\bibnamefont {Ivanov}}, \bibinfo {author} {\bibfnamefont
  {V.}~\bibnamefont {Kamerdzhiev}}, \bibinfo {author} {\bibfnamefont
  {B.}~\bibnamefont {Kamys}}, \bibinfo {author} {\bibfnamefont
  {I.}~\bibnamefont {Keshelashvili}}, \bibinfo {author} {\bibfnamefont
  {A.}~\bibnamefont {Khoukaz}}, \bibinfo {author} {\bibfnamefont
  {I.}~\bibnamefont {Koop}}, \bibinfo {author} {\bibfnamefont {H.-J.}\
  \bibnamefont {Krause}}, \bibinfo {author} {\bibfnamefont {S.}~\bibnamefont
  {Krewald}}, \bibinfo {author} {\bibfnamefont {A.}~\bibnamefont {Kulikov}},
  \bibinfo {author} {\bibfnamefont {A.}~\bibnamefont {Lehrach}}, \bibinfo
  {author} {\bibfnamefont {P.}~\bibnamefont {Lenisa}}, \bibinfo {author}
  {\bibfnamefont {N.}~\bibnamefont {Lomidze}}, \bibinfo {author} {\bibfnamefont
  {B.}~\bibnamefont {Lorentz}}, \bibinfo {author} {\bibfnamefont
  {P.}~\bibnamefont {Maanen}}, \bibinfo {author} {\bibfnamefont
  {G.}~\bibnamefont {Macharashvili}}, \bibinfo {author} {\bibfnamefont
  {A.}~\bibnamefont {Magiera}}, \bibinfo {author} {\bibfnamefont
  {R.}~\bibnamefont {Maier}}, \bibinfo {author} {\bibfnamefont
  {K.}~\bibnamefont {Makino}}, \bibinfo {author} {\bibfnamefont
  {B.}~\bibnamefont {Maria{\'n}ski}}, \bibinfo {author} {\bibfnamefont
  {D.}~\bibnamefont {Mchedlishvili}}, \bibinfo {author} {\bibfnamefont {U.-G.}\
  \bibnamefont {Mei{\ss}ner}}, \bibinfo {author} {\bibfnamefont
  {S.}~\bibnamefont {Mey}}, \bibinfo {author} {\bibfnamefont {A.}~\bibnamefont
  {Nass}}, \bibinfo {author} {\bibfnamefont {G.}~\bibnamefont {Natour}},
  \bibinfo {author} {\bibfnamefont {N.}~\bibnamefont {Nikolaev}}, \bibinfo
  {author} {\bibfnamefont {M.}~\bibnamefont {Nioradze}}, \bibinfo {author}
  {\bibfnamefont {A.}~\bibnamefont {Nogga}}, \bibinfo {author} {\bibfnamefont
  {K.}~\bibnamefont {Nowakowski}}, \bibinfo {author} {\bibfnamefont
  {A.}~\bibnamefont {Pesce}}, \bibinfo {author} {\bibfnamefont
  {D.}~\bibnamefont {Prasuhn}}, \bibinfo {author} {\bibfnamefont
  {J.}~\bibnamefont {Ritman}}, \bibinfo {author} {\bibfnamefont
  {Z.}~\bibnamefont {Rudy}}, \bibinfo {author} {\bibfnamefont {A.}~\bibnamefont
  {Saleev}}, \bibinfo {author} {\bibfnamefont {Y.}~\bibnamefont {Semertzidis}},
  \bibinfo {author} {\bibfnamefont {Y.}~\bibnamefont {Senichev}}, \bibinfo
  {author} {\bibfnamefont {V.}~\bibnamefont {Shmakova}}, \bibinfo {author}
  {\bibfnamefont {A.}~\bibnamefont {Silenko}}, \bibinfo {author} {\bibfnamefont
  {J.}~\bibnamefont {Slim}}, \bibinfo {author} {\bibfnamefont {H.}~\bibnamefont
  {Soltner}}, \bibinfo {author} {\bibfnamefont {A.}~\bibnamefont {Stahl}},
  \bibinfo {author} {\bibfnamefont {R.}~\bibnamefont {Stassen}}, \bibinfo
  {author} {\bibfnamefont {M.}~\bibnamefont {Statera}}, \bibinfo {author}
  {\bibfnamefont {E.}~\bibnamefont {Stephenson}}, \bibinfo {author}
  {\bibfnamefont {H.}~\bibnamefont {Stockhorst}}, \bibinfo {author}
  {\bibfnamefont {H.}~\bibnamefont {Straatmann}}, \bibinfo {author}
  {\bibfnamefont {H.}~\bibnamefont {Str{\"o}her}}, \bibinfo {author}
  {\bibfnamefont {M.}~\bibnamefont {Tabidze}}, \bibinfo {author} {\bibfnamefont
  {R.}~\bibnamefont {Talman}}, \bibinfo {author} {\bibfnamefont
  {P.}~\bibnamefont {Th{\"o}rngren~Engblom}}, \bibinfo {author} {\bibfnamefont
  {A.}~\bibnamefont {Trzci{\'n}ski}}, \bibinfo {author} {\bibfnamefont {{\relax
  Yu}.}~\bibnamefont {Uzikov}}, \bibinfo {author} {\bibfnamefont {{\relax
  Yu}.}~\bibnamefont {Valdau}}, \bibinfo {author} {\bibfnamefont
  {E.}~\bibnamefont {Valetov}}, \bibinfo {author} {\bibfnamefont
  {A.}~\bibnamefont {Vassiliev}}, \bibinfo {author} {\bibfnamefont
  {C.}~\bibnamefont {Weidemann}}, \bibinfo {author} {\bibfnamefont
  {C.}~\bibnamefont {Wilkin}}, \bibinfo {author} {\bibfnamefont
  {A.}~\bibnamefont {Wirzba}}, \bibinfo {author} {\bibfnamefont
  {A.}~\bibnamefont {Wro{\'n}ska}}, \bibinfo {author} {\bibfnamefont
  {P.}~\bibnamefont {W{\"u}stner}}, \bibinfo {author} {\bibfnamefont
  {M.}~\bibnamefont {Zakrzewska}}, \bibinfo {author} {\bibfnamefont
  {P.}~\bibnamefont {Zupra{\'n}ski}},\ and\ \bibinfo {author} {\bibfnamefont
  {D.}~\bibnamefont {Zyuzin}},\ }\bibfield  {title} {\bibinfo {title} {New
  {{Method}} for a {{Continuous Determination}} of the {{Spin Tune}} in
  {{Storage Rings}} and {{Implications}} for {{Precision Experiments}}},\
  }\href {https://doi.org/10.1103/PhysRevLett.115.094801} {\bibfield  {journal}
  {\bibinfo  {journal} {Physical Review Letters}\ }\textbf {\bibinfo {volume}
  {115}},\ \bibinfo {pages} {094801} (\bibinfo {year} {2015})}\BibitemShut
  {NoStop}%
\bibitem [{\citenamefont {Hempelmann}\ \emph {et~al.}(2018)\citenamefont
  {Hempelmann}, \citenamefont {Hejny}, \citenamefont {Pretz}, \citenamefont
  {Soltner}, \citenamefont {Augustyniak}, \citenamefont {Bagdasarian},
  \citenamefont {Bai}, \citenamefont {Barion}, \citenamefont {Berz},
  \citenamefont {Chekmenev}, \citenamefont {Ciullo}, \citenamefont {Dymov},
  \citenamefont {Eversmann}, \citenamefont {Gaisser}, \citenamefont {Gebel},
  \citenamefont {Grigoryev}, \citenamefont {Grzonka}, \citenamefont
  {Guidoboni}, \citenamefont {Heberling}, \citenamefont {Hetzel}, \citenamefont
  {Hinder}, \citenamefont {Kacharava}, \citenamefont {Kamerdzhiev},
  \citenamefont {Keshelashvili}, \citenamefont {Koop}, \citenamefont {Kulikov},
  \citenamefont {Lehrach}, \citenamefont {Lenisa}, \citenamefont {Lomidze},
  \citenamefont {Lorentz}, \citenamefont {Maanen}, \citenamefont
  {Macharashvili}, \citenamefont {Magiera}, \citenamefont {Mchedlishvili},
  \citenamefont {Mey}, \citenamefont {M{\"u}ller}, \citenamefont {Nass},
  \citenamefont {Nikolaev}, \citenamefont {Nioradze}, \citenamefont {Pesce},
  \citenamefont {Prasuhn}, \citenamefont {Rathmann}, \citenamefont {Rosenthal},
  \citenamefont {Saleev}, \citenamefont {Schmidt}, \citenamefont {Semertzidis},
  \citenamefont {Senichev}, \citenamefont {Shmakova}, \citenamefont {Silenko},
  \citenamefont {Slim}, \citenamefont {Stahl}, \citenamefont {Stassen},
  \citenamefont {Stephenson}, \citenamefont {Stockhorst}, \citenamefont
  {Str{\"o}her}, \citenamefont {Tabidze}, \citenamefont {Tagliente},
  \citenamefont {Talman}, \citenamefont {Th{\"o}rngren~Engblom}, \citenamefont
  {Trinkel}, \citenamefont {Uzikov}, \citenamefont {Valdau}, \citenamefont
  {Valetov}, \citenamefont {Vassiliev}, \citenamefont {Weidemann},
  \citenamefont {Wro{\'n}ska}, \citenamefont {W{\"u}stner}, \citenamefont
  {Zupra{\'n}ski},\ and\ \citenamefont {{\.Z}urek}}]{JEDI:2018txsa}%
  \BibitemOpen
  \bibfield  {author} {\bibinfo {author} {\bibfnamefont {N.}~\bibnamefont
  {Hempelmann}}, \bibinfo {author} {\bibfnamefont {V.}~\bibnamefont {Hejny}},
  \bibinfo {author} {\bibfnamefont {J.}~\bibnamefont {Pretz}}, \bibinfo
  {author} {\bibfnamefont {H.}~\bibnamefont {Soltner}}, \bibinfo {author}
  {\bibfnamefont {W.}~\bibnamefont {Augustyniak}}, \bibinfo {author}
  {\bibfnamefont {Z.}~\bibnamefont {Bagdasarian}}, \bibinfo {author}
  {\bibfnamefont {M.}~\bibnamefont {Bai}}, \bibinfo {author} {\bibfnamefont
  {L.}~\bibnamefont {Barion}}, \bibinfo {author} {\bibfnamefont
  {M.}~\bibnamefont {Berz}}, \bibinfo {author} {\bibfnamefont {S.}~\bibnamefont
  {Chekmenev}}, \bibinfo {author} {\bibfnamefont {G.}~\bibnamefont {Ciullo}},
  \bibinfo {author} {\bibfnamefont {S.}~\bibnamefont {Dymov}}, \bibinfo
  {author} {\bibfnamefont {D.}~\bibnamefont {Eversmann}}, \bibinfo {author}
  {\bibfnamefont {M.}~\bibnamefont {Gaisser}}, \bibinfo {author} {\bibfnamefont
  {R.}~\bibnamefont {Gebel}}, \bibinfo {author} {\bibfnamefont
  {K.}~\bibnamefont {Grigoryev}}, \bibinfo {author} {\bibfnamefont
  {D.}~\bibnamefont {Grzonka}}, \bibinfo {author} {\bibfnamefont
  {G.}~\bibnamefont {Guidoboni}}, \bibinfo {author} {\bibfnamefont
  {D.}~\bibnamefont {Heberling}}, \bibinfo {author} {\bibfnamefont
  {J.}~\bibnamefont {Hetzel}}, \bibinfo {author} {\bibfnamefont
  {F.}~\bibnamefont {Hinder}}, \bibinfo {author} {\bibfnamefont
  {A.}~\bibnamefont {Kacharava}}, \bibinfo {author} {\bibfnamefont
  {V.}~\bibnamefont {Kamerdzhiev}}, \bibinfo {author} {\bibfnamefont
  {I.}~\bibnamefont {Keshelashvili}}, \bibinfo {author} {\bibfnamefont
  {I.}~\bibnamefont {Koop}}, \bibinfo {author} {\bibfnamefont {A.}~\bibnamefont
  {Kulikov}}, \bibinfo {author} {\bibfnamefont {A.}~\bibnamefont {Lehrach}},
  \bibinfo {author} {\bibfnamefont {P.}~\bibnamefont {Lenisa}}, \bibinfo
  {author} {\bibfnamefont {N.}~\bibnamefont {Lomidze}}, \bibinfo {author}
  {\bibfnamefont {B.}~\bibnamefont {Lorentz}}, \bibinfo {author} {\bibfnamefont
  {P.}~\bibnamefont {Maanen}}, \bibinfo {author} {\bibfnamefont
  {G.}~\bibnamefont {Macharashvili}}, \bibinfo {author} {\bibfnamefont
  {A.}~\bibnamefont {Magiera}}, \bibinfo {author} {\bibfnamefont
  {D.}~\bibnamefont {Mchedlishvili}}, \bibinfo {author} {\bibfnamefont
  {S.}~\bibnamefont {Mey}}, \bibinfo {author} {\bibfnamefont {F.}~\bibnamefont
  {M{\"u}ller}}, \bibinfo {author} {\bibfnamefont {A.}~\bibnamefont {Nass}},
  \bibinfo {author} {\bibfnamefont {N.~N.}\ \bibnamefont {Nikolaev}}, \bibinfo
  {author} {\bibfnamefont {M.}~\bibnamefont {Nioradze}}, \bibinfo {author}
  {\bibfnamefont {A.}~\bibnamefont {Pesce}}, \bibinfo {author} {\bibfnamefont
  {D.}~\bibnamefont {Prasuhn}}, \bibinfo {author} {\bibfnamefont
  {F.}~\bibnamefont {Rathmann}}, \bibinfo {author} {\bibfnamefont
  {M.}~\bibnamefont {Rosenthal}}, \bibinfo {author} {\bibfnamefont
  {A.}~\bibnamefont {Saleev}}, \bibinfo {author} {\bibfnamefont
  {V.}~\bibnamefont {Schmidt}}, \bibinfo {author} {\bibfnamefont
  {Y.}~\bibnamefont {Semertzidis}}, \bibinfo {author} {\bibfnamefont
  {Y.}~\bibnamefont {Senichev}}, \bibinfo {author} {\bibfnamefont
  {V.}~\bibnamefont {Shmakova}}, \bibinfo {author} {\bibfnamefont
  {A.}~\bibnamefont {Silenko}}, \bibinfo {author} {\bibfnamefont
  {J.}~\bibnamefont {Slim}}, \bibinfo {author} {\bibfnamefont {A.}~\bibnamefont
  {Stahl}}, \bibinfo {author} {\bibfnamefont {R.}~\bibnamefont {Stassen}},
  \bibinfo {author} {\bibfnamefont {E.}~\bibnamefont {Stephenson}}, \bibinfo
  {author} {\bibfnamefont {H.}~\bibnamefont {Stockhorst}}, \bibinfo {author}
  {\bibfnamefont {H.}~\bibnamefont {Str{\"o}her}}, \bibinfo {author}
  {\bibfnamefont {M.}~\bibnamefont {Tabidze}}, \bibinfo {author} {\bibfnamefont
  {G.}~\bibnamefont {Tagliente}}, \bibinfo {author} {\bibfnamefont
  {R.}~\bibnamefont {Talman}}, \bibinfo {author} {\bibfnamefont
  {P.}~\bibnamefont {Th{\"o}rngren~Engblom}}, \bibinfo {author} {\bibfnamefont
  {F.}~\bibnamefont {Trinkel}}, \bibinfo {author} {\bibfnamefont {{\relax
  Yu}.}~\bibnamefont {Uzikov}}, \bibinfo {author} {\bibfnamefont {{\relax
  Yu}.}~\bibnamefont {Valdau}}, \bibinfo {author} {\bibfnamefont
  {E.}~\bibnamefont {Valetov}}, \bibinfo {author} {\bibfnamefont
  {A.}~\bibnamefont {Vassiliev}}, \bibinfo {author} {\bibfnamefont
  {C.}~\bibnamefont {Weidemann}}, \bibinfo {author} {\bibfnamefont
  {A.}~\bibnamefont {Wro{\'n}ska}}, \bibinfo {author} {\bibfnamefont
  {P.}~\bibnamefont {W{\"u}stner}}, \bibinfo {author} {\bibfnamefont
  {P.}~\bibnamefont {Zupra{\'n}ski}},\ and\ \bibinfo {author} {\bibfnamefont
  {M.}~\bibnamefont {{\.Z}urek}},\ }\bibfield  {title} {\bibinfo {title} {Phase
  measurement for driven spin oscillations in a storage ring},\ }\href
  {https://doi.org/10.1103/PhysRevAccelBeams.21.042002} {\bibfield  {journal}
  {\bibinfo  {journal} {Physical Review Accelerators and Beams}\ }\textbf
  {\bibinfo {volume} {21}},\ \bibinfo {pages} {042002} (\bibinfo {year}
  {2018})}\BibitemShut {NoStop}%
\bibitem [{\citenamefont {Saleev}\ \emph {et~al.}(2017)\citenamefont {Saleev},
  \citenamefont {Nikolaev}, \citenamefont {Rathmann}, \citenamefont
  {Augustyniak}, \citenamefont {Bagdasarian}, \citenamefont {Bai},
  \citenamefont {Barion}, \citenamefont {Berz}, \citenamefont {Chekmenev},
  \citenamefont {Ciullo}, \citenamefont {Dymov}, \citenamefont {Eversmann},
  \citenamefont {Gaisser}, \citenamefont {Gebel}, \citenamefont {Grigoryev},
  \citenamefont {Grzonka}, \citenamefont {Guidoboni}, \citenamefont
  {Heberling}, \citenamefont {Hejny}, \citenamefont {Hempelmann}, \citenamefont
  {Hetzel}, \citenamefont {Hinder}, \citenamefont {Kacharava}, \citenamefont
  {Kamerdzhiev}, \citenamefont {Keshelashvili}, \citenamefont {Koop},
  \citenamefont {Kulikov}, \citenamefont {Lehrach}, \citenamefont {Lenisa},
  \citenamefont {Lomidze}, \citenamefont {Lorentz}, \citenamefont {Maanen},
  \citenamefont {Macharashvili}, \citenamefont {Magiera}, \citenamefont
  {McHedlishvili}, \citenamefont {Mey}, \citenamefont {M{\"u}ller},
  \citenamefont {Nass}, \citenamefont {Pesce}, \citenamefont {Prasuhn},
  \citenamefont {Pretz}, \citenamefont {Rosenthal}, \citenamefont {Schmidt},
  \citenamefont {Semertzidis}, \citenamefont {Senichev}, \citenamefont
  {Shmakova}, \citenamefont {Silenko}, \citenamefont {Slim}, \citenamefont
  {Soltner}, \citenamefont {Stahl}, \citenamefont {Stassen}, \citenamefont
  {Stephenson}, \citenamefont {Stockhorst}, \citenamefont {Str{\"o}her},
  \citenamefont {Tabidze}, \citenamefont {Tagliente}, \citenamefont {Talman},
  \citenamefont {Engblom}, \citenamefont {Trinkel}, \citenamefont {Uzikov},
  \citenamefont {Valdau}, \citenamefont {Valetov}, \citenamefont {Vassiliev},
  \citenamefont {Weidemann}, \citenamefont {Wro{\'n}ska}, \citenamefont
  {W{\"u}stner}, \citenamefont {Zupra{\'n}ski},\ and\ \citenamefont
  {Zurek}}]{JEDI:2017wlr}%
  \BibitemOpen
  \bibfield  {author} {\bibinfo {author} {\bibfnamefont {A.}~\bibnamefont
  {Saleev}}, \bibinfo {author} {\bibfnamefont {N.}~\bibnamefont {Nikolaev}},
  \bibinfo {author} {\bibfnamefont {F.}~\bibnamefont {Rathmann}}, \bibinfo
  {author} {\bibfnamefont {W.}~\bibnamefont {Augustyniak}}, \bibinfo {author}
  {\bibfnamefont {Z.}~\bibnamefont {Bagdasarian}}, \bibinfo {author}
  {\bibfnamefont {M.}~\bibnamefont {Bai}}, \bibinfo {author} {\bibfnamefont
  {L.}~\bibnamefont {Barion}}, \bibinfo {author} {\bibfnamefont
  {M.}~\bibnamefont {Berz}}, \bibinfo {author} {\bibfnamefont {S.}~\bibnamefont
  {Chekmenev}}, \bibinfo {author} {\bibfnamefont {G.}~\bibnamefont {Ciullo}},
  \bibinfo {author} {\bibfnamefont {S.}~\bibnamefont {Dymov}}, \bibinfo
  {author} {\bibfnamefont {D.}~\bibnamefont {Eversmann}}, \bibinfo {author}
  {\bibfnamefont {M.}~\bibnamefont {Gaisser}}, \bibinfo {author} {\bibfnamefont
  {R.}~\bibnamefont {Gebel}}, \bibinfo {author} {\bibfnamefont
  {K.}~\bibnamefont {Grigoryev}}, \bibinfo {author} {\bibfnamefont
  {D.}~\bibnamefont {Grzonka}}, \bibinfo {author} {\bibfnamefont
  {G.}~\bibnamefont {Guidoboni}}, \bibinfo {author} {\bibfnamefont
  {D.}~\bibnamefont {Heberling}}, \bibinfo {author} {\bibfnamefont
  {V.}~\bibnamefont {Hejny}}, \bibinfo {author} {\bibfnamefont
  {N.}~\bibnamefont {Hempelmann}}, \bibinfo {author} {\bibfnamefont
  {J.}~\bibnamefont {Hetzel}}, \bibinfo {author} {\bibfnamefont
  {F.}~\bibnamefont {Hinder}}, \bibinfo {author} {\bibfnamefont
  {A.}~\bibnamefont {Kacharava}}, \bibinfo {author} {\bibfnamefont
  {V.}~\bibnamefont {Kamerdzhiev}}, \bibinfo {author} {\bibfnamefont
  {I.}~\bibnamefont {Keshelashvili}}, \bibinfo {author} {\bibfnamefont
  {I.}~\bibnamefont {Koop}}, \bibinfo {author} {\bibfnamefont {A.}~\bibnamefont
  {Kulikov}}, \bibinfo {author} {\bibfnamefont {A.}~\bibnamefont {Lehrach}},
  \bibinfo {author} {\bibfnamefont {P.}~\bibnamefont {Lenisa}}, \bibinfo
  {author} {\bibfnamefont {N.}~\bibnamefont {Lomidze}}, \bibinfo {author}
  {\bibfnamefont {B.}~\bibnamefont {Lorentz}}, \bibinfo {author} {\bibfnamefont
  {P.}~\bibnamefont {Maanen}}, \bibinfo {author} {\bibfnamefont
  {G.}~\bibnamefont {Macharashvili}}, \bibinfo {author} {\bibfnamefont
  {A.}~\bibnamefont {Magiera}}, \bibinfo {author} {\bibfnamefont
  {D.}~\bibnamefont {McHedlishvili}}, \bibinfo {author} {\bibfnamefont
  {S.}~\bibnamefont {Mey}}, \bibinfo {author} {\bibfnamefont {F.}~\bibnamefont
  {M{\"u}ller}}, \bibinfo {author} {\bibfnamefont {A.}~\bibnamefont {Nass}},
  \bibinfo {author} {\bibfnamefont {A.}~\bibnamefont {Pesce}}, \bibinfo
  {author} {\bibfnamefont {D.}~\bibnamefont {Prasuhn}}, \bibinfo {author}
  {\bibfnamefont {J.}~\bibnamefont {Pretz}}, \bibinfo {author} {\bibfnamefont
  {M.}~\bibnamefont {Rosenthal}}, \bibinfo {author} {\bibfnamefont
  {V.}~\bibnamefont {Schmidt}}, \bibinfo {author} {\bibfnamefont
  {Y.}~\bibnamefont {Semertzidis}}, \bibinfo {author} {\bibfnamefont
  {Y.}~\bibnamefont {Senichev}}, \bibinfo {author} {\bibfnamefont
  {V.}~\bibnamefont {Shmakova}}, \bibinfo {author} {\bibfnamefont
  {A.}~\bibnamefont {Silenko}}, \bibinfo {author} {\bibfnamefont
  {J.}~\bibnamefont {Slim}}, \bibinfo {author} {\bibfnamefont {H.}~\bibnamefont
  {Soltner}}, \bibinfo {author} {\bibfnamefont {A.}~\bibnamefont {Stahl}},
  \bibinfo {author} {\bibfnamefont {R.}~\bibnamefont {Stassen}}, \bibinfo
  {author} {\bibfnamefont {E.}~\bibnamefont {Stephenson}}, \bibinfo {author}
  {\bibfnamefont {H.}~\bibnamefont {Stockhorst}}, \bibinfo {author}
  {\bibfnamefont {H.}~\bibnamefont {Str{\"o}her}}, \bibinfo {author}
  {\bibfnamefont {M.}~\bibnamefont {Tabidze}}, \bibinfo {author} {\bibfnamefont
  {G.}~\bibnamefont {Tagliente}}, \bibinfo {author} {\bibfnamefont
  {R.}~\bibnamefont {Talman}}, \bibinfo {author} {\bibfnamefont
  {P.}~\bibnamefont {Engblom}}, \bibinfo {author} {\bibfnamefont
  {F.}~\bibnamefont {Trinkel}}, \bibinfo {author} {\bibfnamefont
  {Y.}~\bibnamefont {Uzikov}}, \bibinfo {author} {\bibfnamefont
  {Y.}~\bibnamefont {Valdau}}, \bibinfo {author} {\bibfnamefont
  {E.}~\bibnamefont {Valetov}}, \bibinfo {author} {\bibfnamefont
  {A.}~\bibnamefont {Vassiliev}}, \bibinfo {author} {\bibfnamefont
  {C.}~\bibnamefont {Weidemann}}, \bibinfo {author} {\bibfnamefont
  {A.}~\bibnamefont {Wro{\'n}ska}}, \bibinfo {author} {\bibfnamefont
  {P.}~\bibnamefont {W{\"u}stner}}, \bibinfo {author} {\bibfnamefont
  {P.}~\bibnamefont {Zupra{\'n}ski}},\ and\ \bibinfo {author} {\bibfnamefont
  {M.}~\bibnamefont {Zurek}},\ }\bibfield  {title} {\bibinfo {title} {Spin tune
  mapping as a novel tool to probe the spin dynamics in storage rings},\
  }\bibfield  {journal} {\bibinfo  {journal} {Physical Review Accelerators and
  Beams}\ }\textbf {\bibinfo {volume} {20}},\ \href
  {https://doi.org/10.1103/PhysRevAccelBeams.20.072801}
  {10.1103/PhysRevAccelBeams.20.072801} (\bibinfo {year} {2017})\BibitemShut
  {NoStop}%
\bibitem [{\citenamefont {Maier}(1997)}]{Maier:1997zj}%
  \BibitemOpen
  \bibfield  {author} {\bibinfo {author} {\bibfnamefont {R.}~\bibnamefont
  {Maier}},\ }\bibfield  {title} {\bibinfo {title} {Cooler synchrotron {{COSY}}
  - {{Performance}} and perspectives},\ }\href
  {https://doi.org/10.1016/S0168-9002(97)00324-0} {\bibfield  {journal}
  {\bibinfo  {journal} {Nuclear Instruments and Methods in Physics Research,
  Section A: Accelerators, Spectrometers, Detectors and Associated Equipment}\
  }\textbf {\bibinfo {volume} {390}},\ \bibinfo {pages} {1} (\bibinfo {year}
  {1997})}\BibitemShut {NoStop}%
\bibitem [{\citenamefont {Weidemann}\ \emph {et~al.}(2015)\citenamefont
  {Weidemann}, \citenamefont {Rathmann}, \citenamefont {Stein}, \citenamefont
  {Lorentz}, \citenamefont {Bagdasarian}, \citenamefont {Barion}, \citenamefont
  {Barsov}, \citenamefont {Bechstedt}, \citenamefont {Bertelli}, \citenamefont
  {Chiladze}, \citenamefont {Ciullo}, \citenamefont {Contalbrigo},
  \citenamefont {Dymov}, \citenamefont {Engels}, \citenamefont {Gaisser},
  \citenamefont {Gebel}, \citenamefont {Goslawski}, \citenamefont {Grigoriev},
  \citenamefont {Guidoboni}, \citenamefont {Kacharava}, \citenamefont
  {Kamerdzhiev}, \citenamefont {Khoukaz}, \citenamefont {Kulikov},
  \citenamefont {Lehrach}, \citenamefont {Lenisa}, \citenamefont {Lomidze},
  \citenamefont {Macharashvili}, \citenamefont {Maier}, \citenamefont {Martin},
  \citenamefont {Mchedlishvili}, \citenamefont {Meyer}, \citenamefont
  {Merzliakov}, \citenamefont {Mielke}, \citenamefont {Mikirtychiants},
  \citenamefont {Mikirtychiants}, \citenamefont {Nass}, \citenamefont
  {Nikolaev}, \citenamefont {Oellers}, \citenamefont {Papenbrock},
  \citenamefont {Pesce}, \citenamefont {Prasuhn}, \citenamefont {Retzlaff},
  \citenamefont {Schleichert}, \citenamefont {Schr{\"o}er}, \citenamefont
  {Seyfarth}, \citenamefont {Soltner}, \citenamefont {Statera}, \citenamefont
  {Steffens}, \citenamefont {Stockhorst}, \citenamefont {Str{\"o}her},
  \citenamefont {Tabidze}, \citenamefont {Tagliente}, \citenamefont {Engblom},
  \citenamefont {Trusov}, \citenamefont {Valdau}, \citenamefont {Vasiliev},\
  and\ \citenamefont {W{\"u}stner}}]{Weidemann:2014qca}%
  \BibitemOpen
  \bibfield  {author} {\bibinfo {author} {\bibfnamefont {C.}~\bibnamefont
  {Weidemann}}, \bibinfo {author} {\bibfnamefont {F.}~\bibnamefont {Rathmann}},
  \bibinfo {author} {\bibfnamefont {H.~J.}\ \bibnamefont {Stein}}, \bibinfo
  {author} {\bibfnamefont {B.}~\bibnamefont {Lorentz}}, \bibinfo {author}
  {\bibfnamefont {Z.}~\bibnamefont {Bagdasarian}}, \bibinfo {author}
  {\bibfnamefont {L.}~\bibnamefont {Barion}}, \bibinfo {author} {\bibfnamefont
  {S.}~\bibnamefont {Barsov}}, \bibinfo {author} {\bibfnamefont
  {U.}~\bibnamefont {Bechstedt}}, \bibinfo {author} {\bibfnamefont
  {S.}~\bibnamefont {Bertelli}}, \bibinfo {author} {\bibfnamefont
  {D.}~\bibnamefont {Chiladze}}, \bibinfo {author} {\bibfnamefont
  {G.}~\bibnamefont {Ciullo}}, \bibinfo {author} {\bibfnamefont
  {M.}~\bibnamefont {Contalbrigo}}, \bibinfo {author} {\bibfnamefont
  {S.}~\bibnamefont {Dymov}}, \bibinfo {author} {\bibfnamefont
  {R.}~\bibnamefont {Engels}}, \bibinfo {author} {\bibfnamefont
  {M.}~\bibnamefont {Gaisser}}, \bibinfo {author} {\bibfnamefont
  {R.}~\bibnamefont {Gebel}}, \bibinfo {author} {\bibfnamefont
  {P.}~\bibnamefont {Goslawski}}, \bibinfo {author} {\bibfnamefont
  {K.}~\bibnamefont {Grigoriev}}, \bibinfo {author} {\bibfnamefont
  {G.}~\bibnamefont {Guidoboni}}, \bibinfo {author} {\bibfnamefont
  {A.}~\bibnamefont {Kacharava}}, \bibinfo {author} {\bibfnamefont
  {V.}~\bibnamefont {Kamerdzhiev}}, \bibinfo {author} {\bibfnamefont
  {A.}~\bibnamefont {Khoukaz}}, \bibinfo {author} {\bibfnamefont
  {A.}~\bibnamefont {Kulikov}}, \bibinfo {author} {\bibfnamefont
  {A.}~\bibnamefont {Lehrach}}, \bibinfo {author} {\bibfnamefont
  {P.}~\bibnamefont {Lenisa}}, \bibinfo {author} {\bibfnamefont
  {N.}~\bibnamefont {Lomidze}}, \bibinfo {author} {\bibfnamefont
  {G.}~\bibnamefont {Macharashvili}}, \bibinfo {author} {\bibfnamefont
  {R.}~\bibnamefont {Maier}}, \bibinfo {author} {\bibfnamefont
  {S.}~\bibnamefont {Martin}}, \bibinfo {author} {\bibfnamefont
  {D.}~\bibnamefont {Mchedlishvili}}, \bibinfo {author} {\bibfnamefont {H.~O.}\
  \bibnamefont {Meyer}}, \bibinfo {author} {\bibfnamefont {S.}~\bibnamefont
  {Merzliakov}}, \bibinfo {author} {\bibfnamefont {M.}~\bibnamefont {Mielke}},
  \bibinfo {author} {\bibfnamefont {M.}~\bibnamefont {Mikirtychiants}},
  \bibinfo {author} {\bibfnamefont {S.}~\bibnamefont {Mikirtychiants}},
  \bibinfo {author} {\bibfnamefont {A.}~\bibnamefont {Nass}}, \bibinfo {author}
  {\bibfnamefont {N.~N.}\ \bibnamefont {Nikolaev}}, \bibinfo {author}
  {\bibfnamefont {D.}~\bibnamefont {Oellers}}, \bibinfo {author} {\bibfnamefont
  {M.}~\bibnamefont {Papenbrock}}, \bibinfo {author} {\bibfnamefont
  {A.}~\bibnamefont {Pesce}}, \bibinfo {author} {\bibfnamefont
  {D.}~\bibnamefont {Prasuhn}}, \bibinfo {author} {\bibfnamefont
  {M.}~\bibnamefont {Retzlaff}}, \bibinfo {author} {\bibfnamefont
  {R.}~\bibnamefont {Schleichert}}, \bibinfo {author} {\bibfnamefont
  {D.}~\bibnamefont {Schr{\"o}er}}, \bibinfo {author} {\bibfnamefont
  {H.}~\bibnamefont {Seyfarth}}, \bibinfo {author} {\bibfnamefont
  {H.}~\bibnamefont {Soltner}}, \bibinfo {author} {\bibfnamefont
  {M.}~\bibnamefont {Statera}}, \bibinfo {author} {\bibfnamefont
  {E.}~\bibnamefont {Steffens}}, \bibinfo {author} {\bibfnamefont
  {H.}~\bibnamefont {Stockhorst}}, \bibinfo {author} {\bibfnamefont
  {H.}~\bibnamefont {Str{\"o}her}}, \bibinfo {author} {\bibfnamefont
  {M.}~\bibnamefont {Tabidze}}, \bibinfo {author} {\bibfnamefont
  {G.}~\bibnamefont {Tagliente}}, \bibinfo {author} {\bibfnamefont {P.~T.}\
  \bibnamefont {Engblom}}, \bibinfo {author} {\bibfnamefont {S.}~\bibnamefont
  {Trusov}}, \bibinfo {author} {\bibfnamefont {{\relax Yu}.}~\bibnamefont
  {Valdau}}, \bibinfo {author} {\bibfnamefont {A.}~\bibnamefont {Vasiliev}},\
  and\ \bibinfo {author} {\bibfnamefont {P.}~\bibnamefont {W{\"u}stner}},\
  }\bibfield  {title} {\bibinfo {title} {Toward polarized antiprotons:
  {{Machine}} development for spin-filtering experiments},\ }\href
  {https://doi.org/10.1103/PhysRevSTAB.18.020101} {\bibfield  {journal}
  {\bibinfo  {journal} {Physical Review Special Topics - Accelerators and
  Beams}\ }\textbf {\bibinfo {volume} {18}},\ \bibinfo {pages} {020101}
  (\bibinfo {year} {2015})}\BibitemShut {NoStop}%
\bibitem [{\citenamefont {Guidoboni}\ \emph {et~al.}(2016)\citenamefont
  {Guidoboni}, \citenamefont {Stephenson}, \citenamefont {Andrianov},
  \citenamefont {Augustyniak}, \citenamefont {Bagdasarian}, \citenamefont
  {Bai}, \citenamefont {Baylac}, \citenamefont {Bernreuther}, \citenamefont
  {Bertelli}, \citenamefont {Berz}, \citenamefont {B{\"o}ker}, \citenamefont
  {B{\"o}hme}, \citenamefont {Bsaisou}, \citenamefont {Chekmenev},
  \citenamefont {Chiladze}, \citenamefont {Ciullo}, \citenamefont
  {Contalbrigo}, \citenamefont {{de Conto}}, \citenamefont {Dymov},
  \citenamefont {Engels}, \citenamefont {Esser}, \citenamefont {Eversmann},
  \citenamefont {Felden}, \citenamefont {Gaisser}, \citenamefont {Gebel},
  \citenamefont {Gl{\"u}ckler}, \citenamefont {Goldenbaum}, \citenamefont
  {Grigoryev}, \citenamefont {Grzonka}, \citenamefont {Hahnraths},
  \citenamefont {Heberling}, \citenamefont {Hejny}, \citenamefont {Hempelmann},
  \citenamefont {Hetzel}, \citenamefont {Hinder}, \citenamefont {Hipple},
  \citenamefont {H{\"o}lscher}, \citenamefont {Ivanov}, \citenamefont
  {Kacharava}, \citenamefont {Kamerdzhiev}, \citenamefont {Kamys},
  \citenamefont {Keshelashvili}, \citenamefont {Khoukaz}, \citenamefont {Koop},
  \citenamefont {Krause}, \citenamefont {Krewald}, \citenamefont {Kulikov},
  \citenamefont {Lehrach}, \citenamefont {Lenisa}, \citenamefont {Lomidze},
  \citenamefont {Lorentz}, \citenamefont {Maanen}, \citenamefont
  {Macharashvili}, \citenamefont {Magiera}, \citenamefont {Maier},
  \citenamefont {Makino}, \citenamefont {Maria{\'n}ski}, \citenamefont
  {Mchedlishvili}, \citenamefont {Mei{\ss}ner}, \citenamefont {Mey},
  \citenamefont {Morse}, \citenamefont {M{\"u}ller}, \citenamefont {Nass},
  \citenamefont {Natour}, \citenamefont {Nikolaev}, \citenamefont {Nioradze},
  \citenamefont {Nowakowski}, \citenamefont {Orlov}, \citenamefont {Pesce},
  \citenamefont {Prasuhn}, \citenamefont {Pretz}, \citenamefont {Rathmann},
  \citenamefont {Ritman}, \citenamefont {Rosenthal}, \citenamefont {Rudy},
  \citenamefont {Saleev}, \citenamefont {Sefzick}, \citenamefont {Semertzidis},
  \citenamefont {Senichev}, \citenamefont {Shmakova}, \citenamefont {Silenko},
  \citenamefont {Simon}, \citenamefont {Slim}, \citenamefont {Soltner},
  \citenamefont {Stahl}, \citenamefont {Stassen}, \citenamefont {Statera},
  \citenamefont {Stockhorst}, \citenamefont {Straatmann}, \citenamefont
  {Str{\"o}her}, \citenamefont {Tabidze}, \citenamefont {Talman}, \citenamefont
  {Th{\"o}rngren~Engblom}, \citenamefont {Trinkel}, \citenamefont
  {Trzci{\'n}ski}, \citenamefont {Uzikov}, \citenamefont {Valdau},
  \citenamefont {Valetov}, \citenamefont {Vassiliev}, \citenamefont
  {Weidemann}, \citenamefont {Wilkin}, \citenamefont {Wro{\'n}ska},
  \citenamefont {W{\"u}stner}, \citenamefont {Zakrzewska}, \citenamefont
  {Zupra{\'n}ski},\ and\ \citenamefont {Zyuzin}}]{JEDI:2016swi}%
  \BibitemOpen
  \bibfield  {author} {\bibinfo {author} {\bibfnamefont {G.}~\bibnamefont
  {Guidoboni}}, \bibinfo {author} {\bibfnamefont {E.}~\bibnamefont
  {Stephenson}}, \bibinfo {author} {\bibfnamefont {S.}~\bibnamefont
  {Andrianov}}, \bibinfo {author} {\bibfnamefont {W.}~\bibnamefont
  {Augustyniak}}, \bibinfo {author} {\bibfnamefont {Z.}~\bibnamefont
  {Bagdasarian}}, \bibinfo {author} {\bibfnamefont {M.}~\bibnamefont {Bai}},
  \bibinfo {author} {\bibfnamefont {M.}~\bibnamefont {Baylac}}, \bibinfo
  {author} {\bibfnamefont {W.}~\bibnamefont {Bernreuther}}, \bibinfo {author}
  {\bibfnamefont {S.}~\bibnamefont {Bertelli}}, \bibinfo {author}
  {\bibfnamefont {M.}~\bibnamefont {Berz}}, \bibinfo {author} {\bibfnamefont
  {J.}~\bibnamefont {B{\"o}ker}}, \bibinfo {author} {\bibfnamefont
  {C.}~\bibnamefont {B{\"o}hme}}, \bibinfo {author} {\bibfnamefont
  {J.}~\bibnamefont {Bsaisou}}, \bibinfo {author} {\bibfnamefont
  {S.}~\bibnamefont {Chekmenev}}, \bibinfo {author} {\bibfnamefont
  {D.}~\bibnamefont {Chiladze}}, \bibinfo {author} {\bibfnamefont
  {G.}~\bibnamefont {Ciullo}}, \bibinfo {author} {\bibfnamefont
  {M.}~\bibnamefont {Contalbrigo}}, \bibinfo {author} {\bibfnamefont {J.-M.}\
  \bibnamefont {{de Conto}}}, \bibinfo {author} {\bibfnamefont
  {S.}~\bibnamefont {Dymov}}, \bibinfo {author} {\bibfnamefont
  {R.}~\bibnamefont {Engels}}, \bibinfo {author} {\bibfnamefont {F.~M.}\
  \bibnamefont {Esser}}, \bibinfo {author} {\bibfnamefont {D.}~\bibnamefont
  {Eversmann}}, \bibinfo {author} {\bibfnamefont {O.}~\bibnamefont {Felden}},
  \bibinfo {author} {\bibfnamefont {M.}~\bibnamefont {Gaisser}}, \bibinfo
  {author} {\bibfnamefont {R.}~\bibnamefont {Gebel}}, \bibinfo {author}
  {\bibfnamefont {H.}~\bibnamefont {Gl{\"u}ckler}}, \bibinfo {author}
  {\bibfnamefont {F.}~\bibnamefont {Goldenbaum}}, \bibinfo {author}
  {\bibfnamefont {K.}~\bibnamefont {Grigoryev}}, \bibinfo {author}
  {\bibfnamefont {D.}~\bibnamefont {Grzonka}}, \bibinfo {author} {\bibfnamefont
  {T.}~\bibnamefont {Hahnraths}}, \bibinfo {author} {\bibfnamefont
  {D.}~\bibnamefont {Heberling}}, \bibinfo {author} {\bibfnamefont
  {V.}~\bibnamefont {Hejny}}, \bibinfo {author} {\bibfnamefont
  {N.}~\bibnamefont {Hempelmann}}, \bibinfo {author} {\bibfnamefont
  {J.}~\bibnamefont {Hetzel}}, \bibinfo {author} {\bibfnamefont
  {F.}~\bibnamefont {Hinder}}, \bibinfo {author} {\bibfnamefont
  {R.}~\bibnamefont {Hipple}}, \bibinfo {author} {\bibfnamefont
  {D.}~\bibnamefont {H{\"o}lscher}}, \bibinfo {author} {\bibfnamefont
  {A.}~\bibnamefont {Ivanov}}, \bibinfo {author} {\bibfnamefont
  {A.}~\bibnamefont {Kacharava}}, \bibinfo {author} {\bibfnamefont
  {V.}~\bibnamefont {Kamerdzhiev}}, \bibinfo {author} {\bibfnamefont
  {B.}~\bibnamefont {Kamys}}, \bibinfo {author} {\bibfnamefont
  {I.}~\bibnamefont {Keshelashvili}}, \bibinfo {author} {\bibfnamefont
  {A.}~\bibnamefont {Khoukaz}}, \bibinfo {author} {\bibfnamefont
  {I.}~\bibnamefont {Koop}}, \bibinfo {author} {\bibfnamefont {H.-J.}\
  \bibnamefont {Krause}}, \bibinfo {author} {\bibfnamefont {S.}~\bibnamefont
  {Krewald}}, \bibinfo {author} {\bibfnamefont {A.}~\bibnamefont {Kulikov}},
  \bibinfo {author} {\bibfnamefont {A.}~\bibnamefont {Lehrach}}, \bibinfo
  {author} {\bibfnamefont {P.}~\bibnamefont {Lenisa}}, \bibinfo {author}
  {\bibfnamefont {N.}~\bibnamefont {Lomidze}}, \bibinfo {author} {\bibfnamefont
  {B.}~\bibnamefont {Lorentz}}, \bibinfo {author} {\bibfnamefont
  {P.}~\bibnamefont {Maanen}}, \bibinfo {author} {\bibfnamefont
  {G.}~\bibnamefont {Macharashvili}}, \bibinfo {author} {\bibfnamefont
  {A.}~\bibnamefont {Magiera}}, \bibinfo {author} {\bibfnamefont
  {R.}~\bibnamefont {Maier}}, \bibinfo {author} {\bibfnamefont
  {K.}~\bibnamefont {Makino}}, \bibinfo {author} {\bibfnamefont
  {B.}~\bibnamefont {Maria{\'n}ski}}, \bibinfo {author} {\bibfnamefont
  {D.}~\bibnamefont {Mchedlishvili}}, \bibinfo {author} {\bibfnamefont {U.-G.}\
  \bibnamefont {Mei{\ss}ner}}, \bibinfo {author} {\bibfnamefont
  {S.}~\bibnamefont {Mey}}, \bibinfo {author} {\bibfnamefont {W.}~\bibnamefont
  {Morse}}, \bibinfo {author} {\bibfnamefont {F.}~\bibnamefont {M{\"u}ller}},
  \bibinfo {author} {\bibfnamefont {A.}~\bibnamefont {Nass}}, \bibinfo {author}
  {\bibfnamefont {G.}~\bibnamefont {Natour}}, \bibinfo {author} {\bibfnamefont
  {N.}~\bibnamefont {Nikolaev}}, \bibinfo {author} {\bibfnamefont
  {M.}~\bibnamefont {Nioradze}}, \bibinfo {author} {\bibfnamefont
  {K.}~\bibnamefont {Nowakowski}}, \bibinfo {author} {\bibfnamefont
  {Y.}~\bibnamefont {Orlov}}, \bibinfo {author} {\bibfnamefont
  {A.}~\bibnamefont {Pesce}}, \bibinfo {author} {\bibfnamefont
  {D.}~\bibnamefont {Prasuhn}}, \bibinfo {author} {\bibfnamefont
  {J.}~\bibnamefont {Pretz}}, \bibinfo {author} {\bibfnamefont
  {F.}~\bibnamefont {Rathmann}}, \bibinfo {author} {\bibfnamefont
  {J.}~\bibnamefont {Ritman}}, \bibinfo {author} {\bibfnamefont
  {M.}~\bibnamefont {Rosenthal}}, \bibinfo {author} {\bibfnamefont
  {Z.}~\bibnamefont {Rudy}}, \bibinfo {author} {\bibfnamefont {A.}~\bibnamefont
  {Saleev}}, \bibinfo {author} {\bibfnamefont {T.}~\bibnamefont {Sefzick}},
  \bibinfo {author} {\bibfnamefont {Y.}~\bibnamefont {Semertzidis}}, \bibinfo
  {author} {\bibfnamefont {Y.}~\bibnamefont {Senichev}}, \bibinfo {author}
  {\bibfnamefont {V.}~\bibnamefont {Shmakova}}, \bibinfo {author}
  {\bibfnamefont {A.}~\bibnamefont {Silenko}}, \bibinfo {author} {\bibfnamefont
  {M.}~\bibnamefont {Simon}}, \bibinfo {author} {\bibfnamefont
  {J.}~\bibnamefont {Slim}}, \bibinfo {author} {\bibfnamefont {H.}~\bibnamefont
  {Soltner}}, \bibinfo {author} {\bibfnamefont {A.}~\bibnamefont {Stahl}},
  \bibinfo {author} {\bibfnamefont {R.}~\bibnamefont {Stassen}}, \bibinfo
  {author} {\bibfnamefont {M.}~\bibnamefont {Statera}}, \bibinfo {author}
  {\bibfnamefont {H.}~\bibnamefont {Stockhorst}}, \bibinfo {author}
  {\bibfnamefont {H.}~\bibnamefont {Straatmann}}, \bibinfo {author}
  {\bibfnamefont {H.}~\bibnamefont {Str{\"o}her}}, \bibinfo {author}
  {\bibfnamefont {M.}~\bibnamefont {Tabidze}}, \bibinfo {author} {\bibfnamefont
  {R.}~\bibnamefont {Talman}}, \bibinfo {author} {\bibfnamefont
  {P.}~\bibnamefont {Th{\"o}rngren~Engblom}}, \bibinfo {author} {\bibfnamefont
  {F.}~\bibnamefont {Trinkel}}, \bibinfo {author} {\bibfnamefont
  {A.}~\bibnamefont {Trzci{\'n}ski}}, \bibinfo {author} {\bibfnamefont {{\relax
  Yu}.}~\bibnamefont {Uzikov}}, \bibinfo {author} {\bibfnamefont {{\relax
  Yu}.}~\bibnamefont {Valdau}}, \bibinfo {author} {\bibfnamefont
  {E.}~\bibnamefont {Valetov}}, \bibinfo {author} {\bibfnamefont
  {A.}~\bibnamefont {Vassiliev}}, \bibinfo {author} {\bibfnamefont
  {C.}~\bibnamefont {Weidemann}}, \bibinfo {author} {\bibfnamefont
  {C.}~\bibnamefont {Wilkin}}, \bibinfo {author} {\bibfnamefont
  {A.}~\bibnamefont {Wro{\'n}ska}}, \bibinfo {author} {\bibfnamefont
  {P.}~\bibnamefont {W{\"u}stner}}, \bibinfo {author} {\bibfnamefont
  {M.}~\bibnamefont {Zakrzewska}}, \bibinfo {author} {\bibfnamefont
  {P.}~\bibnamefont {Zupra{\'n}ski}},\ and\ \bibinfo {author} {\bibfnamefont
  {D.}~\bibnamefont {Zyuzin}},\ }\bibfield  {title} {\bibinfo {title} {How to
  {{Reach}} a {{Thousand-Second}} in-{{Plane Polarization Lifetime}} with
  0.97-{{GeV}}/c {{Deuterons}} in a {{Storage Ring}}},\ }\href
  {https://doi.org/10.1103/PhysRevLett.117.054801} {\bibfield  {journal}
  {\bibinfo  {journal} {Physical Review Letters}\ }\textbf {\bibinfo {volume}
  {117}},\ \bibinfo {pages} {054801} (\bibinfo {year} {2016})}\BibitemShut
  {NoStop}%
\bibitem [{\citenamefont {Hoistad}\ \emph {et~al.}(2004)\citenamefont
  {Hoistad}, \citenamefont {Ritman} \emph {et~al.}}]{WASA-at-COSY:2004mns}%
  \BibitemOpen
  \bibfield  {author} {\bibinfo {author} {\bibfnamefont {B.}~\bibnamefont
  {Hoistad}}, \bibinfo {author} {\bibfnamefont {J.}~\bibnamefont {Ritman}},
  \emph {et~al.},\ }\href {https://doi.org/10.48550/arXiv.nucl-ex/0411038}
  {\bibinfo {title} {Proposal for the {{Wide Angle Shower Apparatus}}
  ({{WASA}}) at {{COSY-Juelich}} - "{{WASA}} at {{COSY}}"}} (\bibinfo {year}
  {2004}),\ \Eprint {https://arxiv.org/abs/nucl-ex/0411038}
  {arxiv:nucl-ex/0411038} \BibitemShut {NoStop}%
\bibitem [{\citenamefont {M{\"u}ller}\ \emph {et~al.}(2020)\citenamefont
  {M{\"u}ller}, \citenamefont {Javakhishvili}, \citenamefont {Shergelashvili},
  \citenamefont {Keshelashvili}, \citenamefont {Mchedlishvili}, \citenamefont
  {Abusaif}, \citenamefont {Aggarwal}, \citenamefont {Barion}, \citenamefont
  {Basile}, \citenamefont {B{\"o}ker}, \citenamefont {Canale}, \citenamefont
  {Ciullo}, \citenamefont {Dymov}, \citenamefont {Felden}, \citenamefont
  {Gagoshidze}, \citenamefont {Gebel}, \citenamefont {Demary}, \citenamefont
  {Grigoryev}, \citenamefont {Grzonka}, \citenamefont {Hahnraths},
  \citenamefont {Hejny}, \citenamefont {Kacharava}, \citenamefont
  {Kamerdzhiev}, \citenamefont {Karanth}, \citenamefont {Kulikov},
  \citenamefont {Lehrach}, \citenamefont {Lenisa}, \citenamefont {Lomidze},
  \citenamefont {Lorentz}, \citenamefont {Macharashvili}, \citenamefont
  {Magiera}, \citenamefont {Metreveli}, \citenamefont {Nass}, \citenamefont
  {Nikolaev}, \citenamefont {Nioradze}, \citenamefont {Pesce}, \citenamefont
  {Poncza}, \citenamefont {Prasuhn}, \citenamefont {Pretz}, \citenamefont
  {Rathmann}, \citenamefont {Saleev}, \citenamefont {Sefzick}, \citenamefont
  {Senichev}, \citenamefont {Shmakova}, \citenamefont {Slim}, \citenamefont
  {Soltner}, \citenamefont {Stephenson}, \citenamefont {Str{\"o}her},
  \citenamefont {Tabidze}, \citenamefont {Tagliente}, \citenamefont {Uzikov},
  \citenamefont {Valdau}, \citenamefont {Wagner}, \citenamefont {Wro{\'n}ska},
  \citenamefont {W{\"u}stner},\ and\ \citenamefont
  {{\.Z}urek}}]{JEDI:2020fzda}%
  \BibitemOpen
  \bibfield  {author} {\bibinfo {author} {\bibfnamefont {F.}~\bibnamefont
  {M{\"u}ller}}, \bibinfo {author} {\bibfnamefont {O.}~\bibnamefont
  {Javakhishvili}}, \bibinfo {author} {\bibfnamefont {D.}~\bibnamefont
  {Shergelashvili}}, \bibinfo {author} {\bibfnamefont {I.}~\bibnamefont
  {Keshelashvili}}, \bibinfo {author} {\bibfnamefont {D.}~\bibnamefont
  {Mchedlishvili}}, \bibinfo {author} {\bibfnamefont {F.}~\bibnamefont
  {Abusaif}}, \bibinfo {author} {\bibfnamefont {A.}~\bibnamefont {Aggarwal}},
  \bibinfo {author} {\bibfnamefont {L.}~\bibnamefont {Barion}}, \bibinfo
  {author} {\bibfnamefont {S.}~\bibnamefont {Basile}}, \bibinfo {author}
  {\bibfnamefont {J.}~\bibnamefont {B{\"o}ker}}, \bibinfo {author}
  {\bibfnamefont {N.}~\bibnamefont {Canale}}, \bibinfo {author} {\bibfnamefont
  {G.}~\bibnamefont {Ciullo}}, \bibinfo {author} {\bibfnamefont
  {S.}~\bibnamefont {Dymov}}, \bibinfo {author} {\bibfnamefont
  {O.}~\bibnamefont {Felden}}, \bibinfo {author} {\bibfnamefont
  {M.}~\bibnamefont {Gagoshidze}}, \bibinfo {author} {\bibfnamefont
  {R.}~\bibnamefont {Gebel}}, \bibinfo {author} {\bibfnamefont
  {N.}~\bibnamefont {Demary}}, \bibinfo {author} {\bibfnamefont
  {K.}~\bibnamefont {Grigoryev}}, \bibinfo {author} {\bibfnamefont
  {D.}~\bibnamefont {Grzonka}}, \bibinfo {author} {\bibfnamefont
  {T.}~\bibnamefont {Hahnraths}}, \bibinfo {author} {\bibfnamefont
  {V.}~\bibnamefont {Hejny}}, \bibinfo {author} {\bibfnamefont
  {A.}~\bibnamefont {Kacharava}}, \bibinfo {author} {\bibfnamefont
  {V.}~\bibnamefont {Kamerdzhiev}}, \bibinfo {author} {\bibfnamefont
  {S.}~\bibnamefont {Karanth}}, \bibinfo {author} {\bibfnamefont
  {A.}~\bibnamefont {Kulikov}}, \bibinfo {author} {\bibfnamefont
  {A.}~\bibnamefont {Lehrach}}, \bibinfo {author} {\bibfnamefont
  {P.}~\bibnamefont {Lenisa}}, \bibinfo {author} {\bibfnamefont
  {N.}~\bibnamefont {Lomidze}}, \bibinfo {author} {\bibfnamefont
  {B.}~\bibnamefont {Lorentz}}, \bibinfo {author} {\bibfnamefont
  {G.}~\bibnamefont {Macharashvili}}, \bibinfo {author} {\bibfnamefont
  {A.}~\bibnamefont {Magiera}}, \bibinfo {author} {\bibfnamefont
  {Z.}~\bibnamefont {Metreveli}}, \bibinfo {author} {\bibfnamefont
  {A.}~\bibnamefont {Nass}}, \bibinfo {author} {\bibfnamefont {N.~N.}\
  \bibnamefont {Nikolaev}}, \bibinfo {author} {\bibfnamefont {M.}~\bibnamefont
  {Nioradze}}, \bibinfo {author} {\bibfnamefont {A.}~\bibnamefont {Pesce}},
  \bibinfo {author} {\bibfnamefont {V.}~\bibnamefont {Poncza}}, \bibinfo
  {author} {\bibfnamefont {D.}~\bibnamefont {Prasuhn}}, \bibinfo {author}
  {\bibfnamefont {J.}~\bibnamefont {Pretz}}, \bibinfo {author} {\bibfnamefont
  {F.}~\bibnamefont {Rathmann}}, \bibinfo {author} {\bibfnamefont
  {A.}~\bibnamefont {Saleev}}, \bibinfo {author} {\bibfnamefont
  {T.}~\bibnamefont {Sefzick}}, \bibinfo {author} {\bibfnamefont
  {Y.}~\bibnamefont {Senichev}}, \bibinfo {author} {\bibfnamefont
  {V.}~\bibnamefont {Shmakova}}, \bibinfo {author} {\bibfnamefont
  {J.}~\bibnamefont {Slim}}, \bibinfo {author} {\bibfnamefont {H.}~\bibnamefont
  {Soltner}}, \bibinfo {author} {\bibfnamefont {E.}~\bibnamefont {Stephenson}},
  \bibinfo {author} {\bibfnamefont {H.}~\bibnamefont {Str{\"o}her}}, \bibinfo
  {author} {\bibfnamefont {M.}~\bibnamefont {Tabidze}}, \bibinfo {author}
  {\bibfnamefont {G.}~\bibnamefont {Tagliente}}, \bibinfo {author}
  {\bibfnamefont {Y.}~\bibnamefont {Uzikov}}, \bibinfo {author} {\bibfnamefont
  {Y.}~\bibnamefont {Valdau}}, \bibinfo {author} {\bibfnamefont
  {T.}~\bibnamefont {Wagner}}, \bibinfo {author} {\bibfnamefont
  {A.}~\bibnamefont {Wro{\'n}ska}}, \bibinfo {author} {\bibfnamefont
  {P.}~\bibnamefont {W{\"u}stner}},\ and\ \bibinfo {author} {\bibfnamefont
  {M.}~\bibnamefont {{\.Z}urek}},\ }\bibfield  {title} {\bibinfo {title} {A new
  beam polarimeter at {{COSY}} to search for electric dipole moments of charged
  particles},\ }\href {https://doi.org/10.1088/1748-0221/15/12/P12005}
  {\bibfield  {journal} {\bibinfo  {journal} {Journal of Instrumentation}\
  }\textbf {\bibinfo {volume} {15}}\bibinfo  {number} { (12)},\ \bibinfo
  {pages} {P12005}}\BibitemShut {NoStop}%
\bibitem [{\citenamefont {Kleines}\ \emph {et~al.}(2006)\citenamefont
  {Kleines}, \citenamefont {Zwoll}, \citenamefont {W\"ustner}, \citenamefont
  {Erven}, \citenamefont {Kammerling}, \citenamefont {Kemmerling},
  \citenamefont {Loevenich}, \citenamefont {Ackens}, \citenamefont {Wolke},
  \citenamefont {Hejny}, \citenamefont {Ohm}, \citenamefont {Sefzick},
  \citenamefont {Nellen}, \citenamefont {Marciniewski}, \citenamefont
  {Fransson}, \citenamefont {Gustafsson}, \citenamefont {Kupsc},\ and\
  \citenamefont {Calen}}]{Kleines:2006cy}%
  \BibitemOpen
\bibfield  {number} {  }\bibfield  {author} {\bibinfo {author} {\bibfnamefont
  {H.}~\bibnamefont {Kleines}}, \bibinfo {author} {\bibfnamefont
  {K.}~\bibnamefont {Zwoll}}, \bibinfo {author} {\bibfnamefont
  {P.}~\bibnamefont {W\"ustner}}, \bibinfo {author} {\bibfnamefont
  {W.}~\bibnamefont {Erven}}, \bibinfo {author} {\bibfnamefont
  {P.}~\bibnamefont {Kammerling}}, \bibinfo {author} {\bibfnamefont
  {G.}~\bibnamefont {Kemmerling}}, \bibinfo {author} {\bibfnamefont {H.-W.}\
  \bibnamefont {Loevenich}}, \bibinfo {author} {\bibfnamefont {A.}~\bibnamefont
  {Ackens}}, \bibinfo {author} {\bibfnamefont {M.}~\bibnamefont {Wolke}},
  \bibinfo {author} {\bibfnamefont {V.}~\bibnamefont {Hejny}}, \bibinfo
  {author} {\bibfnamefont {H.}~\bibnamefont {Ohm}}, \bibinfo {author}
  {\bibfnamefont {T.}~\bibnamefont {Sefzick}}, \bibinfo {author} {\bibfnamefont
  {R.}~\bibnamefont {Nellen}}, \bibinfo {author} {\bibfnamefont
  {P.}~\bibnamefont {Marciniewski}}, \bibinfo {author} {\bibfnamefont
  {K.}~\bibnamefont {Fransson}}, \bibinfo {author} {\bibfnamefont
  {L.}~\bibnamefont {Gustafsson}}, \bibinfo {author} {\bibfnamefont
  {A.}~\bibnamefont {Kupsc}},\ and\ \bibinfo {author} {\bibfnamefont
  {H.}~\bibnamefont {Calen}},\ }\bibfield  {title} {\bibinfo {title} {The new
  {{DAQ}} system for {{WASA}} at {{COSY}}},\ }\href
  {https://doi.org/10.1109/TNS.2006.873305} {\bibfield  {journal} {\bibinfo
  {journal} {IEEE Transactions on Nuclear Science}\ }\textbf {\bibinfo {volume}
  {53}},\ \bibinfo {pages} {893} (\bibinfo {year} {2006})}\BibitemShut
  {NoStop}%
\bibitem [{\citenamefont {Light}(2017)}]{light2017}%
  \BibitemOpen
  \bibfield  {author} {\bibinfo {author} {\bibfnamefont {R.~A.}\ \bibnamefont
  {Light}},\ }\bibfield  {title} {\bibinfo {title} {Mosquitto: Server and
  client implementation of the {{MQTT}} protocol},\ }\href
  {https://doi.org/10.21105/joss.00265} {\bibfield  {journal} {\bibinfo
  {journal} {Journal of Open Source Software}\ }\textbf {\bibinfo {volume}
  {2}},\ \bibinfo {pages} {265} (\bibinfo {year} {2017})}\BibitemShut {NoStop}%
\bibitem [{\citenamefont {Dalesio}\ \emph {et~al.}(1992)\citenamefont
  {Dalesio}, \citenamefont {Kozubal},\ and\ \citenamefont
  {Kraimer}}]{Dalesio:1992fso}%
  \BibitemOpen
  \bibfield  {author} {\bibinfo {author} {\bibfnamefont {L.}~\bibnamefont
  {Dalesio}}, \bibinfo {author} {\bibfnamefont {A.}~\bibnamefont {Kozubal}},\
  and\ \bibinfo {author} {\bibfnamefont {M.}~\bibnamefont {Kraimer}},\
  }\bibfield  {title} {\bibinfo {title} {{{EPICS Architecture}}},\ }in\ \href
  {https://doi.org/10.18429/JACoW-ICALEPCS1991-S07IC03} {\emph {\bibinfo
  {booktitle} {3rd {{International Conference}} on {{Accelerator}} and {{Large
  Experimental Physics Control Systems}} ({{ICALEPCS}}'91), {{Tsukuba}},
  {{Japan}}, 11-15 {{November}} 1991}}}\ (\bibinfo  {publisher} {JACOW
  Publishing, Geneva, Switzerland},\ \bibinfo {year} {1992})\ pp.\ \bibinfo
  {pages} {278--282}\BibitemShut {NoStop}%
\bibitem [{\citenamefont {Nikolaev}\ \emph {et~al.}(2024)\citenamefont
  {Nikolaev}, \citenamefont {Rathmann}, \citenamefont {Slim}, \citenamefont
  {Andres}, \citenamefont {Hejny}, \citenamefont {Nass}, \citenamefont
  {Kacharava}, \citenamefont {Lenisa}, \citenamefont {Pretz}, \citenamefont
  {Saleev}, \citenamefont {Shmakova}, \citenamefont {Soltner}, \citenamefont
  {Abusaif}, \citenamefont {Aggarwal}, \citenamefont {Aksentev}, \citenamefont
  {Alberdi}, \citenamefont {Barion}, \citenamefont {Bekman}, \citenamefont
  {Bey{\ss}}, \citenamefont {B{\"o}hme}, \citenamefont {Breitkreutz},
  \citenamefont {Canale}, \citenamefont {Ciullo}, \citenamefont {Dymov},
  \citenamefont {Fr{\"o}hlich}, \citenamefont {Gebel}, \citenamefont {Gaisser},
  \citenamefont {Grigoryev}, \citenamefont {Grzonka}, \citenamefont {Hetzel},
  \citenamefont {Javakhishvili}, \citenamefont {Kamerdzhiev}, \citenamefont
  {Karanth}, \citenamefont {Keshelashvili}, \citenamefont {Kononov},
  \citenamefont {Laihem}, \citenamefont {Lehrach}, \citenamefont {Lomidze},
  \citenamefont {Lorentz}, \citenamefont {Macharashvili}, \citenamefont
  {Magiera}, \citenamefont {Mchedlishvili}, \citenamefont {Melnikov},
  \citenamefont {M{\"u}ller}, \citenamefont {Pesce}, \citenamefont {Poncza},
  \citenamefont {Prasuhn}, \citenamefont {Shergelashvili}, \citenamefont
  {Shurkhno}, \citenamefont {Siddique}, \citenamefont {Silenko}, \citenamefont
  {Stassen}, \citenamefont {Stephenson}, \citenamefont {Str{\"o}her},
  \citenamefont {Tabidze}, \citenamefont {Tagliente}, \citenamefont {Valdau},
  \citenamefont {Vitz}, \citenamefont {Wagner}, \citenamefont {Wirzba},
  \citenamefont {Wro{\'n}ska}, \citenamefont {W{\"u}stner},\ and\ \citenamefont
  {{\.Z}urek}}]{JEDI:2023trd}%
  \BibitemOpen
  \bibfield  {author} {\bibinfo {author} {\bibfnamefont {N.~N.}\ \bibnamefont
  {Nikolaev}}, \bibinfo {author} {\bibfnamefont {F.}~\bibnamefont {Rathmann}},
  \bibinfo {author} {\bibfnamefont {J.}~\bibnamefont {Slim}}, \bibinfo {author}
  {\bibfnamefont {A.}~\bibnamefont {Andres}}, \bibinfo {author} {\bibfnamefont
  {V.}~\bibnamefont {Hejny}}, \bibinfo {author} {\bibfnamefont
  {A.}~\bibnamefont {Nass}}, \bibinfo {author} {\bibfnamefont {A.}~\bibnamefont
  {Kacharava}}, \bibinfo {author} {\bibfnamefont {P.}~\bibnamefont {Lenisa}},
  \bibinfo {author} {\bibfnamefont {J.}~\bibnamefont {Pretz}}, \bibinfo
  {author} {\bibfnamefont {A.}~\bibnamefont {Saleev}}, \bibinfo {author}
  {\bibfnamefont {V.}~\bibnamefont {Shmakova}}, \bibinfo {author}
  {\bibfnamefont {H.}~\bibnamefont {Soltner}}, \bibinfo {author} {\bibfnamefont
  {F.}~\bibnamefont {Abusaif}}, \bibinfo {author} {\bibfnamefont
  {A.}~\bibnamefont {Aggarwal}}, \bibinfo {author} {\bibfnamefont
  {A.}~\bibnamefont {Aksentev}}, \bibinfo {author} {\bibfnamefont
  {B.}~\bibnamefont {Alberdi}}, \bibinfo {author} {\bibfnamefont
  {L.}~\bibnamefont {Barion}}, \bibinfo {author} {\bibfnamefont
  {I.}~\bibnamefont {Bekman}}, \bibinfo {author} {\bibfnamefont
  {M.}~\bibnamefont {Bey{\ss}}}, \bibinfo {author} {\bibfnamefont
  {C.}~\bibnamefont {B{\"o}hme}}, \bibinfo {author} {\bibfnamefont
  {B.}~\bibnamefont {Breitkreutz}}, \bibinfo {author} {\bibfnamefont
  {N.}~\bibnamefont {Canale}}, \bibinfo {author} {\bibfnamefont
  {G.}~\bibnamefont {Ciullo}}, \bibinfo {author} {\bibfnamefont
  {S.}~\bibnamefont {Dymov}}, \bibinfo {author} {\bibfnamefont {N.-O.}\
  \bibnamefont {Fr{\"o}hlich}}, \bibinfo {author} {\bibfnamefont
  {R.}~\bibnamefont {Gebel}}, \bibinfo {author} {\bibfnamefont
  {M.}~\bibnamefont {Gaisser}}, \bibinfo {author} {\bibfnamefont
  {K.}~\bibnamefont {Grigoryev}}, \bibinfo {author} {\bibfnamefont
  {D.}~\bibnamefont {Grzonka}}, \bibinfo {author} {\bibfnamefont
  {J.}~\bibnamefont {Hetzel}}, \bibinfo {author} {\bibfnamefont
  {O.}~\bibnamefont {Javakhishvili}}, \bibinfo {author} {\bibfnamefont
  {V.}~\bibnamefont {Kamerdzhiev}}, \bibinfo {author} {\bibfnamefont
  {S.}~\bibnamefont {Karanth}}, \bibinfo {author} {\bibfnamefont
  {I.}~\bibnamefont {Keshelashvili}}, \bibinfo {author} {\bibfnamefont
  {A.}~\bibnamefont {Kononov}}, \bibinfo {author} {\bibfnamefont
  {K.}~\bibnamefont {Laihem}}, \bibinfo {author} {\bibfnamefont
  {A.}~\bibnamefont {Lehrach}}, \bibinfo {author} {\bibfnamefont
  {N.}~\bibnamefont {Lomidze}}, \bibinfo {author} {\bibfnamefont
  {B.}~\bibnamefont {Lorentz}}, \bibinfo {author} {\bibfnamefont
  {G.}~\bibnamefont {Macharashvili}}, \bibinfo {author} {\bibfnamefont
  {A.}~\bibnamefont {Magiera}}, \bibinfo {author} {\bibfnamefont
  {D.}~\bibnamefont {Mchedlishvili}}, \bibinfo {author} {\bibfnamefont
  {A.}~\bibnamefont {Melnikov}}, \bibinfo {author} {\bibfnamefont
  {F.}~\bibnamefont {M{\"u}ller}}, \bibinfo {author} {\bibfnamefont
  {A.}~\bibnamefont {Pesce}}, \bibinfo {author} {\bibfnamefont
  {V.}~\bibnamefont {Poncza}}, \bibinfo {author} {\bibfnamefont
  {D.}~\bibnamefont {Prasuhn}}, \bibinfo {author} {\bibfnamefont
  {D.}~\bibnamefont {Shergelashvili}}, \bibinfo {author} {\bibfnamefont
  {N.}~\bibnamefont {Shurkhno}}, \bibinfo {author} {\bibfnamefont
  {S.}~\bibnamefont {Siddique}}, \bibinfo {author} {\bibfnamefont
  {A.}~\bibnamefont {Silenko}}, \bibinfo {author} {\bibfnamefont
  {S.}~\bibnamefont {Stassen}}, \bibinfo {author} {\bibfnamefont {E.~J.}\
  \bibnamefont {Stephenson}}, \bibinfo {author} {\bibfnamefont
  {H.}~\bibnamefont {Str{\"o}her}}, \bibinfo {author} {\bibfnamefont
  {M.}~\bibnamefont {Tabidze}}, \bibinfo {author} {\bibfnamefont
  {G.}~\bibnamefont {Tagliente}}, \bibinfo {author} {\bibfnamefont
  {Y.}~\bibnamefont {Valdau}}, \bibinfo {author} {\bibfnamefont
  {M.}~\bibnamefont {Vitz}}, \bibinfo {author} {\bibfnamefont {T.}~\bibnamefont
  {Wagner}}, \bibinfo {author} {\bibfnamefont {A.}~\bibnamefont {Wirzba}},
  \bibinfo {author} {\bibfnamefont {A.}~\bibnamefont {Wro{\'n}ska}}, \bibinfo
  {author} {\bibfnamefont {P.}~\bibnamefont {W{\"u}stner}},\ and\ \bibinfo
  {author} {\bibfnamefont {M.}~\bibnamefont {{\.Z}urek}},\ }\bibfield  {title}
  {\bibinfo {title} {Spin decoherence and off-resonance behavior of
  radio-frequency-driven spin rotations in storage rings},\ }\href
  {https://doi.org/10.1103/PhysRevAccelBeams.27.111002} {\bibfield  {journal}
  {\bibinfo  {journal} {Physical Review Accelerators and Beams}\ }\textbf
  {\bibinfo {volume} {27}},\ \bibinfo {pages} {111002} (\bibinfo {year}
  {2024})}\BibitemShut {NoStop}%
\bibitem [{\citenamefont {Abusaif}\ \emph {et~al.}(2021)\citenamefont
  {Abusaif}, \citenamefont {Keshelashvili}, \citenamefont {Grigoryev},
  \citenamefont {Mchedlishvili}, \citenamefont {Jorat}, \citenamefont {Pretz},
  \citenamefont {Mei{\ss}ner}, \citenamefont {Kulikov}, \citenamefont {Stahl},
  \citenamefont {Felden}, \citenamefont {Martin}, \citenamefont {Stassen},
  \citenamefont {W{\"u}stner}, \citenamefont {Aksentev}, \citenamefont
  {Javakhishvili}, \citenamefont {Valetov}, \citenamefont {Soltner},
  \citenamefont {{Alberdi-Esuain}}, \citenamefont {Talman}, \citenamefont
  {Shmakova}, \citenamefont {Kacharava}, \citenamefont {H{\"o}lscher},
  \citenamefont {Ciullo}, \citenamefont {M{\"u}ller}, \citenamefont {Silenko},
  \citenamefont {Gebel}, \citenamefont {Lorentz}, \citenamefont {Natour},
  \citenamefont {Grzonka}, \citenamefont {Zurek.}, \citenamefont {Hetzel},
  \citenamefont {Zupranski}, \citenamefont {Siddique}, \citenamefont {Magiera},
  \citenamefont {Ciepa{\l}}, \citenamefont {B{\"o}hme}, \citenamefont
  {Nikolaev}, \citenamefont {Dymov}, \citenamefont {Lehrach}, \citenamefont
  {Gaisser}, \citenamefont {Wro{\'n}ska}, \citenamefont {Heberling},
  \citenamefont {Contalbrigo}, \citenamefont {Kamerdzhiev}, \citenamefont
  {Stephenson}, \citenamefont {Nass}, \citenamefont {Weidemann}, \citenamefont
  {Senichev}, \citenamefont {Schott}, \citenamefont {Ritman}, \citenamefont
  {Wirzba}, \citenamefont {Bey{\ss}}, \citenamefont {Haj~Tahar}, \citenamefont
  {Str{\"o}her}, \citenamefont {Koop}, \citenamefont {Carli}, \citenamefont
  {Lamont}, \citenamefont {Berz}, \citenamefont {B{\"o}ker}, \citenamefont
  {Pesce}, \citenamefont {Tagliente}, \citenamefont {K{\"a}seberg},
  \citenamefont {Saleev}, \citenamefont {Wagner}, \citenamefont {Makino},
  \citenamefont {Aggarwal}, \citenamefont {Slim}, \citenamefont {Prasuhn},
  \citenamefont {Macharashvili}, \citenamefont {Borburgh}, \citenamefont
  {Lomidze}, \citenamefont {Poncza}, \citenamefont {Shergelashvili},
  \citenamefont {Gagoshidze}, \citenamefont {Tabidze}, \citenamefont {Karanth},
  \citenamefont {Laihem}, \citenamefont {Hahnraths}, \citenamefont {Lenisa},
  \citenamefont {Hejny}, \citenamefont {Giese}, \citenamefont {Sefzick},
  \citenamefont {Barion}, \citenamefont {Michaud}, \citenamefont {Nogga},
  \citenamefont {Straatmann}, \citenamefont {Uzikov}, \citenamefont {Basile},
  \citenamefont {Rathmann}, \citenamefont {Atanasov}, \citenamefont
  {Metreveli}, \citenamefont {Rosenthal}, \citenamefont {Valdau}, \citenamefont
  {Simon},\ and\ \citenamefont {De~Conto}}]{abusaif2021}%
  \BibitemOpen
  \bibfield  {author} {\bibinfo {author} {\bibfnamefont {F.}~\bibnamefont
  {Abusaif}}, \bibinfo {author} {\bibfnamefont {I.}~\bibnamefont
  {Keshelashvili}}, \bibinfo {author} {\bibfnamefont {K.}~\bibnamefont
  {Grigoryev}}, \bibinfo {author} {\bibfnamefont {D.}~\bibnamefont
  {Mchedlishvili}}, \bibinfo {author} {\bibfnamefont {L.}~\bibnamefont
  {Jorat}}, \bibinfo {author} {\bibfnamefont {J.}~\bibnamefont {Pretz}},
  \bibinfo {author} {\bibfnamefont {U.-G.}\ \bibnamefont {Mei{\ss}ner}},
  \bibinfo {author} {\bibfnamefont {A.}~\bibnamefont {Kulikov}}, \bibinfo
  {author} {\bibfnamefont {A.}~\bibnamefont {Stahl}}, \bibinfo {author}
  {\bibfnamefont {O.}~\bibnamefont {Felden}}, \bibinfo {author} {\bibfnamefont
  {S.}~\bibnamefont {Martin}}, \bibinfo {author} {\bibfnamefont
  {R.}~\bibnamefont {Stassen}}, \bibinfo {author} {\bibfnamefont
  {P.}~\bibnamefont {W{\"u}stner}}, \bibinfo {author} {\bibfnamefont
  {A.}~\bibnamefont {Aksentev}}, \bibinfo {author} {\bibfnamefont
  {O.}~\bibnamefont {Javakhishvili}}, \bibinfo {author} {\bibfnamefont
  {E.}~\bibnamefont {Valetov}}, \bibinfo {author} {\bibfnamefont
  {H.}~\bibnamefont {Soltner}}, \bibinfo {author} {\bibfnamefont
  {B.}~\bibnamefont {{Alberdi-Esuain}}}, \bibinfo {author} {\bibfnamefont
  {R.}~\bibnamefont {Talman}}, \bibinfo {author} {\bibfnamefont
  {V.}~\bibnamefont {Shmakova}}, \bibinfo {author} {\bibfnamefont
  {A.}~\bibnamefont {Kacharava}}, \bibinfo {author} {\bibfnamefont
  {D.}~\bibnamefont {H{\"o}lscher}}, \bibinfo {author} {\bibfnamefont
  {G.}~\bibnamefont {Ciullo}}, \bibinfo {author} {\bibfnamefont
  {F.}~\bibnamefont {M{\"u}ller}}, \bibinfo {author} {\bibfnamefont
  {A.}~\bibnamefont {Silenko}}, \bibinfo {author} {\bibfnamefont
  {R.}~\bibnamefont {Gebel}}, \bibinfo {author} {\bibfnamefont
  {B.}~\bibnamefont {Lorentz}}, \bibinfo {author} {\bibfnamefont
  {G.}~\bibnamefont {Natour}}, \bibinfo {author} {\bibfnamefont
  {D.}~\bibnamefont {Grzonka}}, \bibinfo {author} {\bibfnamefont
  {M.}~\bibnamefont {Zurek.}}, \bibinfo {author} {\bibfnamefont
  {J.}~\bibnamefont {Hetzel}}, \bibinfo {author} {\bibfnamefont
  {P.}~\bibnamefont {Zupranski}}, \bibinfo {author} {\bibfnamefont
  {S.}~\bibnamefont {Siddique}}, \bibinfo {author} {\bibfnamefont
  {A.}~\bibnamefont {Magiera}}, \bibinfo {author} {\bibfnamefont
  {I.}~\bibnamefont {Ciepa{\l}}}, \bibinfo {author} {\bibfnamefont
  {C.}~\bibnamefont {B{\"o}hme}}, \bibinfo {author} {\bibfnamefont
  {N.}~\bibnamefont {Nikolaev}}, \bibinfo {author} {\bibfnamefont
  {S.}~\bibnamefont {Dymov}}, \bibinfo {author} {\bibfnamefont
  {A.}~\bibnamefont {Lehrach}}, \bibinfo {author} {\bibfnamefont
  {M.}~\bibnamefont {Gaisser}}, \bibinfo {author} {\bibfnamefont
  {A.}~\bibnamefont {Wro{\'n}ska}}, \bibinfo {author} {\bibfnamefont
  {D.}~\bibnamefont {Heberling}}, \bibinfo {author} {\bibfnamefont
  {M.}~\bibnamefont {Contalbrigo}}, \bibinfo {author} {\bibfnamefont
  {V.}~\bibnamefont {Kamerdzhiev}}, \bibinfo {author} {\bibfnamefont
  {E.}~\bibnamefont {Stephenson}}, \bibinfo {author} {\bibfnamefont
  {A.}~\bibnamefont {Nass}}, \bibinfo {author} {\bibfnamefont {C.}~\bibnamefont
  {Weidemann}}, \bibinfo {author} {\bibfnamefont {Y.}~\bibnamefont {Senichev}},
  \bibinfo {author} {\bibfnamefont {M.}~\bibnamefont {Schott}}, \bibinfo
  {author} {\bibfnamefont {J.}~\bibnamefont {Ritman}}, \bibinfo {author}
  {\bibfnamefont {A.}~\bibnamefont {Wirzba}}, \bibinfo {author} {\bibfnamefont
  {M.}~\bibnamefont {Bey{\ss}}}, \bibinfo {author} {\bibfnamefont
  {M.}~\bibnamefont {Haj~Tahar}}, \bibinfo {author} {\bibfnamefont
  {H.}~\bibnamefont {Str{\"o}her}}, \bibinfo {author} {\bibfnamefont
  {I.}~\bibnamefont {Koop}}, \bibinfo {author} {\bibfnamefont {C.}~\bibnamefont
  {Carli}}, \bibinfo {author} {\bibfnamefont {M.}~\bibnamefont {Lamont}},
  \bibinfo {author} {\bibfnamefont {M.}~\bibnamefont {Berz}}, \bibinfo {author}
  {\bibfnamefont {J.}~\bibnamefont {B{\"o}ker}}, \bibinfo {author}
  {\bibfnamefont {A.}~\bibnamefont {Pesce}}, \bibinfo {author} {\bibfnamefont
  {G.}~\bibnamefont {Tagliente}}, \bibinfo {author} {\bibfnamefont
  {C.}~\bibnamefont {K{\"a}seberg}}, \bibinfo {author} {\bibfnamefont
  {A.}~\bibnamefont {Saleev}}, \bibinfo {author} {\bibfnamefont
  {T.}~\bibnamefont {Wagner}}, \bibinfo {author} {\bibfnamefont
  {K.}~\bibnamefont {Makino}}, \bibinfo {author} {\bibfnamefont
  {A.}~\bibnamefont {Aggarwal}}, \bibinfo {author} {\bibfnamefont
  {J.}~\bibnamefont {Slim}}, \bibinfo {author} {\bibfnamefont {D.}~\bibnamefont
  {Prasuhn}}, \bibinfo {author} {\bibfnamefont {G.}~\bibnamefont
  {Macharashvili}}, \bibinfo {author} {\bibfnamefont {J.}~\bibnamefont
  {Borburgh}}, \bibinfo {author} {\bibfnamefont {N.}~\bibnamefont {Lomidze}},
  \bibinfo {author} {\bibfnamefont {V.}~\bibnamefont {Poncza}}, \bibinfo
  {author} {\bibfnamefont {D.}~\bibnamefont {Shergelashvili}}, \bibinfo
  {author} {\bibfnamefont {M.}~\bibnamefont {Gagoshidze}}, \bibinfo {author}
  {\bibfnamefont {M.}~\bibnamefont {Tabidze}}, \bibinfo {author} {\bibfnamefont
  {S.}~\bibnamefont {Karanth}}, \bibinfo {author} {\bibfnamefont
  {K.}~\bibnamefont {Laihem}}, \bibinfo {author} {\bibfnamefont
  {T.}~\bibnamefont {Hahnraths}}, \bibinfo {author} {\bibfnamefont
  {P.}~\bibnamefont {Lenisa}}, \bibinfo {author} {\bibfnamefont
  {V.}~\bibnamefont {Hejny}}, \bibinfo {author} {\bibfnamefont
  {N.}~\bibnamefont {Giese}}, \bibinfo {author} {\bibfnamefont
  {T.}~\bibnamefont {Sefzick}}, \bibinfo {author} {\bibfnamefont
  {L.}~\bibnamefont {Barion}}, \bibinfo {author} {\bibfnamefont
  {J.}~\bibnamefont {Michaud}}, \bibinfo {author} {\bibfnamefont
  {A.}~\bibnamefont {Nogga}}, \bibinfo {author} {\bibfnamefont
  {H.}~\bibnamefont {Straatmann}}, \bibinfo {author} {\bibfnamefont
  {Y.}~\bibnamefont {Uzikov}}, \bibinfo {author} {\bibfnamefont
  {S.}~\bibnamefont {Basile}}, \bibinfo {author} {\bibfnamefont
  {F.}~\bibnamefont {Rathmann}}, \bibinfo {author} {\bibfnamefont
  {A.}~\bibnamefont {Atanasov}}, \bibinfo {author} {\bibfnamefont
  {Z.}~\bibnamefont {Metreveli}}, \bibinfo {author} {\bibfnamefont
  {M.}~\bibnamefont {Rosenthal}}, \bibinfo {author} {\bibfnamefont
  {Y.}~\bibnamefont {Valdau}}, \bibinfo {author} {\bibfnamefont
  {M.}~\bibnamefont {Simon}},\ and\ \bibinfo {author} {\bibfnamefont {J.-M.}\
  \bibnamefont {De~Conto}},\ }\href@noop {} {\emph {\bibinfo {title} {Storage
  Ring to Search for Electric Dipole Moments of Charged Particles:
  {{Feasibility}} Study}}},\ \bibinfo {number} {CERN-2021-003}\ (\bibinfo
  {publisher} {CERN},\ \bibinfo {year} {2021})\BibitemShut {NoStop}%
\bibitem [{\citenamefont {{JEDI Collaboration}}\ \emph
  {et~al.}(2023)\citenamefont {{JEDI Collaboration}}, \citenamefont {Karanth},
  \citenamefont {Stephenson}, \citenamefont {Chang}, \citenamefont {Hejny},
  \citenamefont {Park}, \citenamefont {Pretz}, \citenamefont {Semertzidis},
  \citenamefont {Wirzba}, \citenamefont {Wro{\'n}ska}, \citenamefont {Abusaif},
  \citenamefont {Aggarwal}, \citenamefont {Aksentev}, \citenamefont {Alberdi},
  \citenamefont {Andres}, \citenamefont {Barion}, \citenamefont {Bekman},
  \citenamefont {Bey{\ss}}, \citenamefont {B{\"o}hme}, \citenamefont
  {Breitkreutz}, \citenamefont {{von Byern}}, \citenamefont {Canale},
  \citenamefont {Ciullo}, \citenamefont {Dymov}, \citenamefont {Fr{\"o}hlich},
  \citenamefont {Gebel}, \citenamefont {Grigoryev}, \citenamefont {Grzonka},
  \citenamefont {Hetzel}, \citenamefont {Javakhishvili}, \citenamefont {Jeong},
  \citenamefont {Kacharava}, \citenamefont {Kamerdzhiev}, \citenamefont
  {Keshelashvili}, \citenamefont {Kononov}, \citenamefont {Laihem},
  \citenamefont {Lehrach}, \citenamefont {Lenisa}, \citenamefont {Lomidze},
  \citenamefont {Lorentz}, \citenamefont {Magiera}, \citenamefont
  {Mchedlishvili}, \citenamefont {M{\"u}ller}, \citenamefont {Nass},
  \citenamefont {Nikolaev}, \citenamefont {Pesce}, \citenamefont {Poncza},
  \citenamefont {Prasuhn}, \citenamefont {Rathmann}, \citenamefont {Saleev},
  \citenamefont {Shergelashvili}, \citenamefont {Shmakova}, \citenamefont
  {Shurkhno}, \citenamefont {Siddique}, \citenamefont {Slim}, \citenamefont
  {Soltner}, \citenamefont {Stassen}, \citenamefont {Str{\"o}her},
  \citenamefont {Tabidze}, \citenamefont {Tagliente}, \citenamefont {Valdau},
  \citenamefont {Vitz}, \citenamefont {Wagner},\ and\ \citenamefont
  {W{\"u}stner}}]{JEDI:2022hxaa}%
  \BibitemOpen
  \bibfield  {author} {\bibinfo {author} {\bibnamefont {{JEDI Collaboration}}},
  \bibinfo {author} {\bibfnamefont {S.}~\bibnamefont {Karanth}}, \bibinfo
  {author} {\bibfnamefont {E.~J.}\ \bibnamefont {Stephenson}}, \bibinfo
  {author} {\bibfnamefont {S.~P.}\ \bibnamefont {Chang}}, \bibinfo {author}
  {\bibfnamefont {V.}~\bibnamefont {Hejny}}, \bibinfo {author} {\bibfnamefont
  {S.}~\bibnamefont {Park}}, \bibinfo {author} {\bibfnamefont {J.}~\bibnamefont
  {Pretz}}, \bibinfo {author} {\bibfnamefont {Y.~K.}\ \bibnamefont
  {Semertzidis}}, \bibinfo {author} {\bibfnamefont {A.}~\bibnamefont {Wirzba}},
  \bibinfo {author} {\bibfnamefont {A.}~\bibnamefont {Wro{\'n}ska}}, \bibinfo
  {author} {\bibfnamefont {F.}~\bibnamefont {Abusaif}}, \bibinfo {author}
  {\bibfnamefont {A.}~\bibnamefont {Aggarwal}}, \bibinfo {author}
  {\bibfnamefont {A.}~\bibnamefont {Aksentev}}, \bibinfo {author}
  {\bibfnamefont {B.}~\bibnamefont {Alberdi}}, \bibinfo {author} {\bibfnamefont
  {A.}~\bibnamefont {Andres}}, \bibinfo {author} {\bibfnamefont
  {L.}~\bibnamefont {Barion}}, \bibinfo {author} {\bibfnamefont
  {I.}~\bibnamefont {Bekman}}, \bibinfo {author} {\bibfnamefont
  {M.}~\bibnamefont {Bey{\ss}}}, \bibinfo {author} {\bibfnamefont
  {C.}~\bibnamefont {B{\"o}hme}}, \bibinfo {author} {\bibfnamefont
  {B.}~\bibnamefont {Breitkreutz}}, \bibinfo {author} {\bibfnamefont
  {C.}~\bibnamefont {{von Byern}}}, \bibinfo {author} {\bibfnamefont
  {N.}~\bibnamefont {Canale}}, \bibinfo {author} {\bibfnamefont
  {G.}~\bibnamefont {Ciullo}}, \bibinfo {author} {\bibfnamefont
  {S.}~\bibnamefont {Dymov}}, \bibinfo {author} {\bibfnamefont {N.-O.}\
  \bibnamefont {Fr{\"o}hlich}}, \bibinfo {author} {\bibfnamefont
  {R.}~\bibnamefont {Gebel}}, \bibinfo {author} {\bibfnamefont
  {K.}~\bibnamefont {Grigoryev}}, \bibinfo {author} {\bibfnamefont
  {D.}~\bibnamefont {Grzonka}}, \bibinfo {author} {\bibfnamefont
  {J.}~\bibnamefont {Hetzel}}, \bibinfo {author} {\bibfnamefont
  {O.}~\bibnamefont {Javakhishvili}}, \bibinfo {author} {\bibfnamefont
  {H.}~\bibnamefont {Jeong}}, \bibinfo {author} {\bibfnamefont
  {A.}~\bibnamefont {Kacharava}}, \bibinfo {author} {\bibfnamefont
  {V.}~\bibnamefont {Kamerdzhiev}}, \bibinfo {author} {\bibfnamefont
  {I.}~\bibnamefont {Keshelashvili}}, \bibinfo {author} {\bibfnamefont
  {A.}~\bibnamefont {Kononov}}, \bibinfo {author} {\bibfnamefont
  {K.}~\bibnamefont {Laihem}}, \bibinfo {author} {\bibfnamefont
  {A.}~\bibnamefont {Lehrach}}, \bibinfo {author} {\bibfnamefont
  {P.}~\bibnamefont {Lenisa}}, \bibinfo {author} {\bibfnamefont
  {N.}~\bibnamefont {Lomidze}}, \bibinfo {author} {\bibfnamefont
  {B.}~\bibnamefont {Lorentz}}, \bibinfo {author} {\bibfnamefont
  {A.}~\bibnamefont {Magiera}}, \bibinfo {author} {\bibfnamefont
  {D.}~\bibnamefont {Mchedlishvili}}, \bibinfo {author} {\bibfnamefont
  {F.}~\bibnamefont {M{\"u}ller}}, \bibinfo {author} {\bibfnamefont
  {A.}~\bibnamefont {Nass}}, \bibinfo {author} {\bibfnamefont {N.~N.}\
  \bibnamefont {Nikolaev}}, \bibinfo {author} {\bibfnamefont {A.}~\bibnamefont
  {Pesce}}, \bibinfo {author} {\bibfnamefont {V.}~\bibnamefont {Poncza}},
  \bibinfo {author} {\bibfnamefont {D.}~\bibnamefont {Prasuhn}}, \bibinfo
  {author} {\bibfnamefont {F.}~\bibnamefont {Rathmann}}, \bibinfo {author}
  {\bibfnamefont {A.}~\bibnamefont {Saleev}}, \bibinfo {author} {\bibfnamefont
  {D.}~\bibnamefont {Shergelashvili}}, \bibinfo {author} {\bibfnamefont
  {V.}~\bibnamefont {Shmakova}}, \bibinfo {author} {\bibfnamefont
  {N.}~\bibnamefont {Shurkhno}}, \bibinfo {author} {\bibfnamefont
  {S.}~\bibnamefont {Siddique}}, \bibinfo {author} {\bibfnamefont
  {J.}~\bibnamefont {Slim}}, \bibinfo {author} {\bibfnamefont {H.}~\bibnamefont
  {Soltner}}, \bibinfo {author} {\bibfnamefont {R.}~\bibnamefont {Stassen}},
  \bibinfo {author} {\bibfnamefont {H.}~\bibnamefont {Str{\"o}her}}, \bibinfo
  {author} {\bibfnamefont {M.}~\bibnamefont {Tabidze}}, \bibinfo {author}
  {\bibfnamefont {G.}~\bibnamefont {Tagliente}}, \bibinfo {author}
  {\bibfnamefont {Y.}~\bibnamefont {Valdau}}, \bibinfo {author} {\bibfnamefont
  {M.}~\bibnamefont {Vitz}}, \bibinfo {author} {\bibfnamefont {T.}~\bibnamefont
  {Wagner}},\ and\ \bibinfo {author} {\bibfnamefont {P.}~\bibnamefont
  {W{\"u}stner}},\ }\bibfield  {title} {\bibinfo {title} {First {{Search}} for
  {{Axionlike Particles}} in a {{Storage Ring Using}} a {{Polarized Deuteron
  Beam}}},\ }\href {https://doi.org/10.1103/PhysRevX.13.031004} {\bibfield
  {journal} {\bibinfo  {journal} {Physical Review X}\ }\textbf {\bibinfo
  {volume} {13}},\ \bibinfo {pages} {031004} (\bibinfo {year}
  {2023})}\BibitemShut {NoStop}%
\end{thebibliography}%

\end{document}